\documentclass{emulateapj}
\usepackage{apjfonts}
\usepackage{graphicx}
\usepackage{amsmath}

\def\gtsim {\lower .1ex\hbox{\rlap{\raise .6ex\hbox{\hskip .3ex
        {\ifmmode{\scriptscriptstyle >}\else
                {$\scriptscriptstyle >$}\fi}}}
        \kern -.4ex{\ifmmode{\scriptscriptstyle \sim}\else
                {$\scriptscriptstyle\sim$}\fi}}}
\newcommand{\be}{\begin{equation}}
\newcommand{\ee}{\end{equation}}
\newcommand{\beqa}{\begin{eqnarray}}
\newcommand{\eeqa}{\end{eqnarray}}

\def\Mo{{\rm M_\odot}}
\newcommand{\hMpc}{\ h^{-1}\rm{Mpc}}
\newcommand{\hMsun}{\ h^{-1}\rm{M}_{\odot}}
\newcommand{\hkpc}{\ h^{-1}\rm{kpc}}

\def\kpc{\ {\rm kpc}}
\def\pc{\ {\rm pc}}

\def\kms{{\ }{\rm km}\,{\rm s}^{-1}}
\def\LCDM{$\Lambda$CDM}
\def\degrees{^\circ}

\begin{document}
\submitted{The Astrophysical Journal, submitted}
\vspace{1mm}
\slugcomment{{\it The Astrophysical Journal, submitted}}

\shortauthors{KAZANTZIDIS ET AL.}
\shorttitle{CDM Substructure and Galactic Disks II:}

\title{Cold Dark Matter Substructure and Galactic Disks II: \\  
Dynamical Effects of Hierarchical Satellite Accretion}

\author{Stelios Kazantzidis,\altaffilmark{1}
        Andrew R. Zentner,\altaffilmark{2}
        Andrey V. Kravtsov,\altaffilmark{3}\\
        James S. Bullock,\altaffilmark{4}
        and Victor P. Debattista\altaffilmark{5,6}}

\vspace{2mm}

\begin{abstract}

 We perform a set of fully self-consistent, dissipationless $N$-body
 simulations to elucidate the dynamical response of thin galactic disks
 to bombardment by cold dark matter (CDM) substructure. Our method combines 
 (1) cosmological simulations of the formation of Milky Way (MW)-sized CDM  
 halos to derive the properties of substructure and (2) 
 controlled numerical experiments of consecutive subhalo impacts 
 onto an initially-thin, fully-formed MW type disk galaxy. 
 The present study is the first to account for the evolution of satellite
 populations over cosmic time in such an investigation of disk 
 structure. In contrast to what can be inferred from statistics of 
 the $z=0$ surviving substructure, we find that accretions of massive 
 subhalos onto the central regions of host halos, where the 
 galactic disks reside, since $z \sim 1$ should be common.
 One host halo accretion history is used to initialize the controlled simulations 
 of satellite-disk encounters. The specific merger history involves six dark 
 matter substructures, with initial masses in the range $\sim 20\%-60\%$ of the 
 disk mass and of comparable size to the disk, crossing the central regions 
 of their host host in the past $\sim 8$~Gyr. We show that these accretion events 
 severely perturb the thin galactic disk and produce a wealth of distinctive 
 dynamical signatures on its structure and kinematics.
 These include (1) considerable thickening and heating at
 all radii, with the disk thickness and velocity ellipsoid nearly doubling 
 at the solar radius; (2) prominent flaring associated with an increase in 
 disk thickness greater than a factor of $4$ in the disk outskirts;
 (3) surface density excesses at large radii, beyond $\sim 5$ disk scale lengths, 
 resembling those of observed antitruncated disks; (4) long-lived,
 lopsidedness at levels similar to those measured in observational samples 
 of disk galaxies; and (5) substantial tilting. The interaction with the most 
 massive subhalo in the simulated accretion history drives the disk response while
 subsequent bombardment is much less efficient at disturbing the disk. 
 We also explore a variety of disk and satellite properties that influence 
 these responses. We conclude that substructure-disk encounters of the kind expected 
 in the {\LCDM} paradigm play a significant role in setting the structure of disk
 galaxies and driving galaxy evolution. 

\end{abstract}


\altaffiltext{1}{Center for Cosmology and Astro-Particle Physics; 
        and Department of Physics; and Department of Astronomy, 
        The Ohio State University, 191 West Woodruff Avenue, Columbus, 
        OH 43210 USA; {\tt stelios@mps.ohio-state.edu}.}
        \altaffiltext{2}{Department of Physics and Astronomy,
        University of Pittsburgh, 100 Allen Hall, 3941 O'Hara Street,
        Pittsburgh, PA 15260 USA; {\tt zentner@pitt.edu}.}
        \altaffiltext{3}{Kavli Institute for Cosmological Physics; 
        and Department of Astronomy \& Astrophysics; 
        and the Enrico Fermi Institute, The University of Chicago, 5640 South
        Ellis Avenue, Chicago, IL 60637 USA; {\tt andrey@oddjob.uchicago.edu}.}
        \altaffiltext{4}{Center for Cosmology;
        and Department of Physics \& Astronomy, The University of California at Irvine, 
        4168 Reines Hall, Irvine, CA 92697 USA; {\tt bullock@uci.edu}.}
        \altaffiltext{5}{Center For Astrophysics, University of Central
        Lancashire, Preston PR1 2HE, UK; {\tt vpdebattista@uclan.ac.uk}.}
        \altaffiltext{6}{RCUK Fellow}

\section{Introduction}
\label{section:introduction}

Hierarchical models of cosmological structure formation, such as the currently
favored cold dark matter (CDM) paradigm \citep[e.g.,][]{White_Rees78,Blumenthal_etal84}, 
generically predict substantial amounts of substructure in the form of small, 
dense, self-bound {\it subhalos} orbiting within the virialized regions 
of larger host halos \citep[e.g.,][]{Ghigna_etal98,Tormen_etal98,Moore_etal99,
Klypin_etal99b,Ghigna_etal00}. Observational probes of substructure abundance thus constitute 
fundamental tests of the CDM model. Recently, a growing body of evidence has confirmed 
the hierarchical build-up of galaxy-sized dark matter halos with the discovery 
of tidal streams and complex stellar structures in the Milky Way (MW) 
\citep[e.g.,][]{Ibata_etal94,Yanny_etal00,Ibata_etal01a,Newberg_etal02,Majewski_etal03,Martin_etal04,
Martinez-Delgado_etal05,Grillmair_Dionatos06,Belokurov_etal06},
the Andromeda galaxy (M31) \citep{Ibata_etal01b,Ferguson_etal02,Ferguson_etal05,
Kalirai_etal06,Ibata_etal07}, and beyond the Local Group 
\citep[e.g.,][]{Malin_Hadley97,Shang_etal98,Peng_etal02,Forbes_etal03,Pohlen_etal04}.

A significant fraction of observed galaxies have disk-dominant morphology with 
roughly $70\%$ of Galaxy-sized dark matter halos hosting late-type systems \citep[e.g.,][]
{Weinmann_etal06,Choi_etal07}. Owing to the lack of a significant luminous
component associated with most subhalos in Galaxy-sized host halos 
\citep[e.g.,][]{Klypin_etal99b,Moore_etal99}, information regarding the 
amount of substructure in these systems may be obtained via its 
gravitational influence on galaxy disks. Despite the small contribution of 
substructure to the total mass of the host \citep[e.g.,][]{Ghigna_etal00}, 
a considerable number of subhalos is expected within the virialized region 
of a Galaxy-sized CDM halo at any given epoch. If a large population of satellites 
exists, it may tidally disturb the host galactic disk, possibly leading to the 
imprint of distinctive dynamical signatures on its structure and kinematics. 

Theoretical studies set within the CDM paradigm have convincingly shown 
that the accretion of massive substructures is commonplace 
during the formation of Galaxy-sized halos \citep[e.g.,][]{Lacey_Cole93, 
Zentner_Bullock03,Purcell_etal07,Stewart_etal08} and that typical subhalo 
orbits are highly eccentric \citep[e.g.,][]{Ghigna_etal98,Tormen_etal98,
Zentner_etal05a,Benson05}. These facts suggest that passages of massive 
satellites near the center of the host potential, where the galactic disks
reside, should be common in CDM models. Such 
accretion events are expected to perturb the fragile circular 
orbits of disk stars by depositing large amounts of orbital energy into 
random stellar motions, gradually heating the disk and increasing its 
scale height. 

Yet, many disk galaxies are observed to be cold and thin, with average 
axial ratios of radial scale lengths to vertical scale heights, 
$R_{d}/z_{d}$, in the range $\sim 4-5$ \citep[e.g.,][]{de_Grijs98,Bizyaev_Mitrova02,
Kregel_etal02,Yoachim_Dalcanton06}. In addition, recent studies of edge-on 
disk galaxies using the Sloan Digital Sky Survey (SDSS) database have 
revealed a notable fraction of ``super-thin'' bulgeless disks with 
much larger axial ratios \citep{Kautsch_etal06}. 

In the case of the MW, measurements of the disk scale height 
at the solar radius obtained using a variety of methods, including star 
counts and mass modeling, indicate that the Galaxy comprises a thin, stellar disk 
with an exponential scale height of $h_z \simeq 300 \pm 50$~pc 
\citep[e.g.,][]{Kent_etal91,Dehnen_Binney98,
Mendez_Guzman98,Larsen_Humphreys03,Widrow_Dubinski05,Juric_etal08}. 
Furthermore, the age-velocity dispersion relation of 
disk stars in the solar neighborhood suggests that a significant fraction
of the thin disk of the MW was in place by $z \sim 1$ \citep[e.g.,][]{Wyse01,
Quillen_Garnet01,Nordstrom_etal04,Seabroke_Gilmore07}. 
The existence of such an old, thin stellar disk, as established by the 
age distribution of disk stars, may imply an absence of satellite 
accretion events over the past $\sim 8$~Gyr. Such an extended period of 
quiescent dynamical evolution is difficult to reconcile with the 
hierarchical assembly of structure prescribed by the {\LCDM} cosmological 
model. Although hydrodynamical simulations of disk galaxy formation have offered 
some insights into accommodating observational facts that challenge the CDM paradigm 
\citep[e.g.,][]{Abadi_etal03}, the detailed dynamical response of galactic 
disks to halo substructure in a cosmological context remains poorly
understood. 

Significant theoretical effort, including both semi-analytic modeling
\citep{Toth_Ostriker92,Benson_etal04,Hopkins_etal08} and  numerical simulations
\citep{Quinn_Goodman86,Quinn_etal93,Walker_etal96,
Huang_Carlberg97,Sellwood_etal98,Velazquez_White99,Font_etal01,Ardi_etal03,Gauthier_etal06,
Hayashi_Chiba06,Read_etal08,Villalobos_Helmi08,Purcell_etal08} has been devoted 
to quantifying the resilience of galactic disks to infalling satellites.
Although valuable in several respects, these earlier investigations could 
not capture fully the amount of global dynamical evolution induced in {\it
  thin} disk galaxies by substructure in the context of the {\LCDM} model.
While a detailed comparison to previous work 
is presented in \S~\ref{sec:comparison}, we mention at the outset that each of
the aforementioned numerical studies suffered from at least one critical
shortcoming by either: 
(1) not being fully self-consistent, modeling various components of the 
primary disk galaxy and/or the satellites as rigid potentials, 
a choice which leads to overestimating the damage caused to the disk; 
(2) focusing on experiments with infalling systems
on nearly circular orbits that are poor approximations of the highly
eccentric orbits typical of CDM substructure; 
(3) adopting galactic disks that are much thicker compared to typical 
thin disks including the old, thin stellar disk of the MW;
(4) modeling the compact, baryonic cores of accreting systems 
exclusively and neglecting the more diffuse and extended dark matter 
component; and 
(5) considering the encounters of individual satellites with galactic 
disks, despite the CDM expectations of numerous accretion events 
over the history of a galaxy.

Regarding the final point, notable exceptions were the studies 
by \citet{Font_etal01} and \citet{Gauthier_etal06} which examined the dynamical 
evolution of a stellar disk in the presence of a large {\it ensemble} of dark matter 
subhalos. Both contributions reported negligible tidal effects on the global 
structure of the disks. However, these investigations had the drawback of adopting 
the $z=0$ surviving substructure present in a Galaxy-sized CDM halo, 
and thus not accounting for past encounters of systems with the galactic disk 
during the evolution of the satellite populations. This point is 
critical because as subhalos on highly eccentric orbits continuously lose 
mass, the number of massive satellites with small orbital pericenters that are most 
capable of severely perturbing the disk declines with redshift so that few would 
be present at $z=0$ \citep[e.g.,][]{Zentner_Bullock03,Kravtsov_etal04,Gao_etal04}.
Establishing the role of halo substructure in shaping the fine 
structure of disk galaxies clearly requires a more realistic treatment of its 
evolution over cosmic time. 

Our aim is to improve upon these shortcomings and extend consideration to 
the rich structure of perturbed galactic disks. This paper is the second 
in a series elucidating the effects of halo substructure on thin disk 
galaxies in the context of the prevailing CDM paradigm. 
\citet[][hereafter Paper I]{Kazantzidis_etal08} focused on the generic 
{\it morphological} signatures induced in the disk 
by a typical {\LCDM}-motivated satellite accretion history, while the 
present paper discusses the {\it dynamical} response of the galactic disk subject 
to bombardment by the same population of dark matter subhalos. We implement 
a two-step strategy in an effort to overcome some of the drawbacks of 
past studies. First, we analyze {\it dissipationless}, cosmological 
simulations of the formation of Galaxy-sized CDM halos
to derive the accretion histories and properties of substructure populations. 
This information is subsequently used to seed {\it collisionless},
controlled numerical experiments of consecutive satellite impacts onto 
$N$-body realizations of fully-formed, thin disk galaxies.
Given the outstanding issues regarding disk galaxy formation in 
CDM cosmogonies \citep[e.g.,][]{Mayer_etal08} as well as the inadequacies 
of the current generation of cosmological simulations to resolve all dynamical scales 
and physical processes relevant to satellite-disk interactions, 
this strategy is most appropriate. As in Paper I, we model the infalling 
systems as pure dark matter subhalos and focus exclusively on the evolution 
of the stellar material in the disk itself.

Our contribution improves upon earlier studies in several important 
respects. First and foremost, we examine the response of galactic disks 
to subhalo populations that are truly representative of those accreted and 
possibly destroyed in the past. As such, we mitigate the biases 
in the incidence of accretion events and properties of infalling 
satellites induced by considering only the $z=0$ substructure 
of a CDM halo. Specifically, we extract merger histories of host 
halos since $z \sim 1$ and study the ramifications of such accretion 
events for disk structure. As we illustrate below, our methodology 
results in a substantially larger number of potentially damaging 
subhalo-disk encounters than previously considered. Second, we 
construct self-consistent satellite models whose properties 
are culled directly from the same cosmological simulations of 
Galaxy-sized CDM halos. This obviates the need for arbitrary 
assumptions regarding the numbers, masses, internal structures, 
orbital parameters, and accretion times of the infalling subhalos.

Lastly, we employ multi-component disk galaxy models that are derived 
from explicit distribution functions, are motivated by the {\LCDM} paradigm, and 
are flexible enough to permit detailed modeling of actual galaxies such as the MW 
and M31 by incorporating a wide range of observational constraints. 
These properties in synergy with the adopted high mass and force 
resolution enable us to construct realistic, equilibrium $N$-body models 
of thin disk galaxies and study their dynamical response to encounters 
with satellites. For our primary simulation set, we use a best-fit model for the present-day 
structure of the MW and employ an initial disk scale height of $z_d=400\pc$ 
which is consistent with that of the old, thin stellar disk of the Galaxy.
Although we utilize a model for the MW, we do emphasize that our simulation
campaign is neither designed to follow the evolution of nor to draw specific 
conclusions about the Galaxy or any other particular system. Given the complex
interplay of effects (e.g., gas cooling, star formation, chemical evolution) 
relevant to the formation and evolution of spiral galaxies, our collisionless 
simulations aim for generic features of the evolution of a thin galactic 
disk subject to bombardment by CDM substructure

Our work establishes that the types of merger histories expected in {\LCDM} 
can substantially perturb the structure of a cold, stellar disk. 
We demonstrate that cosmological halo assembly via multiple accretion events is 
capable of generating a wealth of distinctive dynamical signatures in the structural 
and kinematic properties of disk stars. These include pronounced thickening 
and heating, prominent flaring and tilting, surface density excesses 
that develop in the outskirts similar to those of observed antitruncated disks,
and lopsidedness at levels similar to those measured in observed galaxies. 
Our findings suggest that details of the galaxy assembly 
process may be imprinted on the dynamics of stellar populations and corroborate the 
concept of accretion-induced galaxy evolution. We also show that the global dynamical
response of a galactic disk to interactions with satellites depends sensitively on 
a variety of parameters including the initial disk thickness, the presence of a 
bulge component in the primary disk, the internal density distribution 
of the infalling systems, and the relative orientation of disk and satellite 
angular momenta.

The remainder of this paper is organized as follows. 
In \S~\ref{sec:methods} we describe the methods employed 
in this study. \S~\ref{sec:results} contains the results 
regarding the dynamical signatures induced in a thin galactic disk 
by a typical {\LCDM} accretion history. In \S~\ref{sec:effects}, 
we study various factors that could influence the response of disk 
galaxies to satellite accretion events. Implications and extensions 
of our findings along with a comparison to previous work are presented in 
\S~\ref{sec:discussion} and \S~\ref{sec:comparison}. Finally, in 
\S~\ref{sec:summary}, we summarize our main results and conclusions. 
Throughout this work we use the terms ``satellites,'' 
``subhalos,'' and ``substructures'' interchangeably to indicate 
the distinct, gravitationally bound objects that we use as the 
basis for the controlled satellite-disk encounter simulations.

\section{Methods}
\label{sec:methods}

A thorough description of the adopted methodology is given in Paper I. 
For completeness, we provide a summary of our approach here and refer 
the reader to Paper I for details.

\subsection{Hierarchical Cosmological Simulations}
\label{sec:cosmo_sim}

We analyze high-resolution, collisionless cosmological simulations of the formation of
four Galaxy-sized halos in a flat {\LCDM} cosmological model with parameters 
$(\Omega_m, \Omega_{\Lambda}, \Omega_b, h, \sigma_8)=(0.3, 0.7, 0.043, 0.7, 0.9)$. 
The simulations were performed with the Adaptive Refinement Tree (ART) 
$N$-body \citep{Kravtsov_etal97,Kravtsov99}. The host halos that we considered 
come from two different simulations. Halos ``G$_1$,'' ``G$_2$,'' 
and ``G$_3$'' all formed within a cubic box of length $25 \hMpc$ on a side, 
while halo ``G$_4$'' formed in a cubic volume of $20 \hMpc$ on a side. 
The mass and force resolution of these simulations as well as various 
properties of halos G$_1$-G$_4$ and their substructure were discussed in Paper I 
and in previous literature \citep{Klypin_etal01,Kravtsov_etal04,Zentner_etal05b,
Prada_etal06,Gnedin_Kravtsov06}. We identify halos (outside hosts G$_1$-G$_4$) 
and substructures using a variant of the Bound Density Maxima algorithm 
\citep{Klypin_etal99a} and we have constructed detailed accretion histories for 
each host halo and orbital tracks for all subhalos. The specifics of this 
analysis can be found in Paper I and are based largely on \citet{Kravtsov_etal04}. 

Our conventions are as follows. We adopt a mean overdensity of $337$
to define the virial radius, $r_{\rm vir}$, and the corresponding virial 
mass, $M_{\rm vir}$, for each host halo at $z=0$. Halos G$_1$ through G$_4$ 
have $r_{\rm vir} \simeq (234,215,216,230) \hkpc$ and $M_{\rm vir} 
\simeq (1.5,1.1, 1.1, 1.4) \times 10^{12} \hMsun$, respectively. 
All four of these halos accrete only a small fraction of their final 
mass and experience no major mergers at $z \lesssim 1$, and are thus 
likely to host a disk galaxy. Moreover, their accretion histories are 
typical of systems in this mass range \citep[e.g.,][]{Wechsler_etal02}.
We have chosen G$_1$ as our fiducial case for the satellite-disk interaction 
experiments described in \S~\ref{sub:control_sims}.


\begin{figure}[t]
\centerline{\epsfxsize=3.5in \epsffile{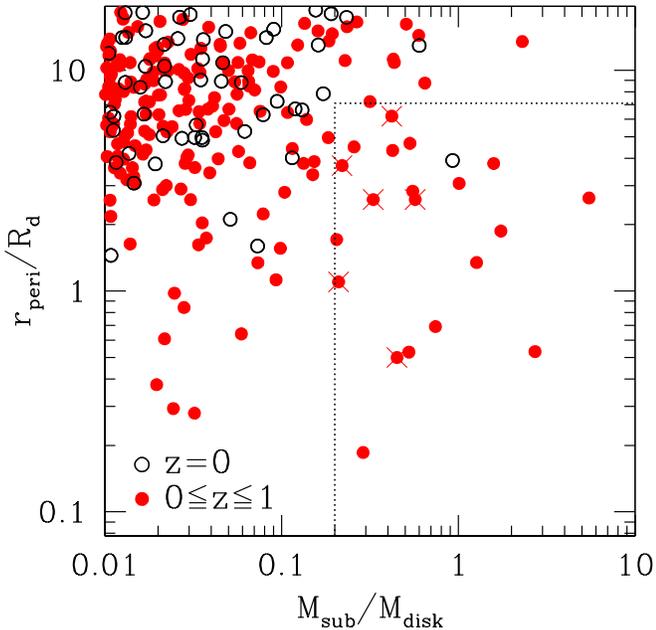}}
\caption{A scatter plot of mass versus pericentric distance for subhalos
  identified in four Galaxy-sized dark matter halos formed in the {\LCDM} cosmology. Results 
  are shown after scaling the virial quantities of all host halos to the corresponding 
  values of the primary disk galaxy model used in the controlled satellite-disk encounter 
  simulations of \S~\ref{sub:control_sims}. Subhalo masses and pericenters are assigned 
  according to the description in the text and are presented in units of the mass, 
  $M_{\rm disk}=3.53 \times 10^{10} \Mo$, and scale length, $R_d = 2.82\kpc$, respectively, 
  of the disk in the same galaxy model. {\it Filled} symbols refer to systems
  that cross within a (scaled) infall radius of $r_{\rm inf}=50\kpc$ from 
  their host halo center since a redshift $z=1$. Symbols with crosses 
  correspond to the specific subhalos in host halo G$_1$ used 
  to seed the controlled simulations of satellite-disk interactions. 
  {\it Open} symbols refer to the $z=0$ surviving substructures. {\it Dotted}
  lines mark the so-called ``danger zone'' with $M_{\rm sub} \gtrsim 0.2 M_{\rm disk}$ and 
  $r_{\rm peri} \lesssim 20\kpc$ corresponding to infalling subhalos that are 
  capable of substantially perturbing the disk. Close encounters between massive 
  substructures and galactic disks since $z=1$ should be common occurrences in {\LCDM}. 
  In contrast, very few satellites in present-day subhalo populations are likely to have 
  a significant dynamical impact on the disk structure. 
\label{fig1}}
\end{figure}


\subsection{Interactions Between Substructures and Disks in CDM}
\label{sec:sat_disk_encounters}

We investigate the dynamical response of thin galactic disks to interactions
with CDM substructure incorporating for the first time a model that accounts 
for its evolution over cosmic time. Previous related studies 
\citep[e.g.,][]{Font_etal01,Gauthier_etal06} only utilized subhalo 
populations at $z=0$, rather than a complete merger history, to seed 
simulations of satellite-disk encounters in a cosmological context.
This procedure has the drawback of eliminating from consideration those 
massive satellites that, prior to $z=0$, cross through the central regions of 
their hosts, where the galactic disks reside.
Such systems can potentially produce strong tidal effects on the disk, 
but are unlikely to constitute effective perturbers at $z=0$ as they 
suffer substantial mass loss (or even become disrupted) during their
orbital evolution precisely because of their forays into the 
central halo \citep[e.g.,][]{Zentner_Bullock03,
Kravtsov_etal04,Gao_etal04,Zentner_etal05a,Benson05}. Accounting for 
the impact of these relatively short-lived objects on the global 
dynamical response of galactic disks is the major improvement 
we introduce in the present study.

Figure~\ref{fig1} serves as a dramatic illustration of this element 
of our modeling. This figure is a scatter plot of mass versus pericentric 
distance for two different substructure populations within 
host halos G$_1$-G$_4$. The masses and pericenters of all 
subhalos have been scaled to the mass, $M_{\rm disk} = 3.53 \times 10^{10} \Mo$,
and radial scale length, $R_d = 2.82\kpc$, of the disk in the primary  
galaxy model used in the controlled satellite-disk encounter simulations
of \S~\ref{sub:control_sims}. For the purposes of this presentation, we have also scaled 
the virial quantities ($M_{\rm vir}$ and $r_{\rm vir}$) of all four 
Galaxy-sized host halos to the {\it total} mass, $M_h=7.35 \times 10^{11} \Mo$, 
and {\it tidal} radius, $R_h=244.5 \kpc$, of the parent dark matter halo 
in the same galaxy model. The full list of parameters pertaining to the 
primary disk galaxy will be listed in \S~\ref{sub:galaxy_models}.

The first substructure population of Figure~\ref{fig1} comprises systems 
that cross within a (scaled) infall radius of $r_{\rm inf}=50\kpc$ from their 
host halo center since a redshift $z=1$. This selection is fixed 
empirically to identify orbiting satellites that approach the central regions 
of the host potential and are thus likely to have a significant 
dynamical impact on the disk structure (Paper I). We assign masses to the satellites
of this group at the simulation output time nearest to the first inward crossing 
of $r_{\rm inf}$. As we discuss below, we define this to be the epoch that our controlled 
simulations initiate. The corresponding pericenters are computed from 
the orbit of a test particle in a static \citet[][hereafter NFW]{Navarro_etal96} 
potential whose properties match those of the host CDM halo at the time of $r_{\rm inf}$. 
We note that a single distinct object of this population may be recorded 
multiple times as one subhalo may undergo several passes through the 
central regions of its host with different masses and pericenters. 
Many of these satellites suffer substantial mass loss or even become tidally 
disrupted prior to $z=0$.

The second subhalo population consists of the $z=0$ surviving substructures.
Their pericenters are also estimates based on the orbit of a test 
particle in a static NFW potential whose properties match those of the host 
CDM halo at $z=0$. The dotted line in Figure~\ref{fig1} encloses an area in the 
$M_{\rm sub}- r_{\rm peri}$ plane corresponding to satellites more massive than $0.2 M_{\rm disk}$
with pericenters of $r_{\rm peri} \lesssim 20\kpc$ ($r_{\rm peri} \lesssim 7 R_d$). We refer to 
this area as the ``danger zone''. Subhalos within this area should 
be effective perturbers, but we intend this as a rough criterion to aid in 
illustrating our point.

Figure~\ref{fig1} demonstrates that the $z=0$ substructure populations contain
very few massive systems on potentially damaging orbits. In fact, in all four of 
our host halos only {\it one} satellite can be identified inside the 
danger zone at $z=0$. As a consequence, the present-day substructure in
a CDM halo likely plays only a minor role in driving the 
dynamical evolution of galactic disks. This conclusion is supported by 
\citet{Font_etal01} and \citet{Gauthier_etal06} who did consider the 
$z=0$ subhalo populations and reported negligible effects on the global
structure of the disk.  

On the other hand, the danger zone contains numerous satellites that have 
penetrated deeply into their host halo since $z=1$: on average, $\sim 5$ systems 
more massive than $0.2M_{\rm disk}$ pass close to the center of their hosts
with $r_{\rm peri} \lesssim 20\kpc$ in the past $\sim 8$~Gyr. We also stress
that three of the host halos have accreted at least one satellite
more massive than the disk itself since $z=1$. This finding is corroborated 
by analysis of simulations with much better statistics \citep{Stewart_etal08}.
Overall, the results in Figure~\ref{fig1} indicate that close encounters 
between massive subhalos and galactic disks since $z=1$ should be common 
occurrences in the {\LCDM} cosmological model. It is thus important 
to quantify and account for such interactions when the goal is to investigate 
the cumulative dynamical effects of halo substructure on galactic disks. 

\subsection{Initial Conditions for Satellite-Disk Encounters}
\label{sub:initial_conditions}

Initializing the controlled experiments of subhalo-disk encounters involves: 
(1) identifying relevant substructures in the cosmological simulations 
and recording their properties (mass functions, internal structures, orbital parameters, 
and accretion times); and (2) constructing $N$-body realizations of disk galaxy
and satellite models. In what follows, we describe each of these steps.

\subsubsection{Subhalo Selection Criteria}
\label{sub:sat_criteria}

In practice, it is not feasible to simulate the encounters of all orbiting 
satellites present in host halo G$_1$ with the galactic disk at sufficient 
numerical resolution. This, however, is not needed as the dynamical effects 
of substructure on galactic disks are expected to be dominated by the few 
most massive subhalos that penetrate deeply into the central regions of 
their hosts. To limit computational expense as well as to ensure that only
relevant encounters will be considered, we impose two criteria for selecting 
target satellites for re-simulation (see also Paper I).

We limit our search to satellites that approach the central regions 
of halo G$_1$ with (scaled) pericenters of $r_{\rm peri} \lesssim 20 \kpc$ 
since $z=1$. This choice is motivated by the fact that subhalos with small orbital 
pericenters are expected to substantially perturb the galactic disk. 
We initiate our controlled re-simulations at the 
epoch when each selected subhalo first crossed a (scaled) infall radius 
of $r_{\rm inf}=50\kpc$ from the host halo center. In what follows and unless otherwise 
explicitly stated all satellite properties (obviously except for those at $z=0$) 
are recorded at the simulation output closest to the first inward crossing of 
$r_{\rm inf}$. Likewise, we select only subhalos that are a significant fraction 
of the disk mass, with (scaled) masses of $M_{\rm sub} \gtrsim 0.2 M_{\rm disk}$,
as these will have the largest impact on the disk structure.  

Figure~\ref{fig1} shows that, on average, $\sim 5$ systems meet these two criteria for each
host halo. Although at least one subhalo more massive than the disk in
our controlled simulations is expected to cross the central regions of their
hosts since $z\sim 1$ (Figure~\ref{fig1}; \citealt{Stewart_etal08}),
we explicitly ignore these interactions with $M_{\rm sub} \gtrsim M_{\rm
  disk}$. 0ur goal is to investigate the tidal effects of substructure 
on a well-preserved disk galaxy while such encounters may destroy the 
disk upon impact \citep{Purcell_etal08}.

The aforementioned criteria result in a merger history of six accretion events, 
which we denote S1-S6, for subsequent re-simulation over a $\approx 8$~Gyr 
period. These substructures correspond to the filled symbols with crosses seen in 
Figure~\ref{fig1} and their properties are summarized in Table~\ref{table:sat_param} 
below. A critical reader may note that the internal properties and orbital parameters
of subhalos in collisionless cosmological simulations, as those of the present study,
may differ substantially from those in simulations with gasdynamics. Indeed, in hydrodynamical
simulations of disk galaxy formation, the average satellite crossing the central regions 
of its host experiences additional mass loss due to disk shocking and both circularization 
of its orbit and decrease of its orbital inclination via dynamical friction against the disk 
\citep[e.g.,][]{Quinn_Goodman86,Penarrubia_etal02,Meza_etal05}. These
considerations do not apply to subhalos S1-S6 as these systems were identified in the 
cosmological simulation and their properties were recorded {\it before} crossing the center 
of their host halo for the first time. 

Table~\ref{table:sat_param} contains the main orbital and structural properties 
of cosmological satellites S1-S6. The parameters of each subhalo in this table 
are scaled so that they correspond to the same fractions of virial quantities 
($M_{\rm rvir}$, $r_{\rm rvir}$, and $V_{\rm vir}$) in both host halo G$_1$ 
and in the primary galaxy model. Columns (2)-(9) list properties measured 
in the cosmological simulation. Columns (10)-(12) present structural 
parameters computed directly from the $N$-body realizations of satellite models 
constructed for the controlled subhalo-disk encounter experiments. 

The properties of satellites S1-S6 have been discussed extensively in Paper I. 
Nonetheless, a few parameters are worthy of explicit note. 
First, the masses of these six substructures correspond to the upper limit of the 
mass function of observed satellites in the Local Group. For reference, the mass 
of the Large Magellanic Cloud \citep[e.g.,][]{Schommer_etal92} is similar 
to that of our most massive subhalo S2. Moreover, satellites S1-S6 are spatially 
extended with their tidal radii, $r_{\rm tid}$, encompassing a significant 
fraction of the disk itself ($r_{\rm tid} \gtrsim 7 R_d$). Because of that the 
entire disk will be subject to potential fluctuations, although of different 
magnitude in different places, during the encounters with these subhalos.
Thus, the energy imparted by typical cosmological substructures
will likely not be deposited locally at the point of impact, as it was assumed by 
\citet{Toth_Ostriker92}, but rather globally across the entire disk. Lastly, 
we note that most subhalos S1-S6 are on highly eccentric orbits and that 
the simulated accretion history includes nearly polar (S1,S2), prograde (S3,S5), 
and retrograde (S4,S6) encounters.


\begin{table*}
\caption{Parameters of the Satellite Models}
\begin{center}
\begin{tabular}{lcccccccccccc}
\tableline\tableline 
\\
\multicolumn{1}{c}{}              &
\multicolumn{1}{c}{}              &
\multicolumn{1}{c}{}              &
\multicolumn{1}{c}{}              &
\multicolumn{1}{c}{}              &
\multicolumn{1}{c}{}              &
\multicolumn{1}{c}{$\theta$}      &
\multicolumn{1}{c}{$V_{\rm sub}$}  &
\multicolumn{1}{c}{}              &
\multicolumn{1}{c}{}              &
\multicolumn{1}{c}{$r_s$}         &
\multicolumn{1}{c}{}              
\\
\multicolumn{1}{c}{Model}&
\multicolumn{1}{c}{$z_{\rm acc}$} &
\multicolumn{1}{c}{$M_{\rm sub}/M_{\rm disk}$} &
\multicolumn{1}{c}{$r_{\rm tid}/R_d$} &
\multicolumn{1}{c}{$r_{\rm peri}/R_d$} &
\multicolumn{1}{c}{$r_{\rm apo}/r_{\rm peri}$} &
\multicolumn{1}{c}{($\degrees$)}&
\multicolumn{1}{c}{(${\rm km s^{-1}}$)}&
\multicolumn{1}{c}{$\epsilon_J$} &
\multicolumn{1}{c}{$(\alpha,\beta,\gamma)$} &
\multicolumn{1}{c}{(${\rm kpc}$)}&
\multicolumn{1}{c}{$c$} &
\\
\multicolumn{1}{c}{(1)} &
\multicolumn{1}{c}{(2)} &
\multicolumn{1}{c}{(3)} &
\multicolumn{1}{c}{(4)} &
\multicolumn{1}{c}{(5)} &
\multicolumn{1}{c}{(6)} &
\multicolumn{1}{c}{(7)} &
\multicolumn{1}{c}{(8)} &
\multicolumn{1}{c}{(9)} &
\multicolumn{1}{c}{(10)} &
\multicolumn{1}{c}{(11)}&
\multicolumn{1}{c}{(12)}
\\
\\
\tableline\tableline
\\
S1 & $ 0.96 $ & $ 0.33 $ & $ 8.8 $ & $ 2.6 $ & $ 6.8 $  & $ 93.3 $  & $ 72.9  $ & $ 0.41 $  & 
$(0.12,3.85,1)$ & $ 2.1 $ & $ 11.6 $  \\
S2 & $ 0.89 $ & $ 0.57 $ & $ 7.6 $ & $ 2.6 $ & $ 6.0 $  & $ 86.6 $  & $ 70.9  $ & $ 0.46 $  & 
$(0.21,4.02,1)$ & $ 2.6 $ & $ 8.1 $ \\
S3 & $ 0.54 $ & $ 0.42 $ & $ 8.2 $ & $ 6.2 $ & $ 3.2 $  & $ 45.1 $  & $ 158.2 $ & $ 0.72 $ & 
$(0.38,3.72,1)$ & $ 2.2 $ & $ 10.6 $ \\
S4 & $ 0.32 $ & $ 0.45 $ & $ 7.0 $ & $ 0.5 $ & $ 20.6 $ & $ 117.7 $ & $ 19.6  $ & $ 0.16 $ & 
$(0.25,4.18,1)$ & $ 1.6 $ & $ 12.5 $ \\
S5 & $ 0.20 $ & $ 0.22 $ & $ 9.7 $ & $ 3.7 $ & $ 9.3 $  & $ 59.9  $ & $ 171.3 $ & $ 0.35 $ & 
$(0.16,3.94,1)$ & $ 1.8 $ & $ 15.2 $ \\
S6 & $ 0.11 $ & $ 0.21 $ & $ 8.2 $ & $ 1.1 $ & $ 19.6 $ & $ 144.5 $ & $ 89.6  $ & $ 0.17 $ & 
$(0.29,4.09,1)$ & $ 1.3 $ & $ 18.3 $ \\
\tableline
\end{tabular}
\end{center}
{\sc Notes.}--- Columns (2)-(9) record satellite properties at the epoch closest to 
when each subhalo first crossed within a (scaled) infall radius 
of $r_{\rm inf}=50\kpc$ from the center of host halo G$_1$. Columns (10)-(12) list structural 
parameters computed from the $N$-body realizations of satellite models used in the 
controlled subhalo-disk encounter simulations of \S~\ref{sub:control_sims}. All entries 
are listed after scaling the virial quantities of halo G$_1$ to the corresponding values 
of the parent halo in the fiducial disk galaxy model used in the controlled experiments.
Note that values of orbital circularities and apocenter-to-pericenter ratios may deviate from 
those measured directly in the controlled simulations because the former are estimated 
from the satellite orbits in the potential of host halo G$_1$.
Col. (1): Satellite model.
Col. (2): Redshift at which the properties of cosmological satellites were recorded.
Col. (3): Bound satellite mass in units of the mass of the disk, 
          $M_{\rm disk}=3.53 \times 10^{10} \Mo$, in the fiducial 
          galaxy model used in the controlled encounter simulations.
Col. (4): Satellite tidal radius in units of the radial scale length
          of the fiducial disk, $R_d = 2.82\kpc$.
Col. (5): Pericenter of the satellite orbit in units of $R_d$.
Col. (6): Satellite orbital apocenter-to-pericenter ratio.
Col. (7): Angle between the initial angular momenta of the satellite
          and the disk in degrees. 
          This angle is defined so that $ 0 \degrees < \theta < 90 \degrees$
          corresponds to a prograde orbit and $ 90 \degrees < \theta < 180 \degrees$
          corresponds to a retrograde orbit. The cases where $\theta = 0 \degrees$,
          $\theta = 90 \degrees$, and $\theta = 180 \degrees$ represent a coplanar prograde,
          a polar, and a coplanar retrograde orbit, respectively.
Col. (8): Satellite 3D orbital velocity in ${\rm km s^{-1}}$.
Col. (9): Circularity of the orbit. This parameter is defined as 
          $\epsilon_J \equiv  J/J_{\rm circ}(E)$, where $J$ is the satellite 
          angular momentum and $J_{\rm circ}(E)$ is the corresponding angular momentum 
          for a circular orbit of the same energy $E$.
Col. (10): Intermediate, outer, and inner slopes of the satellite density 
          profile. Inner and outer slopes correspond to the asymptotic values.
Col. (11): Scale radius of the satellite density profile in $\rm {kpc}$. This 
          radius is defined as the distance where the logarithmic slope is the 
          average of the inner and outer slopes, $\rm d \ln \rho(r)/\rm d \ln r 
          = -(\gamma + \beta)/2$.
Col. (12): Satellite concentration defined as $c\equiv r_{\rm tid}/r_{\rm s}$.
\label{table:sat_param}
\end{table*}



\begin{figure}[t]
\centerline{\epsfxsize=3.5in\epsffile{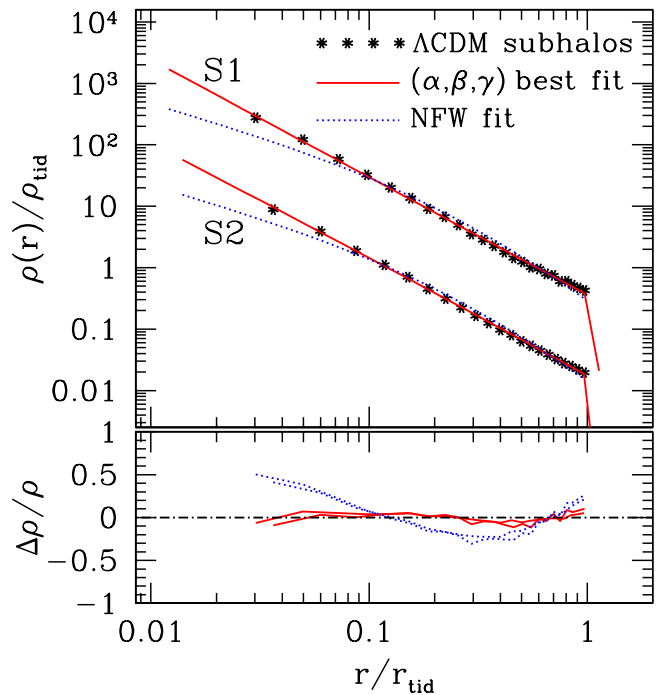}}
\caption{{\it Upper panel:} Spherically-averaged density profiles, $\rho(r)$,
  for representative cosmological satellites S1 and S2 as a function of radius 
  in units of the tidal radius of each system, $r_{\rm tid}$. {\it Stars}
  correspond to the subhalo profiles extracted directly from the cosmological 
  simulation of host halo G$_1$. {\it Solid} lines present fits to the density 
  structure using a multi-parameter $(\alpha, \beta, \gamma)$ density law, 
  while {\it dotted} lines show corresponding NFW fits. The solid lines include 
  an exponential cutoff at $r > r_{\rm tid}$. All curves are plotted from 
  the adopted force resolution ($2\epsilon_{\rm sub}=300\pc$) outward and densities 
  are normalized to the mean density within the tidal radius of each satellite. For 
  clarity, the density profiles corresponding to the lower curves are vertically 
  shifted downward by a factor of $0.05$. The $(\alpha, \beta, \gamma)$ density 
  law provides an accurate description of the internal structure of cosmological 
  subhalos at all radii, while the NFW functional form substantially underestimates subhalo 
  densities in the innermost regions. {\it Bottom panel:} Residuals for the density 
  profile fits, $\Delta \rho\,/\rho \equiv \rho_{\rm sub}-\rho_{\rm fit}/\rho_{\rm sub}$,
  where $\rho_{\rm sub}$ is the true subhalo density computed in the cosmological 
  simulation and $\rho_{\rm fit}$ denotes the fitted density.
\label{fig2}}
\end{figure}


\subsubsection{Constructing $N$-body Realizations of Satellites}
\label{sub:sat_models}

Our controlled experiments employ satellite models which are constructed 
to match the internal structure of cosmological subhalos S1-S6. The density 
profiles of these systems are extracted at the simulation output closest to 
each subhalo's first inward crossing of $r_{\rm inf}$. 
We model all cosmological satellites with the 
general density profile law \citep{Zhao96,Kravtsov_etal98}
\be
   \rho(r) = 
   \frac{\rho_{\rm s}} {(r/r_{\rm s})^\gamma [1+(r/r_{\rm s})^\alpha]^
   {(\beta-\gamma)/\alpha}} \qquad\hbox{($r \leq r_{\rm tid}$)} \ ,
   \label{general_density}
\ee
where $\rho_s$ sets the normalization of the density profile and $r_{\rm tid}$ 
is the limiting radius of the satellite imposed by the host halo tidal field. 
The exponents $\gamma$ and $\beta$ denote the asymptotic inner and outer slopes of the 
profile, respectively. Exponent $\alpha$ parametrizes the transition between the 
inner and outer profiles, with larger values of $\alpha$ corresponding to sharper 
transitions. Lastly, at the scale radius $r_s$, the logarithmic slope is 
the average of the inner and outer slopes, $\rm d \ln \rho(r)/\rm d \ln r = -(\gamma + \beta)/2$.

In practice, all subhalos are described well by Eq.~(\ref{general_density}) 
with $\gamma=1$, so we fix this value and fit their structure by 
varying the other parameters. Moreover, we truncate subhalo profiles 
beyond the tidal radius with an exponential law 
\be
   \rho(r) = \rho(r_{\rm tid}) \left(\frac{r}{r_{\rm tid}}\right)^{\kappa}
\exp\left(-\frac{r-r_{\rm tid}}{r_{\rm decay}}\right) \qquad\hbox{($r > r_{\rm tid}$)} \ ,
\label{exp_cutoff}
\ee
where $\kappa$ is fixed by the requirement that $\rm d \ln \rho(r)/\rm d \ln r$
is continuous at $r_{\rm tid}$. This procedure is necessary because sharp truncations 
result in subhalo models that are not in equilibrium \citep{Kazantzidis_etal04a}, 
but it results in additional bound mass beyond $r_{\rm tid}$. The precise amount 
of additional mass depends upon the model parameters, but is roughly 
$\sim 2\%$ of the bound mass for each subhalo that we consider.

Figure~\ref{fig2} demonstrates the efficacy of our procedure for modeling the 
internal structure of cosmological subhalos. This figure presents
spherically-averaged density profiles, $\rho(r)$, of two 
representative cosmological satellites along with two different fits to their 
density distributions and the associated residuals, $\Delta \rho/\rho$.  
The first is a fit to the multi-parameter $(\alpha, \beta, \gamma)$ functional 
form introduced above [Eq.~(\ref{general_density})], while the second is a 
fit to the NFW density profile with $(\alpha, \beta, \gamma)=(1,3,1)$.

Figure~\ref{fig2} shows that $(\alpha, \beta, \gamma)$ models 
provide accurate representations of the density structures of 
cosmological subhalos, while the NFW profile much less so, 
as one might expect because it has less parameter freedom. 
Indeed, the density residuals with respect to Eq.~(\ref{general_density}) 
are $\lesssim 10\%$ for both satellites over the entire range 
of radii, while the residuals to NFW fits are much larger 
reaching $\sim 50\%$ in the innermost parts of the profile.
We note that these findings were confirmed in the remaining 
subhalos S3-S6. Overall, Figure~\ref{fig2} 
demonstrates that the NFW law underestimates the normalization of the density 
profile and consequently the concentration of these cosmological subhalos.  
This is important because, as we illustrate in \S~\ref{sub:internal.density}, 
the dynamical response of galactic disks to encounters with satellites 
depends sensitively upon the precise density distribution of the infalling 
systems.

We list fit parameters for each subhalo in Table~\ref{table:sat_param}; however, 
we note two general features. First, in all cases the asymptotic outer slopes $\beta$ 
are found to be much steeper than $3$. Best-fit values vary from $\beta = 3.7$ 
to $\beta = 4.2$ suggesting that these satellites are better 
approximated by the \citet{Hernquist90} profile at large radii 
\citep[see also][]{Ghigna_etal98}. Second, the intermediate slopes $\alpha$
range from $\alpha = 0.1-0.4$ indicating a much more gradual transition
between inner and outer asymptotic power-laws compared to that of the NFW
profile with $\alpha = 1$. The steep outer profiles and variety in structural
parameters is, at least in part, due to the strong dynamical evolution that
subhalos experience within their host 
potential \citep[e.g.,][]{Moore_etal96,Klypin_etal99a,Hayashi_etal03,Kazantzidis_etal04b,
Kravtsov_etal04,Mayer_etal07}.  

$N$-body realizations of satellites are constructed from a distribution 
function (DF) that self-consistently reproduces the density structures of 
selected subhalos S1-S6. Because substructures 
in cosmological simulations are nearly spherical in both configuration and 
velocity space \citep{Moore_etal04,Kazantzidis_etal06,Kuhlen_etal07}, 
we assume that the DF depends only upon the energy per unit mass and 
calculate it through an Abel transform \citep{Kazantzidis_etal04a}.

Lastly, each satellite is represented by $N_{\rm sub} = 10^6$ particles.
The gravitational softening length is set to $\epsilon_{\rm sub}=150 \pc$ 
which allows us to resolve the structure of subhalos to $\sim 1\%$ of their tidal 
radii. The adopted mass and force resolution are adequate 
to resolve mass loss processes experienced by orbiting satellite halos 
\citep{Kazantzidis_etal04b}.

\vspace{2cm}

\subsubsection{Disk Galaxy Models}
\label{sub:galaxy_models}

We construct $N$-body realizations of multi-component primary disk galaxies 
using the method of \citet{Widrow_Dubinski05} as described in detail in Paper I. 
The \citet{Widrow_Dubinski05} models are derived from explicit DFs 
and thus represent axisymmetric, equilibrium solutions 
to the coupled collisionless Boltzmann and Poisson equations. Because of that,
they are ideal for studying complex dynamical processes associated with the intrinsic 
fragility of galactic disks such as gravitational interactions with 
infalling satellites. The galaxy models consist of an exponential stellar 
disk, a Hernquist bulge \citep{Hernquist90}, 
and an NFW dark matter halo, and are characterized by $15$ free parameters 
that may be tuned to fit a wide range of observational data for actual 
galaxies. 

For the majority of satellite-disk encounter simulations, we use the specific 
parameter choices of model ``MWb'' in \citet{Widrow_Dubinski05}, which 
satisfies a broad range of observational constraints on the MW galaxy. 
The stellar disk has a mass of $M_{\rm disk} = 3.53 \times 10^{10}\Mo$, 
a radial scale length of $R_d=2.82\kpc$, and a sech$^2$ scale height of 
$z_d=400\pc$. The latter is consistent with that inferred for the old, 
thin stellar disk of the MW \citep[e.g.,][]{Kent_etal91,Dehnen_Binney98,
Mendez_Guzman98,Larsen_Humphreys03,Juric_etal08}. 
It is reasonable to initialize the galactic disk with such thickness 
as observational evidence \citep[e.g.,][]{Quillen_Garnet01,Nordstrom_etal04} indicates 
that the scale height of the thin disk of the MW has not changed significantly 
over the period modeled in this study ($z \lesssim 1$). The bulge 
has a mass and a scale radius of $M_b=1.18 \times 10^{10}\Mo$ and 
$a_b=0.88\kpc$, respectively. The dark matter halo has a tidal 
radius of $R_h=244.5 \kpc$, a mass of $M_h=7.35 \times 10^{11} \Mo$, 
and a scale radius of $r_h=8.82\kpc$. The total circular velocity 
of the galaxy model at the solar radius, $R_{\odot} \simeq 8 \kpc$, 
is $V_c(R_\odot)=234.1\kms$ and the Toomre disk stability parameter 
is equal to $Q = 2.2$ at $R=2.5 R_d$. We note that direct numerical simulations 
of the evolution of model MWb in isolation confirm its stability against bar 
formation for $10$~Gyr. Therefore, any significant bar growth and associated disk 
evolution identified during the satellite-disk encounter experiments should be 
the result of subhalo bombardment. In what follows, we refer to this galaxy model as 
``D1.'' 

In addition to our primary simulation set, we also address the dependence 
of disk response to encounters with infalling subhalos upon initial disk thickness 
and the presence of a bulge. Thicker disks have larger vertical velocity 
dispersions so we might expect them to be more robust to satellite accretion
events. In addition, a central bulge component may act to reduce the amount 
of damage done to the structure of the inner disk by the infalling subhalos. 
Correspondingly, we initialize two additional disk galaxy models. 

The first modified galaxy model was constructed with the same parameter 
set as D1, except that it was realized with a scale height 
that was larger by a factor of $2.5$ ($z_d=1\kpc$). We refer to 
this ``thick'' disk galaxy model as ``D2.'' Except from disk thickness, all 
of the other gross properties of the three galactic components of model 
D2 are within a few percent of the corresponding values for D1.
The second modified galaxy model is the same as D1, but constructed 
without a bulge component. We refer to this ``bulgeless'' disk galaxy model 
as ``D3.'' We stress that a bulgeless version of D1 constructed using 
the \citet{Widrow_Dubinski05} method differs significantly from 
D1. This is because the DF of the composite galaxy model is related to 
the individual density distributions and DFs of each galactic 
component in a non-trivial way. To mitigate such differences 
which make model comparison cumbersome, we have chosen to 
realize disk model D3 by adiabatically evaporating the bulge 
component from galaxy model D1. The timescale of evaporation was 
set to $500$~Myr to minimize the response of the disk and halo to the 
removal of the bulge. During the evaporation process, while the main
properties of the inner disk and halo (e.g., density profiles, 
velocity ellipsoids) have changed in response to the decrease of 
the central potential, the disk thickness has remained largely 
unmodified. It is also worth mentioning that when evolved 
in isolation model D3 develops a bar inside $\sim 4\kpc$ at 
time $t \sim 2$~Gyr.

For each disk galaxy model, we generated an $N$-body realization 
containing $N_d=10^{6}$ particles in the disk and $N_h=2\times10^{6}$ 
particles in the dark matter halo. In experiments with models D1 and D2, 
bulges were represented with $N_b=5\times10^{5}$ particles. 
The gravitational softening lengths for the three components were set 
to $\epsilon_d=50 \pc$, $\epsilon_h=100 \pc$, and $\epsilon_b=50 \pc$,
respectively. Mass and force resolution are sufficient to resolve the 
vertical structure of our galactic disks as well as minimize 
artificial heating of the disk particles through interactions with 
the much more massive halo particles. Lastly, in all cases we oriented the 
primary galaxy models such that the disk and host halo G$_1$ angular momenta 
were aligned \citep[e.g.,][]{Libeskind_etal07}.

\subsection{Satellite-Disk Encounter Simulations}
\label{sub:control_sims}

All controlled simulations of satellite-disk interactions were carried 
out with the PKDGRAV multi-stepping, parallel, tree $N$-body code 
\citep{Stadel01} as described in Paper I. We treated impacts of cosmological
subhalos S1-S6 onto the disk as a sequence of encounters. Recall that the 
internal properties and orbital parameters of these satellites
were scaled so that they correspond to the same fractions of virial 
quantities in both host halo G$_1$ and in the primary disk galaxy model. 
Starting with subhalo S1 we included subsequent systems at the epoch when 
they were recorded in the cosmological simulation (Table~\ref{table:sat_param}). 
This procedure is described in more detail in Paper I.
We note here that the sequence of accretion events is such that S1 and S2 undergo 
simultaneous interactions with the disk. We have conducted an additional 
experiment in which satellite S2 was introduced only after the interaction with 
S1 was completed and confirmed that the disk dynamical response is similar in the 
two cases. 

Due to their considerable mass and size, the infalling subhalos were not introduced 
in the simulations directly; instead, they were grown adiabatically in their 
orbits. This method ensures that the disk does not suffer substantial perturbations 
due to the sudden presence of the satellite and change in potential at its vicinity.
Specifically, each encounter simulation was performed using 
the following procedure: (i) Insert a {\it massless} particle  
realization of each satellite at the distance at which it was recorded in the 
cosmological simulation of host halo G$_1$. (ii) Increase the mass 
of this distribution to its final value linearly over a timescale that ranges 
between $\sim 150$ and $\sim 400$~Myr depending on subhalo mass. During this 
period, the satellite remained rigid and its particles are fixed in place, while 
the dark and baryonic components of the primary disk galaxy were allowed to achieve 
equilibrium with the subhalo as its mass grows. (iii) Initialize the ``live''
$N$-body satellite model by setting its internal kinematics 
according to the method described in \S~\ref{sub:sat_models}, placing it on the 
desired orbit, and switching on its self-gravity.
We note that after the satellite models are grown to full mass and 
have become self-gravitating, the tidal field of the primary disk galaxy 
acts to truncate their outer parts establishing new tidal radii that are 
smaller than the nominal values in Table~\ref{table:sat_param}.
However, this difference is less than $10\%$ in all cases suggesting that the 
primary disk galaxy model and the host CDM halo G$_1$ have similar densities 
at $r_{\rm inf}$.

To limit computational cost, subhalos were removed from the controlled 
simulations once they reached their maximum distances from the disk after 
crossing, so satellites were not permitted to begin a second passage. 
Uniformly, these distances were only slightly smaller compared to the starting 
radii of the orbits ($\sim 50\kpc$). This is because dynamical friction is very 
efficient for these massive satellites causing their orbit to decay very 
rapidly. In all cases, satellites lost $\gtrsim 80 \%$ of their mass after 
completing their first passage. This justifies our decision to neglect the 
dynamically insignificant subsequent crossing events. We emphasize
that only the {\it self-bound} core of each satellite was extracted from the 
simulations at the subsequent apocenter and not the unbound material, a significant 
fraction of which remained in the vicinity of the disk. Removing the latter 
would result in potential fluctuations that could throw the disk out of its relaxed 
state, and thus interfere with the interpretation of the results.

We present results for the disk structure after allowing the disk to relax from 
the previous interaction. Consequently, our findings are relevant to systems that exhibit 
no obvious, ongoing encounters. Due to the complexity of the interaction, we determined 
this timescale empirically by monitoring basic properties of the disk structure 
(e.g., surface density, velocity dispersion, thickness) as a function of 
time. When these quantities stopped evolving significantly within radii of interest 
(changes of the order of $5-10\%$ in disk properties were considered acceptable), 
the encounter was deemed complete. Typical ``settling'' timescales were found to be 
in the range $\sim 500-600$~Myr. The limiting radius was chosen to be equal to $7R_{\rm d}$ 
which contains $\sim 99\%$ of the initial mass of the disk. It is necessary to adopt some 
limiting radius because dynamical times grow with radius and the outer disk regions continue 
to show signs of evolution well after each satellite passage. These transient structures 
can be readily identified in the edge-on views of the disk presented in the 
next section.

When time intervals between subhalo passages were larger than the 
timescale needed for the disk to relax after the previous interaction, we 
introduced the next satellite immediately after the disk had settled from 
the previous encounter. This eliminates the computational expense of 
simulating the disk during the quiet intervals between interactions.

We compute all disk properties and show all visualizations of the disk 
morphology after rotating the disk to the coordinate frame defined by the three 
principal axes of the {\it total} disk inertia tensor and centering it to its center 
of mass. The motivation behind performing these actions is twofold. First, infalling 
satellites tilt the disk through the transfer of angular momentum (see \S~\ref{sub:tilting}). 
Introducing the new coordinate frame is important because in the original coordinate frame, 
rotation of stars in a tilted disk would appear as vertical motion interfering with 
the interpretation of the results. Of course, the tilting angles of the inner and 
outer parts of the disk may differ from each other and from the tilting angle of 
the disk determined by the principal axes of the {\it total} disk inertia tensor.  
However, these differences are typically small ($\lesssim 2\degrees$; see Figure
~\ref{fig8}) and we have verified that they only have a minor influence on measured 
disk properties. In what follows, we quote all results relative to the coordinate 
frame defined by the three principal axes of the total disk inertia tensor. 
Second, the masses of the simulated satellites are a substantial fraction of the
disk mass. As a result, the disk center of mass may drift from its initial position at 
the origin of the coordinate frame due to the encounters with the infalling 
subhalos.

\section{Results: Dynamical Signatures of Hierarchical Satellite Accretion}
\label{sec:results}

In this section we examine the response of the thin, disk galaxy
model D1 to interactions with cosmological subhalos S1-S6 of host halo G$_1$, 
which are designed to mimic a typical {\it central} accretion history for a 
Galaxy-sized CDM halo over the past $\sim 8$~Gyr. The ``final'' disk discussed 
in the next sections has experienced the S1-S6 encounters and was further evolved in
isolation for $\sim 4.3$~Gyr after the last interaction, so that the disk
evolution is followed from $z=1$ to $z=0$. We reiterate that 
though our numerical experiments employ a best-fit model for the present-day 
structure of the MW, our aim is to study the general, dynamical evolution 
of a disk galaxy subject to a {\LCDM}-motivated satellite accretion history 
since $z \sim 1$. 

The focus of this study is exclusively on the evolution of the disk 
material, so we do not consider the bulge component in any of the analysis presented 
below. Lastly, while our simulation program mimics the accretion events in halo G$_1$, 
the similarity of subhalo populations in all four Galaxy-sized host halos we analyzed 
suggests that the results presented next should be regarded as 
fairly general. We discuss in turn disk global morphology, thickening, velocity 
structure, surface density, lopsidedness, and tilting.  


\begin{figure*}[t]
\begin{center}
 \includegraphics[scale=0.37]{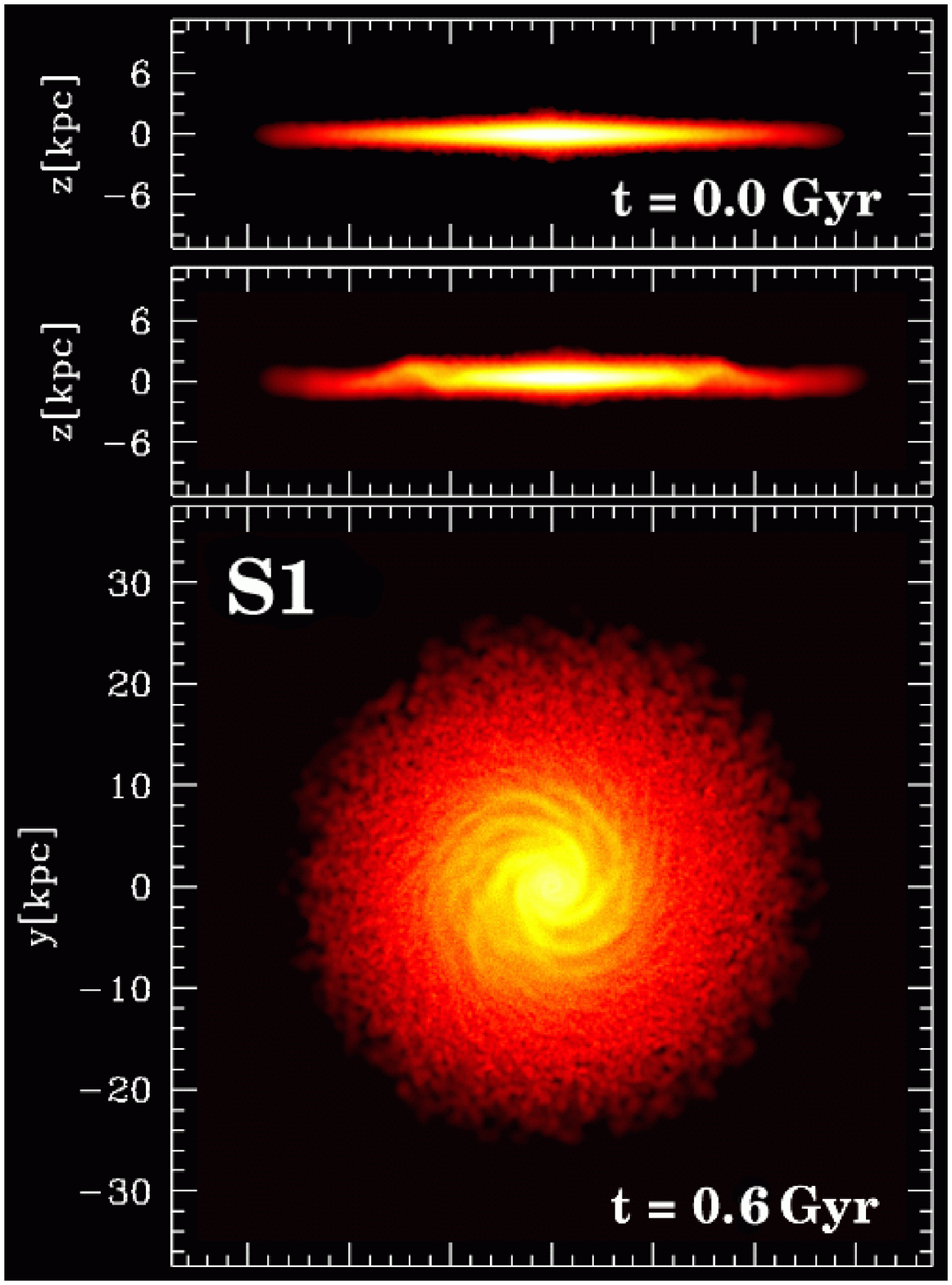}\vspace{-0.1cm} \hspace{-0.2cm}
  \includegraphics[scale=0.37]{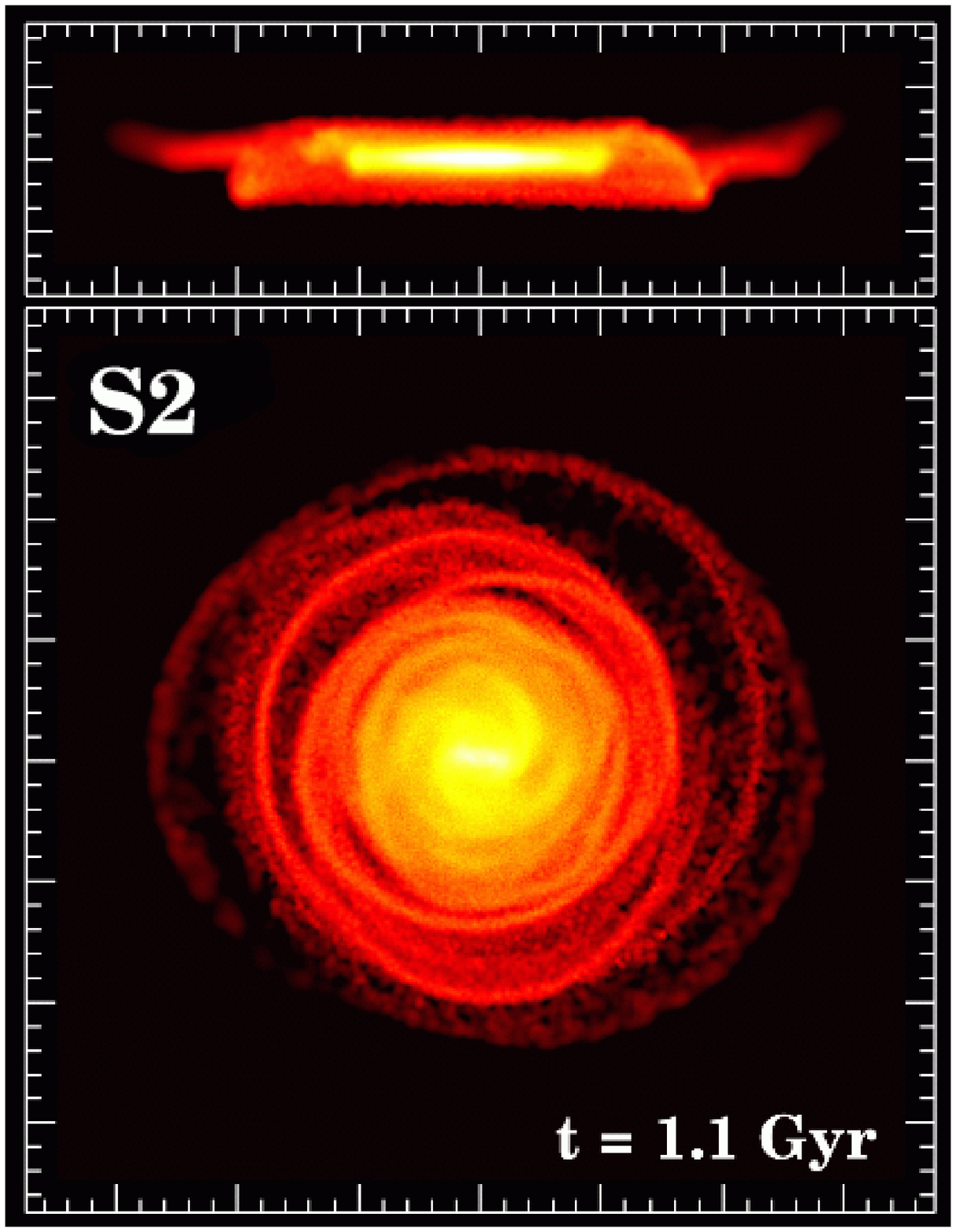} \hspace{-0.2cm} 
   \includegraphics[scale=0.37]{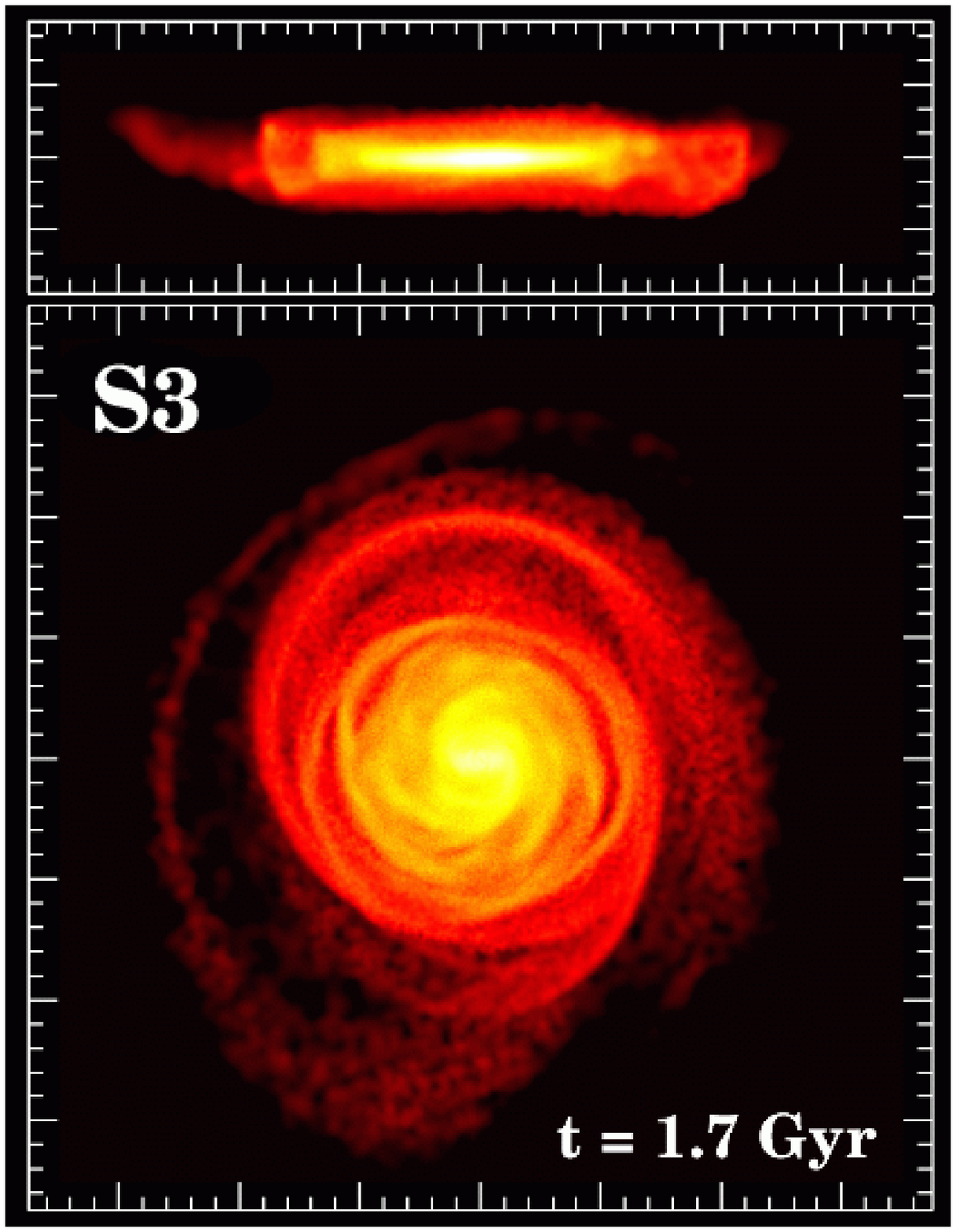} 
    \includegraphics[scale=0.37]{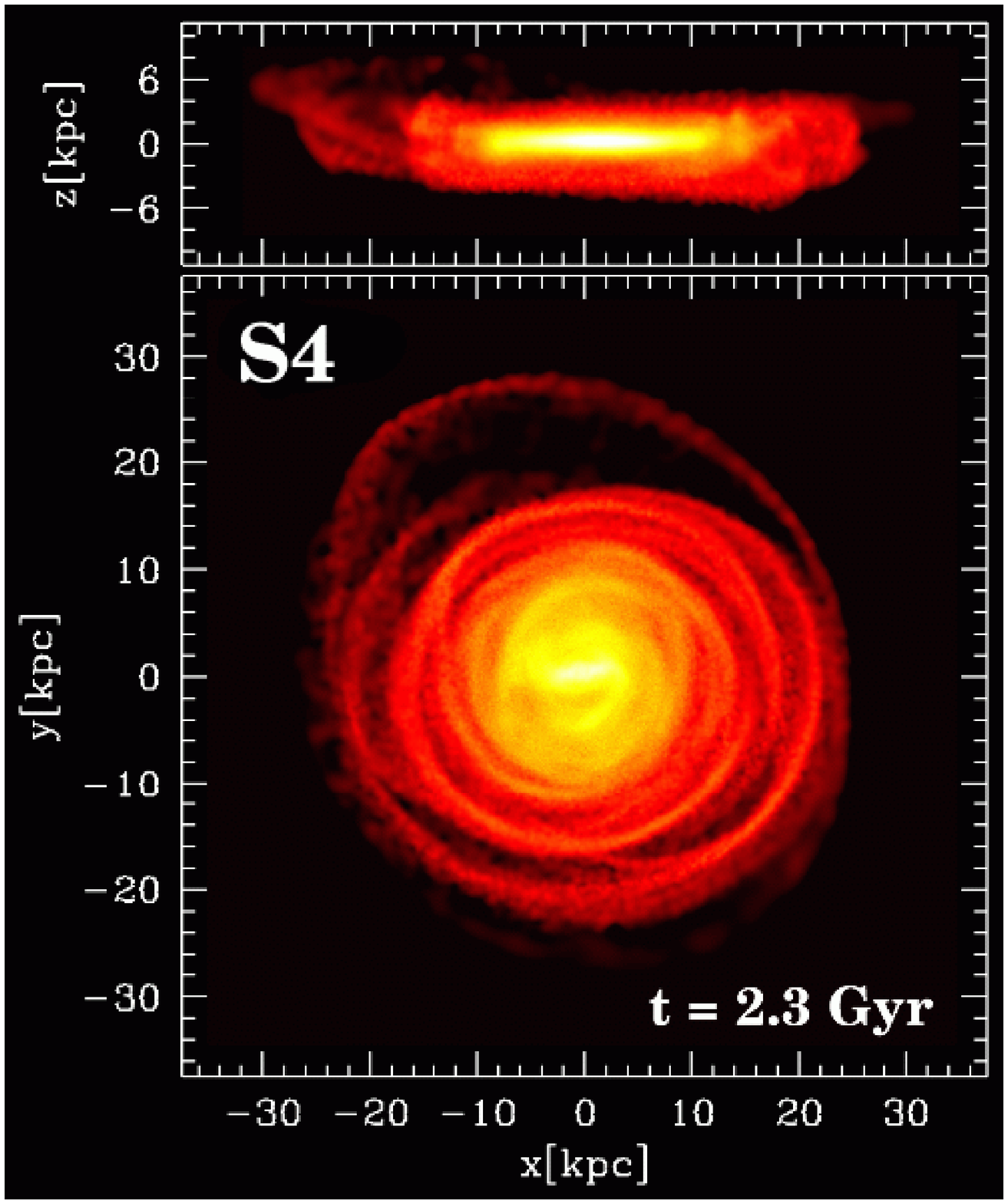} \hspace{-0.2cm}
     \includegraphics[scale=0.37]{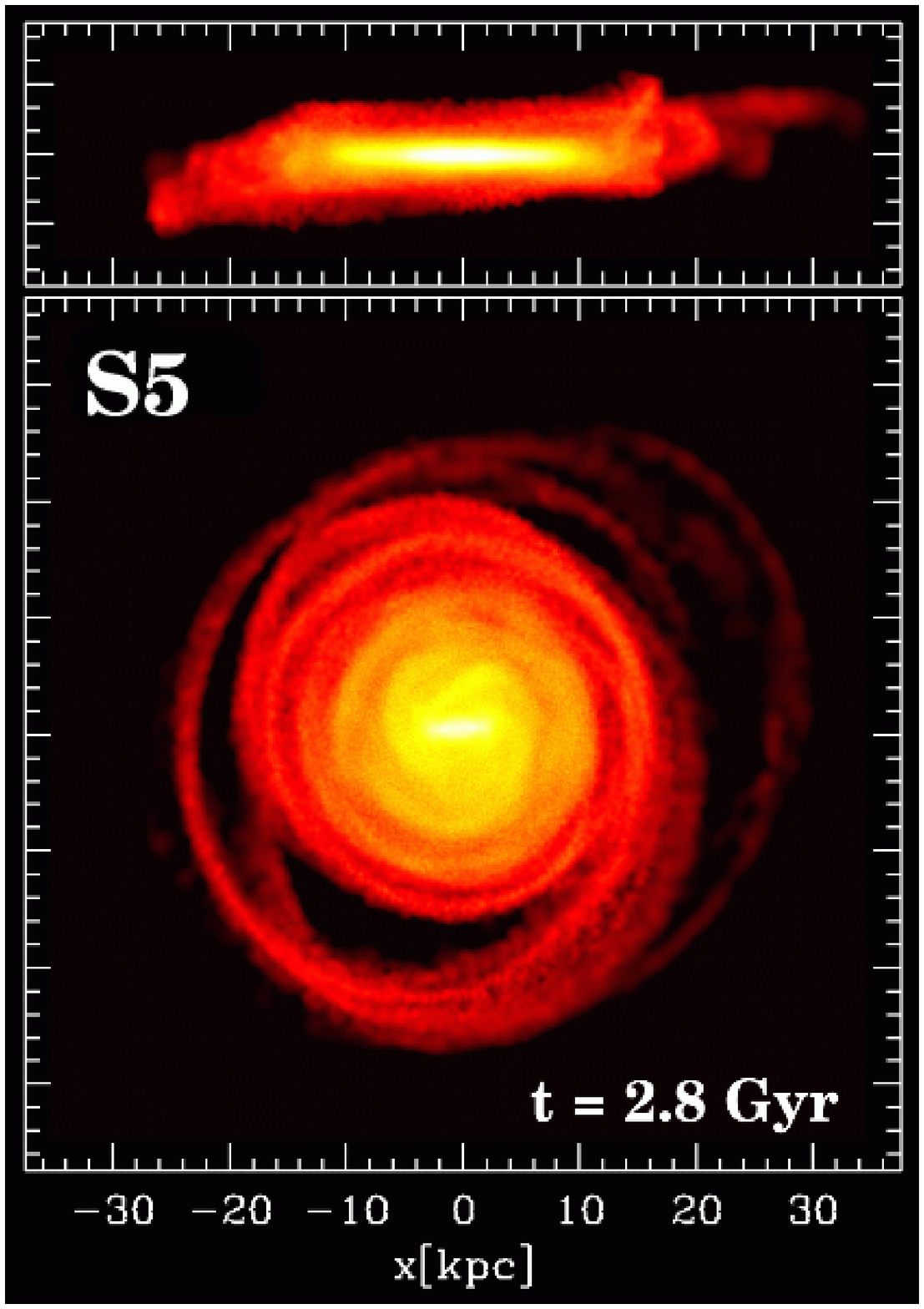} \hspace{-0.2cm}
      \includegraphics[scale=0.37]{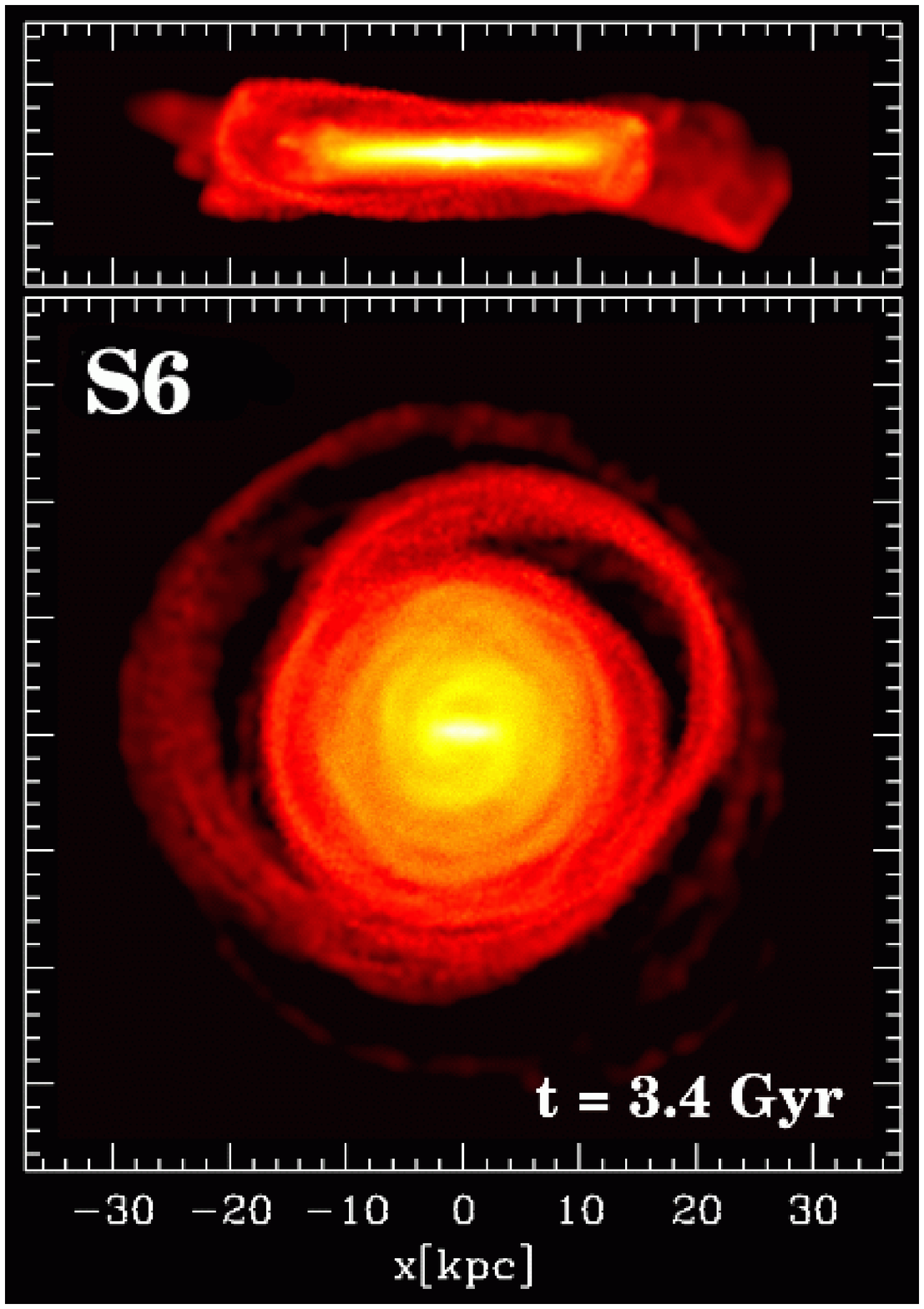} 
\end{center}    
\caption{Density maps illustrating the global morphological evolution of 
  the disk in galaxy model D1 subject to a typical {\LCDM}-motivated accretion 
  history expected for a Galaxy-sized dark matter halo since $z \sim 1$. 
  Particles are color-coded on a logarithmic scale with brighter colors indicating 
  regions of higher stellar density. Local density is calculated using an SPH smoothing 
  kernel of $32$ particles. The face-on ({\it bottom panels}) and edge-on ({\it upper panels}) 
  distributions of disk stars are shown in each frame and to aid comparison the 
  first panel also includes the edge-on view of the initial disk. The labels for 
  individual satellite passages from S1 to S6 and the time corresponding to each snapshot 
  are indicated in the {\it upper left-hand} and {\it lower right-hand} corners of each 
  bottom panel. Results are presented after centering the disk to its center of mass and 
  rotating it to a new coordinate frame defined by the three principal axes of the total 
  disk inertia tensor. The first satellite passage generates a conspicuous warp 
  while the second encounter, which involves the most massive subhalo S2, causes 
  substantial thickening of the disk and excites a moderately strong bar 
  and extended ring-like features in the outskirts of the disk. Accretion histories of 
  the kind expected in {\LCDM} models play a substantial role in setting the global 
  structure of galactic disks and driving their morphological evolution.
\label{fig3}}
\end{figure*}


\subsection{Global Disk Morphology}
\label{sub:morphology}

Figure~\ref{fig3} illustrates the global response of the disk to the 
infalling subhalos. The encounter with the first substructure (S1) generates a 
conspicuous warp beyond $\sim 12$~kpc. The impact of the second most massive 
satellite (S2) has a dramatic effect on the global disk structure. The entire disk 
visually becomes considerably thicker compared to the initial model after this accretion 
event. In addition, this interaction excites extended ring-like features in the 
outskirts of the disk and a moderately strong bar (Paper I), both of 
which indicate that the axisymmetry of the initial disk is destroyed 
as a result of the first two accretion events. We stress that the bar is
induced in response to the subhalo passages, not by amplified noise. It has a 
semi-major axis varying between $\sim 3-4\kpc$, within the range of values inferred 
for the bar in the MW \citep[e.g.,][]{Bissantz_Gerhard02}. Throughout its evolution 
the bar fails to form a boxy bulge and remains rather thin. We note that we evolved 
the disk galaxy in isolation for $\sim 5$~Gyr after the encounter with S1 
and verified that the first satellite alone is not capable of exciting a bar on these 
timescales. Fourier decomposition of the final disk also reveals that the radial variation of 
the amplitude of the observed spiral structure is similar to that of normal 
spiral galaxies in the near infrared \citep[e.g.,][]{Grosbol_Patsis98}. 

Non-axisymmetric structures such as the bar and the rings drive continued evolution by 
redistributing mass and angular momentum in the disk \citep[e.g.,][]{Debattista_etal06}. 
Thus, infalling satellites affect galactic disks not only {\it directly} by impulsively 
shocking the orbits of disk stars, but also {\it indirectly} by exciting global 
instabilities. In fact, angular-momentum transport from one part of the disk
to another causes the disk to expand radially during the encounters
(see \S~\ref{sub:surface.density}). 
However, we expect this expansion to be significantly less pronounced compared 
to that in the vertical direction as rotational energy dominates over random motions 
in the plane of the disk.

Apart from changes associated with the structural details of the bar and the location and 
extent of the ring-like features, the disk structure does not evolve appreciably 
during subsequent satellite passages (S3-S6). In broad terms, the global morphological
evolution of the galactic disk is driven by the interaction with the most
massive subhalo of the accretion history. We will return to this point
repeatedly in subsequent discussions. Lastly, the face-on projections of the
disk show that the tidally induced bar and other structures are non-transient features 
that survive long after the initial perturbations. Indeed, we confirmed their presence 
in the final disk some $\sim 4.3$~Gyr after the last encounter 
(see also Figure~\ref{fig6} in Paper I).


\begin{figure*}[t]
\centerline{\epsfxsize=7.2in \epsffile{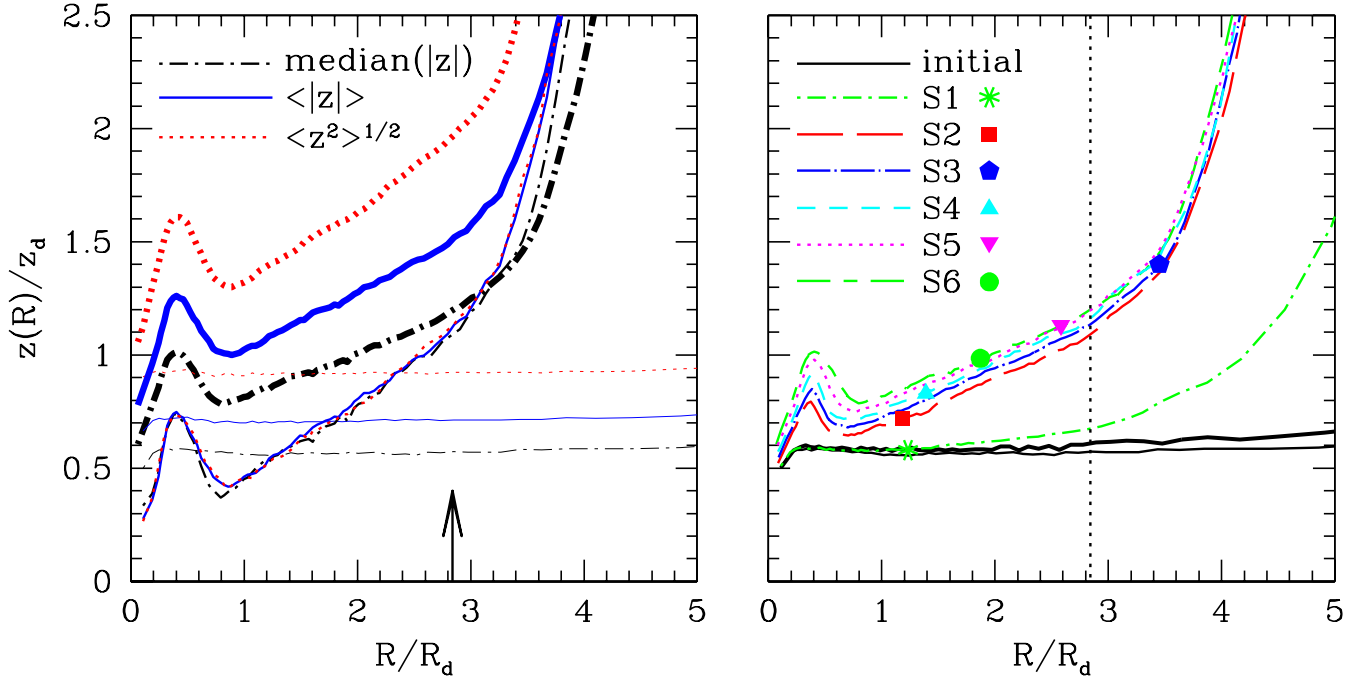}}
\caption{Disk thickening. Thickness profiles, $z(R)$, of the disk in galaxy model 
  D1 viewed edge-on. Thicknesses and radii are normalized to the scale height,
  $z_d$, and radial scale length, $R_d$, of the initial disk.
  {\it Left:} Thickness profiles for the initial ({\it thin lines}) and final disk 
  ({\it thick lines}). Lines of {\it intermediate} thickness show the fractional 
  increases in disk thickness defined as 
  $ [\rm{median}(|z|)_f - \rm{median}(|z|)_i]/ \rm{median}(|z|)_i$, 
  where $\rm{median}(|z|)_f$ and $\rm{median}(|z|)_i$ denote the thickness 
  of the final and initial disk, respectively. {\it Dot-dashed}, {\it solid}, 
  and {\it dotted} lines correspond to thicknesses measured by the median 
  of the absolute value, $\rm{median}(|z|)$, the mean of the absolute value, $<|z|>$, 
  and the dispersion, $<z^{2}>^{1/2}$, of disk particle height above the midplane, 
  respectively. The initial disk is constructed with a constant scale height, 
  explaining why the corresponding curves are flat. The arrow indicates the location 
  of the solar radius, $R_\odot$. The initial thin disk thickens considerably at 
  all radii as a result of the encounters with CDM substructure and a conspicuous 
  flare is evident in the final disk beyond $R \gtrsim 4 R_d$. {\it Right:} Evolution 
  of disk thickness profiles. Different lines show results for individual satellite passages 
  from S1 to S6 and thicknesses are measured as $\rm{median}(|z|)$. The vertical dotted 
  line indicates $R_\odot$ and various symbols correspond to the pericenters
  of the infalling subhalos. The {\it thick} solid line corresponds to the initial 
  disk evolved in isolation for a timescale equal to that of the combined satellite 
  passages. The first two satellite passages thicken the disk considerably at all radii
  and cause substantial flaring in the outskirts. The combined effect of the remaining 
  subhalos (S3-S6) is much less dramatic, indicating that the second, most
  massive accretion event is responsible for setting the scale height of the final disk. 
\label{fig4}}
\end{figure*}


\vspace{2cm}
\subsection{Disk Thickening}
\label{sub:thickening}

Among the most intriguing dynamical effects of the subhalo impacts 
is the pronounced increase in disk thickness, which is evident in the edge-on views 
of the disk in Figure~\ref{fig3}. A more quantitative measure of the thickening of 
the disk in response to the satellite accretion events is presented in 
Figure~\ref{fig4}. This figure shows thickness profiles of the disk, $z(R)$, 
viewed edge-on. 

The thickening of the disk can be formally described by the increase of its 
vertical scale height. In the present study we quantify disk thickness at any given 
radius by the median of the absolute value of disk particle height above the disk midplane, 
$\rm{median}(|z|)$. The motivation behind this choice is threefold. First, it permits direct 
comparison with earlier work because similar estimators of disk thickness have been employed 
by previous authors \citep[e.g.,][]{Quinn_etal93,Walker_etal96,
Velazquez_White99,Gauthier_etal06}. Second, the gravitational interaction between 
satellites and disks is capable of heating individual disk stars at quite large radii 
above the plane of the disk. The choice of the {\it median} value of the particle height 
mitigates the influence of distant outlier particles and thus produces conservative 
estimates for the increase of disk thickness.  Third, disk scale heights must be formally 
derived by means of fitting the particle distribution to an appropriate functional form 
(e.g., exponential or a sech$^2$ law). In contrast, quantities such as $\rm{median}(|z|)$ 
do not require a fit to any functional form and so they are unambiguous and require no 
assumptions for their interpretation.  

The left panel of Figure~\ref{fig4} shows thickness profiles for the initial 
and final disk together with the fractional increase in thickness caused by 
the infalling satellites. For convenience, in Fig.~\ref{fig4} we compare the 
$\rm{median}(|z|)$ to two additional estimators 
of disk thickness used extensively in the literature, namely the mean of 
the absolute value, $<|z|>$, and the dispersion, $<z^{2}>^{1/2}$, of disk 
particle height above the midplane. While all different estimators yield nearly 
identical fractional increase in disk thickness, 
$\rm{median}(|z|)$ leads to a smaller amount of disk thickening at large radii 
($R \gtrsim 3 R_d$). This is true for all of our analyses and so we address
only $\rm{median}(|z|)$ in the remaining of the paper.

We note that the initial disk exhibits a small departure from
a pure exponential profile near its center. This feature develops when constructing 
the composite DF and potential of the multicomponent galaxy model. As a result, 
the stellar disk is somewhat thinner in the middle though the resulting density is 
fully self-consistent within the context of the epicyclic and third integral approximations 
that are built into the disk DF \citep{Widrow_Dubinski05}. This deviation is
insignificant for our purposes because the influence of this effect on disk
thickness is quite moderate, and is confined to a fraction of a scale length 
$R_d$. The main goal of the present study is to assess the {\it global} dynamical 
response of a galactic disk to cosmologically-motivated satellite accretion events.

The left panel of Fig.~\ref{fig4} demonstrates that the initial, thin disk thickens 
considerably at all radii as a result of the interactions with subhalos S1-S6. More 
specifically, disk thickness near the solar radius has increased in excess of 
a factor of $2$. We emphasize that the thickening does not occur uniformly as a function 
of radius suggesting that the inner and outer disk regions respond differently to the 
accretion events. The outer disk is much more susceptible to damage by the accreting
satellites. Indeed, at $R = R_d$ the thickness increases by $\sim 50\%$ compared 
to a factor of $\sim 3$ rise at $R \sim 4 R_d$. The larger binding energy of the inner 
{\it exponential} disk and the presence of a massive, central bulge 
($M_ b \simeq 0.3 M_{\rm disk})$ which deepens the central potential 
are likely responsible for the robustness of the inner disk.
Given the fact that the infalling subhalos are spatially extended and 
the self-gravity of the disk grows weaker as a function of distance from the center, it 
is not unexpected that the scale height of the disk should increase with radius. Indeed, 
by making the simplest assumption that the accreting satellites deposit their orbital 
energy evenly in the disk, it can be shown that the disk scale height increases as 
$\Delta z(R) \propto \Sigma_d^{-2}(R)$, where $\Sigma_d(R)$ is the disk surface density
(Paper I). It is also worth emphasizing that a significant part of the
evolution seen in the inner parts of the disk ($R \lesssim R_d$), including
the peak in the thickness profiles at $R \simeq 0.5 R_d$, can be attributable
to the tidally induced bar.

The right panel of Figure~\ref{fig4} presents the {\it evolution} of disk thickness 
caused by individual satellite passages S1-S6. After the S1 encounter, the thickness 
of the outer disk ($R \gtrsim 3.5 R_d$) rises considerably leading to a 
a distinct flare. Specifically, at $R = 4 R_d$ the thickness increases 
by $\sim 60\%$ compared to a $\sim 20\%$ increases at $R_\odot$. In contrast, the inner 
disk ($R \lesssim 1.5 R_d$) appears unaffected by the accretion event despite the 
significant initial mass ($M_{\rm sub} \simeq 0.3 M_{\rm disk}$) and small pericenter 
($r_{\rm peri} \simeq 1.2 R_d)$ of S1. 

The second accretion event involves the most massive satellite S2 
($M_{\rm sub} \simeq 0.6 M_{\rm disk})$ and is responsible for generating a much more 
pronounced flare in the outer parts as well as causing the entire disk to thicken. 
However, the inner disk regions still exhibit substantial resilience to the encounter. 
Indeed, disk thickness at $R = 1.5 R_d$ rises only by $\sim 40\%$ compared to a factor 
of $\sim 3.6$ increase at $R = 4 R_d$. The corresponding increase at $R_\odot$ is 
a factor of $\sim 1.9$. 

Though the masses of satellites S1 and S2 differ by less than a factor of $2$, subhalo S2 
constitutes a far more efficient perturber compared to satellite S1. Taking into account 
that both subhalos have similar internal properties (Table~\ref{table:sat_param}) and 
are on approximately the same polar orbit, this suggests that disk thickening is very 
sensitive to perturber mass. In fact, using $N$-body simulations of 
satellite-disk encounters \citet{Hayashi_Chiba06} found that the increase of
the disk vertical scale height,$\Delta z_d$, 
is proportional to the square of the subhalo mass, $\Delta z_d/R_d \propto M_{\rm sub}^2$.
Our results suggest that such correlation may be even stronger. 

The remaining satellite passages have a markedly less dramatic effect in perturbing 
the vertical structure of the disk. This fact confirms that subhalo interactions 
with already thickened disks induce much smaller relative changes in disk thickness 
compared to the initial passages (\citealt{Quinn_etal93}; Paper I). It also suggests 
that thicker disks may exhibit enhanced resilience to encounters with substructure. 
We return to this point in \S~\ref{sub:initial.thickness}. The combined effect of subhalos S3-S6 
acts to increases the disk thickness at intermediate ($R_\odot$) and 
large ($R = 4 R_d$) radii by only $\sim 10\%$ compared to that after passage S2.
The larger differences observed at small radii ($R \lesssim 1.5 R_d$) 
are attributable to the bar, which grows stronger as a function of time.
Finally, we remark that the thickness of the same disk galaxy 
evolved in isolation for a timescale equal to that of all satellite passages 
grows by only $\sim 10\%$ indicating the excellent quality of the initial 
conditions and adequate resolution of the simulations. 


\begin{figure*}[t]
\centerline{\epsfxsize=6.8in \epsffile{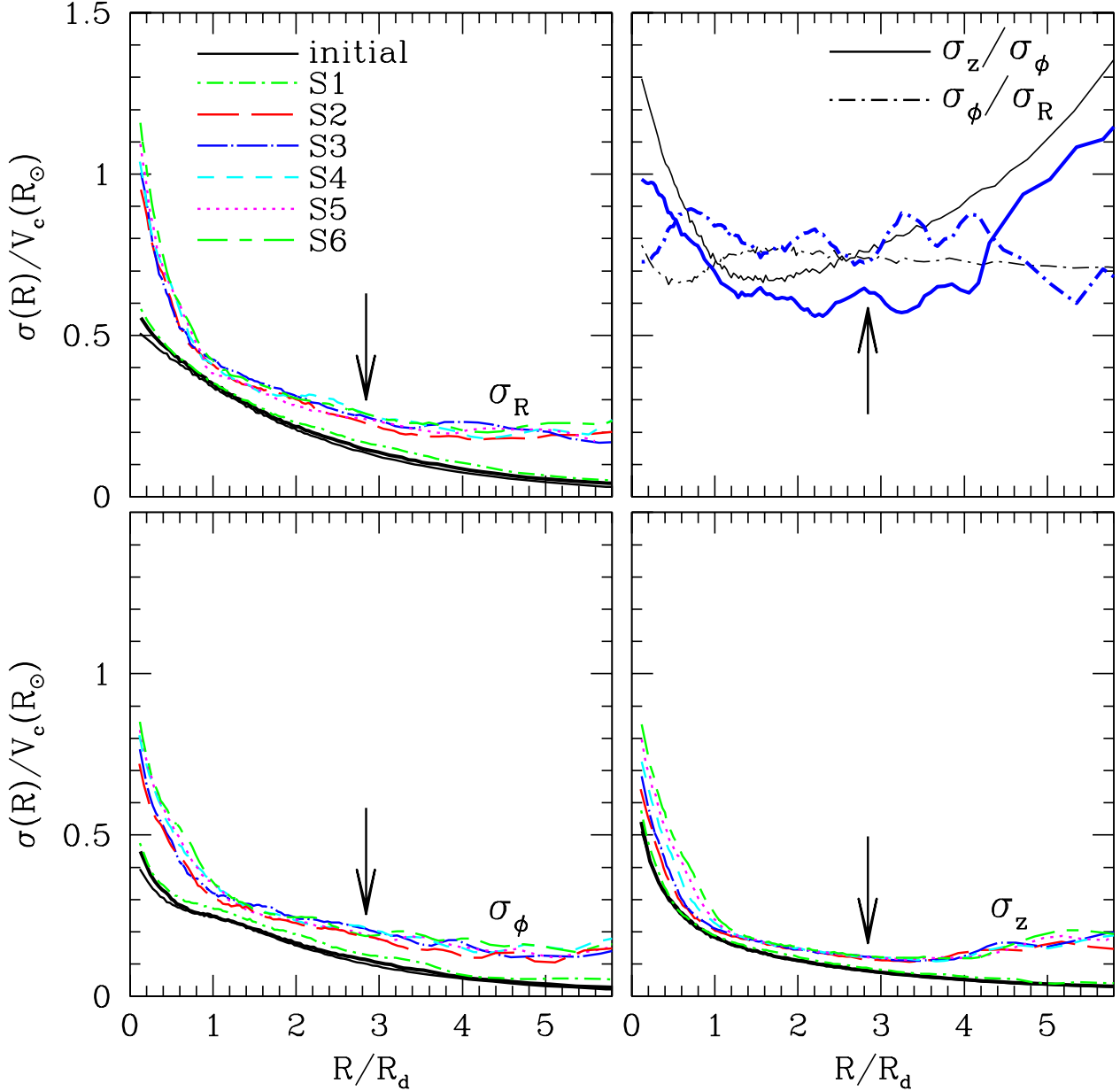}}
\caption{Disk heating. Velocity dispersion profiles of the disk in galaxy 
  model D1 as a function of projected radius in units of the radial scale 
  length of the initial disk, $R_d$. Counterclockwise from the upper left, 
  the panels display the evolution of 
  the radial, $\sigma_R$, azimuthal, $\sigma_\phi$, and vertical, $\sigma_z$,
  velocity dispersions. All profiles are normalized to the total 
  circular velocity of galaxy model D1 at the solar radius, 
  $V_{\rm c}(R_\odot)=234.1\kms$. Different lines correspond to individual
  satellite passages from S1 to S6 and line types are the same as in the right panel 
  of Figure~\ref{fig4}. {\it Thick} lines present the kinematical
  properties of the primary disk galaxy evolved in isolation for a timescale 
  equal to that of the combined satellite passages. In the upper right panel
  the velocity dispersion ratios $\sigma_z/\sigma_\phi$ ({\it solid lines}) and 
  $\sigma_\phi/\sigma_R$ ({\it dot-dashed lines}) are shown for the initial 
  ({\it thin lines}) and final ({\it thick lines}) disk. In all 
  panels arrows indicate the location of the solar radius, $R_\odot$. 
  Bombardment by CDM substructure heats the thin galactic disk considerably 
  in all three directions and causes its velocity ellipsoid to become 
  more anisotropic. 
\label{fig5}}
\end{figure*}


\subsection{Disk Velocity Structure and Heating}
\label{sub:heating}

The dynamical response of a thin, galactic disk to infalling satellites 
manifests itself in a variety of ways. In this section we address the 
degree to which the disk velocity structure evolves as a result of the simulated 
accretion history. We characterize the disk kinematical response to accretion 
events through the evolution of the disk velocity ellipsoid ($\sigma_R, \sigma_\phi, 
\sigma_z$), where $\sigma_R$, $\sigma_\phi$, and $\sigma_z$ correspond to the 
radial, azimuthal, and vertical velocity dispersions, respectively. 
In the isothermal sheet approximation, we expect that $\sigma_z^2 \propto z_d$, 
and so the evolution of the disk vertical velocity dispersion should be less 
pronounced than the evolution of disk thickness. The relevant analysis is 
presented in Figure~\ref{fig5}.

This figure reveals that the disk velocity ellipsoid grows significantly in 
all three directions as a result of the gravitational interactions with the 
infalling satellites. The velocity structure of the disk evolved in isolation
for an equivalent period shows virtually no evolution in the vertical 
direction. Some minimal evolution in the radial and azimuthal components 
of the ellipsoid is observed because spiral structure arising from the 
swing amplification of discreteness noise leads to angular-momentum transport 
in the plane of the disk. Nevertheless, this increase is negligible compared 
to the net effect caused by the satellite encounters. 

As with disk thickening, the velocity structure is 
influenced most dramatically by S2. Both the direct deposition of energy by the 
subhalos and global instabilities such as the bar and spiral structure which 
are excited during these encounters are responsible for the observed evolution. 
The latter phenomena act predominantly in the plane of the disk and cause the 
planar components of the disk velocity ellipsoid to evolve significantly. 

The disk velocity ellipsoid at the solar radius $R_\odot$ 
increases from $(\sigma_R, \sigma_\phi, \sigma_z) = (31, 24, 17)\kms$ to 
$(\sigma_R,\sigma_\phi,\sigma_z) \simeq (61, 49, 31)\kms$. 
For reference, the velocity ellipsoid of the thick disk of the MW 
at $R_\odot$ is estimated to be $\sim (46, 50, 35)\kms$ by
\citet{Chiba_Beers00} and $\sim (63,39,39)\kms$ by \citet{Soubiran_etal03}. 
However, there are at least two reasons to exercise caution interpreting the details
of these results. First, the final dispersions are calculated considering 
all stars, rather than only those that belong to the thick disk component (see Paper I).  
Second, we have neglected any stellar components in the infalling satellites 
and thus we are unable to address their contribution to the final disk. 
This is relevant in light of theoretical \citep[e.g.,][]{Abadi_etal03} as well
as observational studies \citep{Yoachim_Dalcanton06} suggesting that stars
formed in accreted satellite galaxies play a significant role in the formation
of thick disks.

Figure~\ref{fig5} shows that subhalo impacts heat the galactic disk in a 
non-uniform way. At small radii ($R \lesssim R_d$), the planar velocity dispersions 
exhibit a steep increase which is again primarily a consequence of the bar. 
The central values of $\sigma_R$ and $\sigma_\phi$
grow by a factor of $\sim 2.2$, while $\sigma_z$ increases by a factor 
of $\sim 1.6$. As with disk thickness, the effect of the accretion events 
on the velocity ellipsoid grows stronger with distance for $R \gtrsim R_d$.
While the initial disk is designed so that its velocity ellipsoid decreases 
monotonically as a function of projected radius, after the satellite passages 
all dispersions become nearly constant outside $\sim 3 R_d$. Such flat 
velocity dispersion profiles are in very good agreement with results from recent 
kinematic studies of planetary nebulae in the extreme outskirts of spiral galaxies 
\citep{Herrmann_etal09}.

In addition, though the initial dispersion profiles are smooth, at radii 
$R \gtrsim R_d$, the planar components $\sigma_R$ and $\sigma_\phi$ of the velocity
ellipsoid exhibit wave-like features which are associated with the stellar rings 
in the disk plane seen in Figure~\ref{fig3}. These structures result from the 
satellite encounters and do not dissipate even after several Gyr of evolution 
subsequent to the last accretion event. 

The upper-right panel of Figure~\ref{fig5} shows that the ratio of $\sigma_z$ to 
$\sigma_\phi$ decreases at all radii as a result of the subhalo impacts. This 
finding, in conjunction with the fact that $\sigma_\phi/\sigma_R$ appears not 
to be affected in any noteworthy way, indicates that both planar
components of the disk velocity ellipsoid respond more strongly to the accretion
events compared to the vertical component. In other words, infalling
satellites cause the disk velocity ellipsoid to become more anisotropic. 
The larger increase of $\sigma_R$ and $\sigma_\phi$ 
relative to $\sigma_z$ is due to the presence of the bar and the action of 
spiral structure that transports angular momentum through the disk. 
It is worth noting that though the resultant heating is substantially larger in the
plane the disk spreads primarily in the vertical direction. This is 
because rotational energy dominates over random motions in the plane of the disk.


\begin{figure}[t]
\centerline{\epsfxsize=3.5in \epsffile{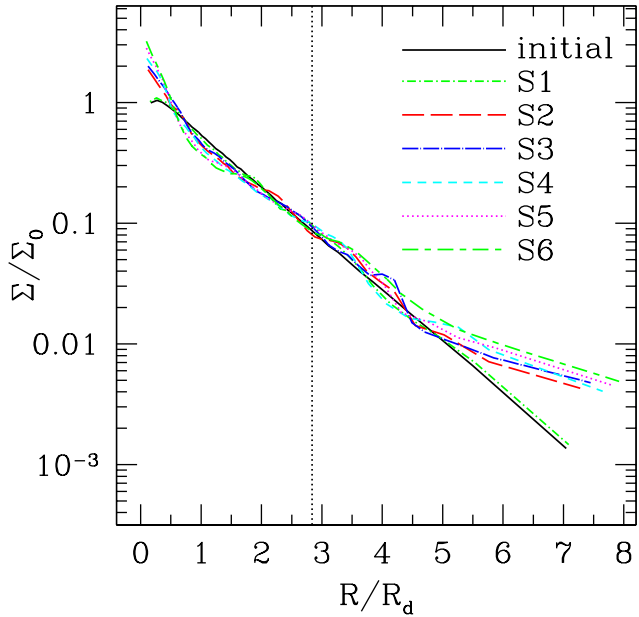}}
\caption{Disk antitruncation. Evolution of the surface density profiles, $\Sigma(R)$, 
  of the disk in galaxy model D1 viewed face-on as a function of projected
  radius, $R$, in units of the radial scale length of the initial disk,
  $R_d$. Different lines show results for both 
  the disk initially ({\it solid line}) and the disk after each subhalo impact as 
  indicated in the {\it upper right-hand corner}. All profiles are normalized to the surface 
  density of the initial disk at the origin, $\Sigma_{0}$, and the vertical dotted line 
  indicates the location of the solar radius, $R_\odot$. Encounters with CDM substructure 
  can generate surface density excesses in galactic disks similar to those seen
  in the light profiles of observed antitruncated disk galaxies.
\label{fig6}}
\end{figure}


\subsection{Disk Surface Density and Antitruncation}
\label{sub:surface.density}

In is interesting to investigate how the simulated subhalo accretion history would 
influence the disk surface density distribution. Figure~\ref{fig6} 
shows the evolution of the {\it face-on} surface density profile of the disk as a 
function of satellite passages S1-S6. By construction, the surface density profile 
of the initial disk follows an exponential distribution in cylindrical radius $R$. 

As with both disk thickness and velocity structure, the disk surface density is minimally 
affected by the encounter with the first satellite S1. The response of the surface density 
distribution to subsequent accretion events is notable and in agreement with the results of 
the previous subsections it is driven by the most massive subhalo S2.

At small radii ($R \lesssim R_d$), the surface density profile steepens considerably, 
a feature that is again due to the bar and the transport of disk material inwards by the 
spiral patterns and ring-like features. The central disk surface brightness increases 
by $\Delta \mu = 1$ mag arcsec$^{-2}$ during the course of the interactions with subhalos 
S1-S6. At intermediate radii ($R_d \lesssim R \lesssim 5 R_d$), the surface density profiles 
are very similar to that of the initial disk. Additionally, these profiles are not smooth but 
display strong wave-like features which are associated with the rings of disk material 
(Figure~\ref{fig3}). 

More interestingly, the disk spreads radially, and beyond $R \gtrsim 5 R_d$, there is a clear 
excess of surface density compared to that of the initial disk. The subhalo encounters modify 
the face-on structure of a single-component, exponential disk to producing a
system with two distinct components. Beyond the very inner disk regions where the bar
dominates and within approximately $5$ scale lengths of the initial disk 
($5 R_d \sim 14\kpc$), the final surface density is exponential with a scale 
length roughly equal to that of the initial disk. Indeed, fitting an exponential profile
to the mass surface density of the final disk at $R_d \le R \le 5 R_d$ yields a best-fit scale 
length of $R_{d,\rm inner} \approx 3.05\kpc$, or $\sim 8\%$ larger than the initial scale length.  
At larger radii the disk exhibits a flatter surface density profile, which is again crudely 
exponential. Fitting another exponential profile to the outer final disk ($R > 5R_d$) 
yields a scale length of $R_{d, \rm outer} \approx 5\kpc \sim 1.8 R_d \sim 1.6
R_{d, \rm inner}$. This excess surface density at large radii relative to the
exponential profile of the inner disk is interesting in the context of the
so-called ``antitruncated'' stellar disks whose surface brightness profiles
display similar excesses at roughly $4-6$ disk scale lengths from galaxy 
centers \citep[e.g.,][]{Erwin_etal05,Pohlen_Trujillo06,Pohlen_etal07,Erwin_etal08}. 

Our simulations lack the physical effect of star formation, so this disk antitruncation 
arises from {\it old} stars that resided in the initial thin disk and migrated outward.  
The driver behind the expansion of the disk and the excess surface density is 
redistribution of disk angular momentum (and hence disk mass) caused both directly 
and indirectly by the subhalo impacts. Infalling satellites directly deposit kinetic 
energy and angular momentum into the disk during the gravitational interactions. 
Additionally, non-axisymmetric features such as the bar and spiral structure 
transport angular momentum and stellar mass to large radii. In response 
to the net transfer of angular momentum between its inner and outer regions, the disk 
expands. Indeed, we find that the {\it magnitude} of the total angular momentum at 
$R \gtrsim 5 R_d$ grows by $\sim 80\%$ over the course of the simulated satellite 
accretion history. The transport of disk material outwards in radius also leads to 
the excess surface density observed at large projected radii. 

\subsection{Disk Lopsidedness}
\label{sub:lopsidednes}

The face-on views of the simulated disk in Figure~\ref{fig3} illustrate that encounters 
with infalling satellites are responsible for generating large-scale, long-lived 
asymmetries. An intriguing phenomenon observed in many disk galaxies is the 
existence of significant lopsided asymmetries in their neutral hydrogen 
and/or stellar mass distributions \citep[e.g.,][]{Baldwin_etal80,Richter_Sancisi94,
Rix_Zaritsky95,Zaritsky_Rix97,Matthews_etal98,Haynes_etal98,Bournaud_etal05,
Reichard_etal08}. Systematic attempts to quantify the frequency of such asymmetries in 
stellar disks have revealed the presence of lopsidedness in a substantial fraction of 
disk galaxies. The fraction of lopsided disks varies between $\sim 20\%$ and $\sim 30\%$ 
for strongly lopsided systems and may exceed $50\%$ for moderately lopsided disks 
\citep[e.g.,][]{Rix_Zaritsky95,Zaritsky_Rix97,Rudnick_Rix98,Bournaud_etal05}.  
It is therefore interesting to quantify the degree to which the simulated disk 
exhibits deviations from axisymmetry. 

A standard characterization of asymmetries in galaxies is via Fourier decomposition. 
We express the surface density of the stellar disk as a Fourier series 
\be
   \Sigma(R,\phi) = a_0(R) + \sum_{m=1}^{\infty}\ a_m(R)
   e^{im[\phi-\phi_m(R)]} \ ,
   \label{Fourier}
\ee
where $A_m(R)\equiv a_m(R)/a_0(R)$ and $\phi_m(R)$ denote the normalized strength 
and the phase of the Fourier component $m$, and quantify lopsidedness by the ratio 
of the amplitudes of the $m=1$ to $m=0$ (azimuthally-averaged surface density) 
Fourier coefficients, $A_1(R)\equiv a_1(R)/a_0(R)$ \citep{Rix_Zaritsky95}.
We performed the Fourier decomposition after centering the system to the peak 
of the density distribution \citep{Rix_Zaritsky95}, whereas using centroids 
\citep{Debattista_Sellwood00} gave similar results. 

\begin{figure}[t]
\centerline{\epsfxsize=3.5in \epsffile{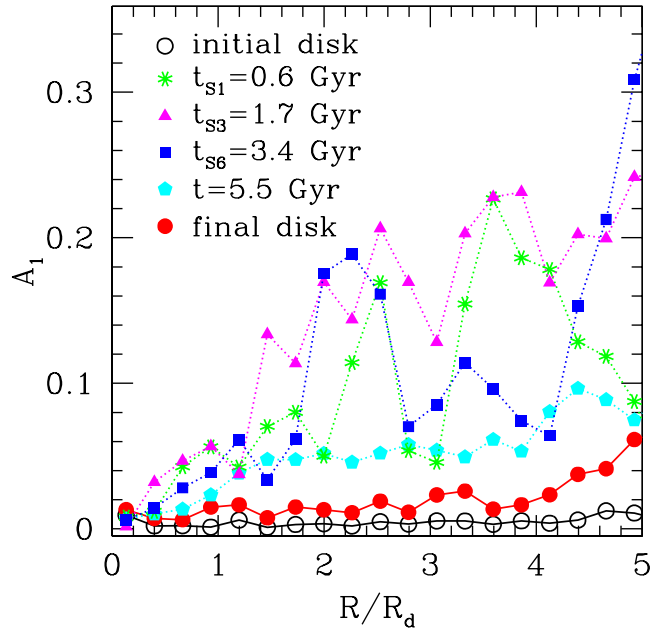}}
\caption{Disk lopsidedness. Evolution of the $A_1$ parameter, the 
  normalized amplitude of the $m=1$ Fourier component of the 
  disk surface density, of the disk in galaxy model D1 as a function 
  of projected radius in units of the radial scale length of the initial 
  disk, $R_d$. Results are presented for both the initial 
  ({\it open circles}) and final distribution of disk stars ({\it filled circles}), 
  and for four characteristic times in the simulated accretion history of 
  host halo G$_1$ (after each of the S1, S3, and S6 encounters, and $\sim 2$~Gyr 
  subsequent to the last accretion event). {\LCDM}-motivated satellite accretion 
  histories are responsible for triggering as well as maintaining for a significant 
  fraction of the cosmic time lopsidedness in stellar disks at levels similar to those 
  in observed galaxies. 
\label{fig7}}
\end{figure}

Figure~\ref{fig7} shows the variation of the lopsidedness parameter $A_1$ 
as a function of projected radius from the center of the disk. Results are presented 
for both the initial and final distribution of disk stars and for four characteristic 
timescales in the simulated accretion history of halo G$_1$. The initial disk is 
constructed to be axisymmetric explaining why the corresponding curve is flat. 
Fig.~\ref{fig7} clearly demonstrates that encounters with CDM substructures 
trigger lopsidedness in the galactic disk. Furthermore, the induced lopsidedness 
is not constant as a function of radius nor are changes monotonic. 
Different regions of the disk may thus become lopsided to different 
degrees by the infalling satellites. 

During the simulated accretion history, typical values of $A_1$ span the range 
$0.1 \lesssim A_1 \lesssim 0.2$ ($1.5 R_d \lesssim R \lesssim 4.5 R_d$), but 
more prominent lopsided asymmetries reaching amplitudes of $\sim 0.3$ is observed at larger 
radii ($R \gtrsim 5 R_d$). These values are consistent with observational estimates 
of lopsidedness from various samples of stellar disks \citep[e.g.,][]{Rix_Zaritsky95,Zaritsky_Rix97,
Rudnick_Rix98,Bournaud_etal05}. The outer disk regions are more susceptible to the 
tidal perturbations that generate these asymmetries, and because they are characterized 
by longer dynamical times, significant lopsidedness persists there. 
Though the bombardment by CDM substructure has ceased by $\sim 3.5$~Gyr, 
lopsidedness of moderate amplitude is still imprinted in the stellar disk 
$\sim 2$~Gyr after the last accretion event. The final disk which corresponds 
to $\sim 4.3$~Gyr after the last satellite impact exhibits virtually no signs 
of lopsided asymmetries at $R \lesssim 4.5 R_d$ and very weak lopsidedness 
at larger radii. The lifetimes of the reported asymmetries 
($\sim 1$~Gyr) are in agreement with estimates from phase mixing and winding 
arguments \citep[e.g.,][]{Baldwin_etal80,Rix_Zaritsky95} and direct 
analysis of $N$-body simulations \citep[e.g.,][]{Zaritsky_Rix97,Bournaud_etal05,
Mapelli_etal08}. A more thorough discussion regarding the longevity of lopsidedness 
excited by halo substructure is beyond the scope of the present paper and is 
deferred to future work. 

The main implication of the results reported in Figure~\ref{fig7} is that the 
{\it continuous} accretion of satellites over a galaxy's lifetime 
constitutes a significant source of external perturbations that 
excite as well as maintain lopsidedness in stellar disks at observed levels. 
This might be important as many observational studies find no 
correlation between the presence of nearby companions and lopsidedness 
in disk galaxies \citep[e.g.,][]{Zaritsky_Rix97,Wilcots_Prescott04,Bournaud_etal05}.


\begin{figure*}[t]
\hspace{0.1cm}
\begin{tabular}{c}
 \includegraphics[scale=0.55]{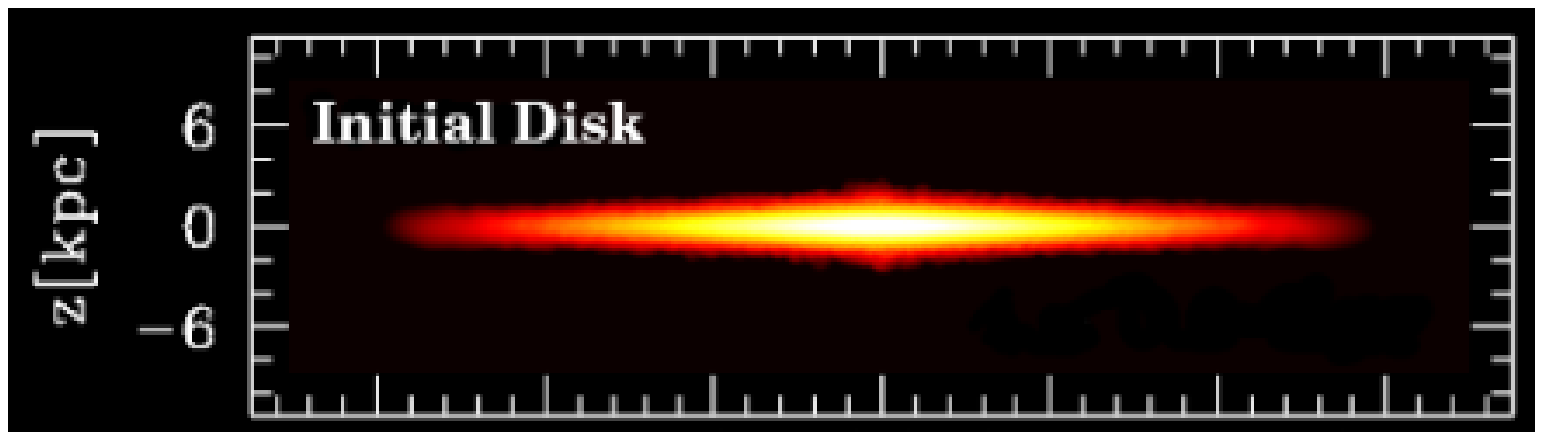}\vspace{-0.1cm}\\
  \includegraphics[scale=0.55]{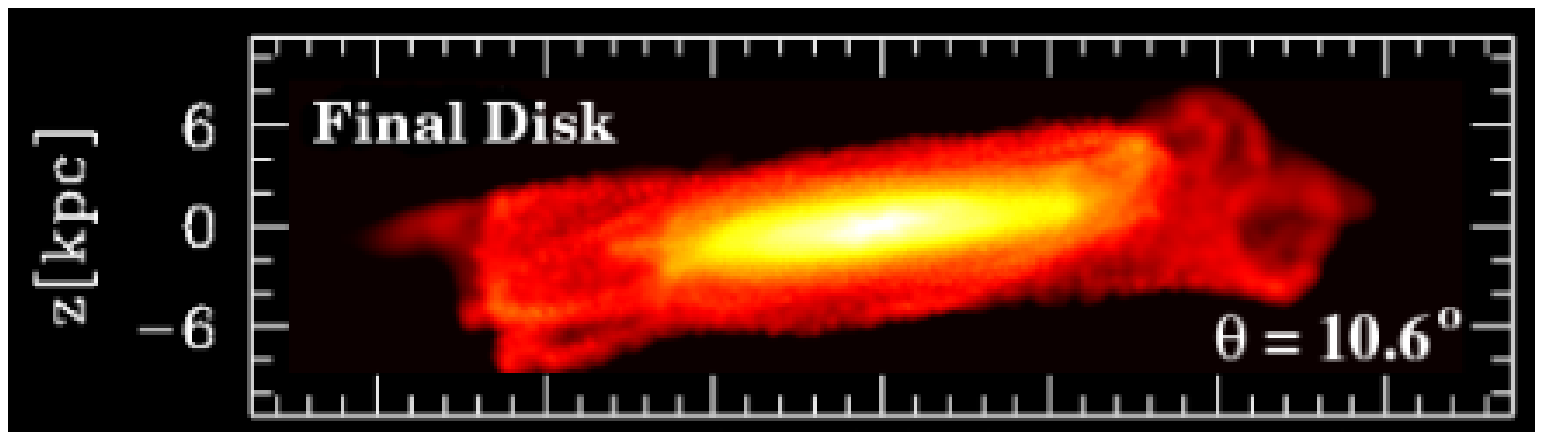}\vspace{-0.1cm}\\
   \includegraphics[scale=0.55]{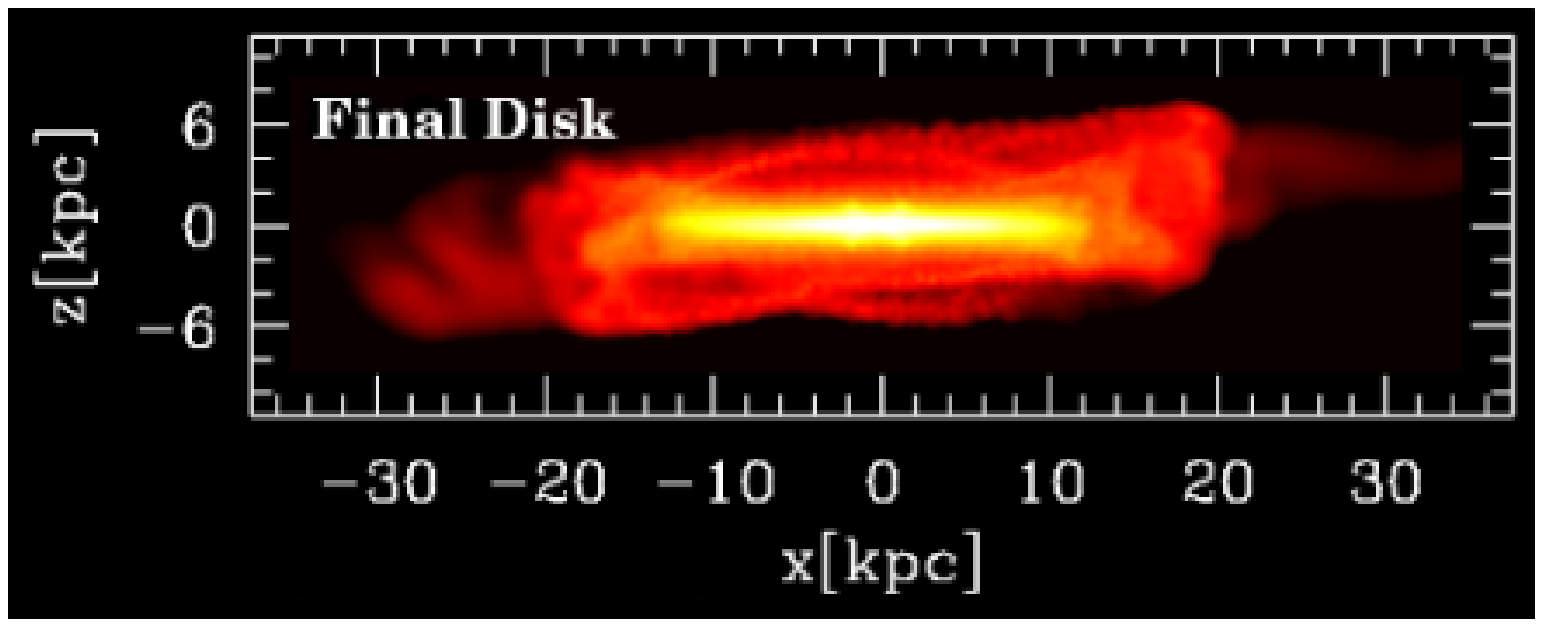} 
\end{tabular}
\hspace{-0.3cm}
\begin{tabular}{c}
        \includegraphics[scale=0.43]{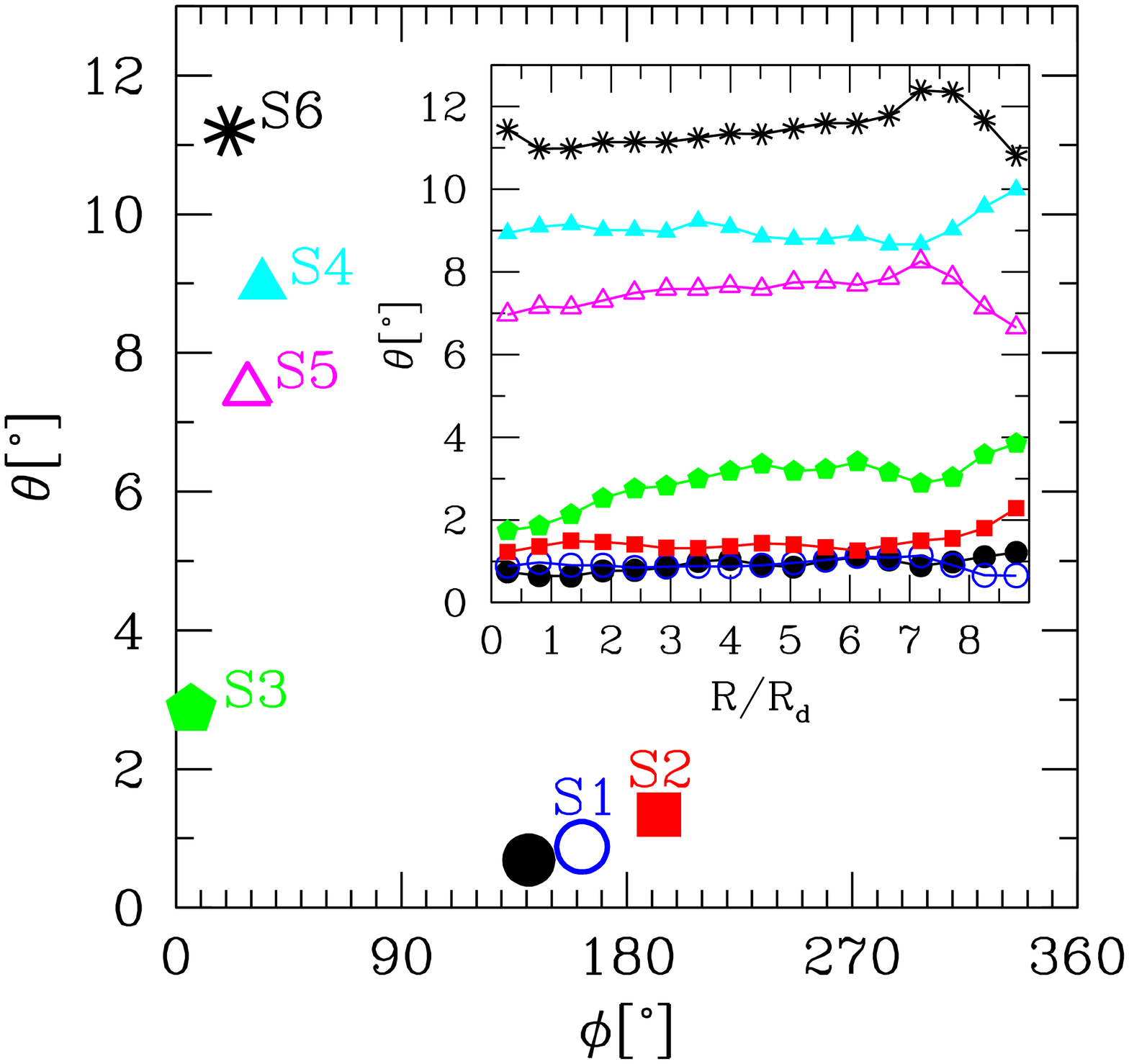} 
\end{tabular}
\caption{Disk tilting. {\it Left:} Density maps of the disk in galaxy 
  model D1 viewed edge-on. The {\it upper} panel shows the initial disk, 
  whereas the {\it middle} and {\it bottom} panels depict the final 
  disk in the initial coordinate frame and the frame defined by the 
  three principal axes of its total inertia tensor, respectively.
  The zenith angle $\theta$ between the two coordinate frames is indicated 
  in the middle panel. {\it Right:} A scatter plot of the evolution 
  of the angular position of disk pole relative to initial disk pole, 
  where $\phi$ denotes the azimuthal angle. Different symbols correspond to 
  individual satellite passages from S1 to S6. Filled circles show results 
  for the initial disk evolved in isolation for a timescale equal to that 
  of the combined satellite passages. The inset presents the 
  evolution of angle $\theta$ as a function of radius from the center of 
  the disk in units of the disk truncation radius, $R_{\rm out}=30\kpc$. 
  Interactions with infalling satellites of the kind expected in {\LCDM} 
  models can drive substantial tilting in galactic disks.
\label{fig8}}
\end{figure*}



\begin{figure*}[t]
\centerline{\epsfxsize=7.2in \epsffile{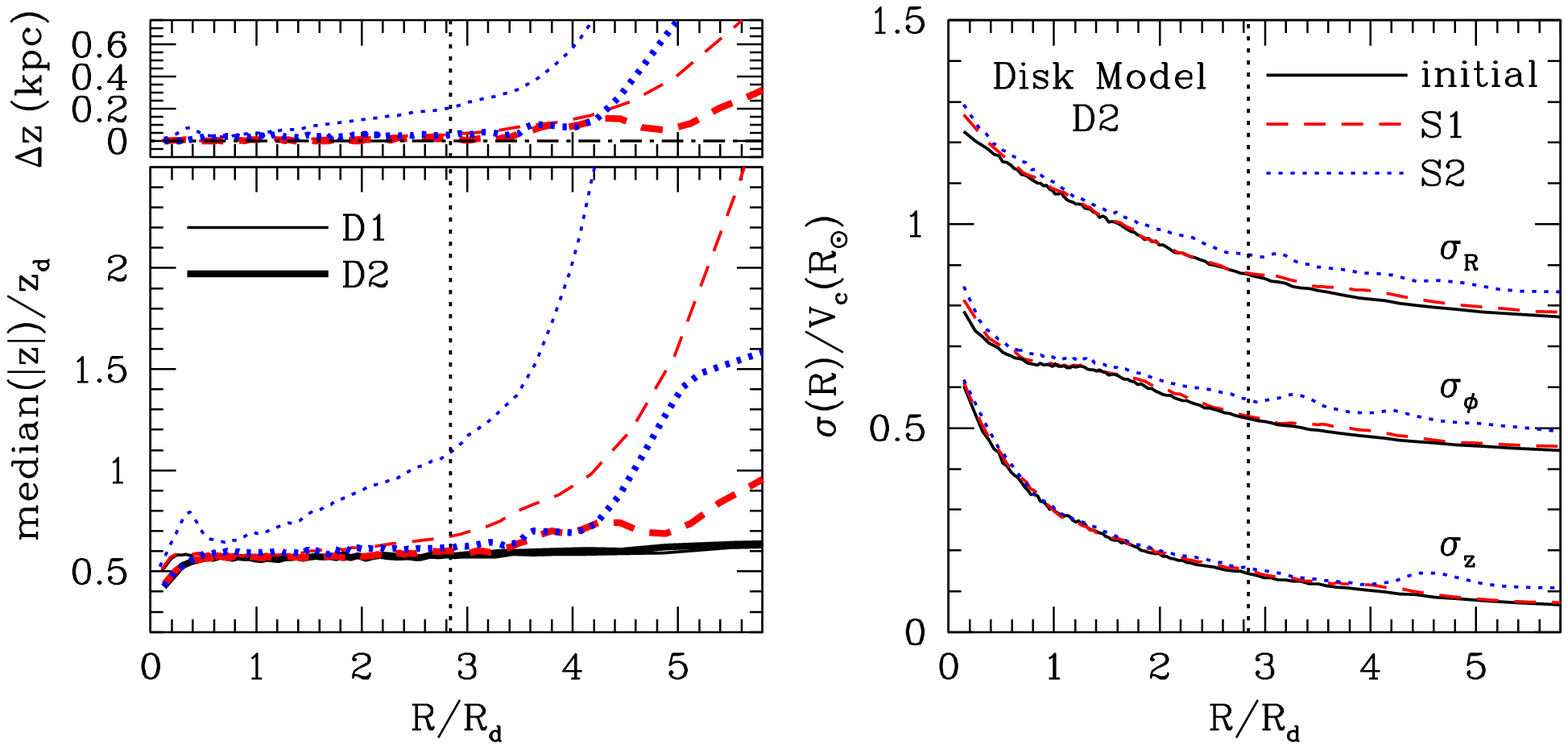}}
\caption{The effect of the initial disk thickness on the response of galactic disks
  to encounters with satellites. {\it Bottom left panel:} Thickness profiles 
  as a function of projected radius in units of the radial scale 
  length of the initial disk, $R_d$. {\it Thin} lines show the evolution of thickness 
  in the standard disk galaxy model D1 ($z_d = 0.4\kpc$) after the interactions with 
  subhalos S1 and S2. {\it Thick} lines present results for the same impacts 
  onto the much thicker disk galaxy model D2 ($z_d = 1\kpc$). {\it Dashed} 
  and {\it dotted} lines correspond to satellites S1 and S2, respectively, 
  while the {\it solid} curves show results for the initial disks.
  All curves are normalized to the initial scale height of each corresponding disk. 
  {\it Upper left panel:} Thickness increase, $\Delta z \equiv
  [\rm{median}(|z|)_{\rm final} - \rm{median}(|z|)_{\rm
    initial}]$. Here $\rm{median}(|z|)_{\rm initial}$ and
  $\rm{median}(|z|)_{\rm final}$ denote the thicknesses of the disk 
  initially and after each subhalo impact, respectively. 
  {\it Right:} Evolution of the disk velocity ellipsoid in galaxy model D2. 
  Bottom to top: $\sigma_z$, $\sigma_\phi + 100\kms$, and $\sigma_R + 175\kms$. 
  The line types are as in the left panel, but the results for galaxy 
  model D1 are not shown. All profiles are normalized to the total circular
  velocity of model D2 at the solar radius, $V_{\rm c}(R_\odot)=235.6\kms$. 
  The vertical dotted lines in all panels indicate the location of the solar 
  radius, $R_\odot$. Thicker disks are less susceptible to damage by infalling
  subhalos compared to their thin counterparts.
\label{fig9}}
\end{figure*}


\subsection{Disk Tilting}
\label{sub:tilting}

As discussed earlier, angular momentum exchange between the satellites and 
the disk is expected to tilt the disk relative to its initial orientation.
The left panel of Figure~\ref{fig8} presents edge-on density maps of the initial 
and final disk. The final disk is 
shown relative to both the initial coordinate frame and the frame defined by 
the principal axes of the total disk inertia tensor. 

The panel reveals that the simulated subhalo impacts cause a significant amount 
of disk tilting. For this particular accretion history, the angle by which the
disk is tilted from its original plane is $\theta \sim 11\degrees$, implying 
a significant transfer of angular momentum from the infalling satellites to the host 
galactic disk. This indicates that not all of the orbital energy associated with 
the vertical motion of the subhalos is converted into random motions of disk 
stars causing disk thickening. When viewed in the original coordinate frame, the 
final disk appears visually much thicker, which is simply a consequence of the fact 
that the disk is substantially tilted by the orbiting satellites. Employing 
the tilted coordinate frame for all computations of disk properties is required as, 
in the original coordinate frame, quantities such as the scale height, $z(R)$, and 
the vertical kinetic energy, $T_z$, will be artificially inflated.
Before continuing, we note that in the absence of satellites there is some 
exchange of angular momentum between the disk and the dark matter halo of the
primary galaxy. This exchange results in tilts of $\lesssim 1\degrees$ when
the galaxy is evolved in isolation for a timescale equal to that of the
combined subhalo passages.

The right panel of Figure~\ref{fig8} shows a scatter plot of the evolution of the 
angular position of disk pole, defined by the azimuthal and zenith angles 
$(\phi,\theta)$, relative to initial disk pole as a function of satellite 
passages S1-S6. These angles are computed
considering all disk particles within $15\kpc$ from the disk center, 
but we note that the results do not depend sensitively 
upon this cutoff. The combined action of the first two satellites S1 and S2 serves to tilt 
the galactic disk by a very small amount $\theta \lesssim 1.5\degrees$.
Both subhalos are on nearly polar 
orbits (Table~\ref{table:sat_param}) and thus transfer of angular momentum to the
disk is expected to be minimal. In this case, virtually all of the orbital 
energy associated with the $z$ motion of the infalling satellites is converted
into random vertical stellar motions causing thickening of the disk. The next four 
interactions involve subhalos on both prograde (S3,S5) and retrograde (S4,S6) 
orbits, so angular momentum exchange can take place. While these impacts 
result in $\theta \sim 11\degrees$, they keep $\phi$ between $(0-35)\degrees$, 
so they tilt the disk in almost the opposite direction compared to passages S1 
and S2. It is interesting that these four accretion events continue to tilt
the disk in this small range of $\phi$. Lastly, the disk response to subhalo passages
S3-S6 suggests that, in addition to the mass, the orbital orientation of the 
infalling satellite is crucial in determining the amount of disk tilting. 
Retrograde encounters appear to be associated with more pronounced tilting
compared to their prograde counterparts. We address this issue explicitly
in \S~\ref{sub:orbit.orient}. 

The inset in the right panel of Figure~\ref{fig8} presents the evolution of
the angle $\theta$ between the original and tilted coordinate frames as a 
function of radius from the center of the disk. The maximum radius we consider
for this calculation is $r=0.85 R_{\rm out}=25.5\kpc$, where $R_{\rm out}$ 
denotes the disk truncation radius \citep{Widrow_Dubinski05}, which is set by the
requirement that each radial bin should contain at least $1000$ particles. 
This is an empirical criterion which ensures a robust determination of 
the inertia tensor in each bin. The inset illustrates that the angle $\theta$ between the
original and tilted coordinate frames is not constant as a function of radius 
and changes are not monotonic. Different regions of the disk may thus be tilted to different 
degrees by the infalling satellites. Interestingly, the dense inner disk ($r/R_{\rm out} \lesssim 0.5$), 
which contains most of the disk mass, responds nearly as a rigid body to the accretion events 
\citep[see also][]{Shen_Sellwood06}. While the tilting angles of the inner and outer 
parts of the disk are different from each other, which is indicative of the presence of 
warping, the maximum difference is relatively small ($\Delta \theta \lesssim 2\degrees$).
Furthermore, as inertia increases quadratically with distance, the direction of the axes in the tilted 
coordinate frame is mainly determined by the outer parts of the disk. 
These facts justify our decision to employ the coordinate frame defined by the 
principal axes of the total inertia tensor of the tilted disk to compute disk
properties rather than to adopt the appropriate local coordinate frame for
each radius within the tilted disk. 

\section{Sensitivity of the Disk Dynamical Response to Modeling Choices}
\label{sec:effects}

In this section we explore various factors that could influence the dynamical 
response of galactic disks to infalling satellites. We discuss in turn 
initial disk thickness, the presence of a bulge component in the primary disk, 
the internal density distribution of the infalling systems, and the relative
orientation of disk and satellite angular momenta.

\subsection{Initial Disk Thickness}
\label{sub:initial.thickness}

The results in \S~\ref{sub:thickening} and \S~\ref{sub:heating}
suggest that subsequent subhalo encounters with already thickened disks
produce smaller changes in disk thickness and disk velocity ellipsoid 
compared to the initial accretion events. In what follows we investigate
the effect of initial disk thickness on the thickening and heating 
of a galactic disk by infalling satellites in a more controlled manner. 
For this reason we constructed a self-consistent disk galaxy identical to 
the standard thin disk model but with a scale height that 
was larger by a factor of $2.5$, $z_d = 1\kpc$ (model D2). We then repeated 
the impacts of satellites S1 and S2 onto this thicker disk model.  

Figure~\ref{fig9} shows the evolution of disk thickness as quantified by 
$\rm{median}(|z|)$, and disk velocity ellipsoid $(\sigma_R, \sigma_\phi,
\sigma_z)$, after each subhalo passage. There is little additional thickening 
in model D2 within roughly four scale lengths. Even beyond this radius, D2 is 
much less flared compared to disk model D1. More specifically, D2 thickens
by less than $50$~pc at $R_\odot$, while the thickness of D1 increases 
by more than $200$~pc at the same radius. Likewise, we find that 
the velocity structure of the thicker initial disk is significantly less 
influenced by the perturbers. The velocity ellipsoid of model D2 increases 
by less than $\sim 15\%$ at the solar radius due to the 
interactions with subhalos S1 and S2. The corresponding increase for 
galaxy model D1 is slightly less than a factor of $2$ (Figure~\ref{fig5}). 
It is also worth noting that model D2 does not form a bar as a result of 
the subhalo impacts and is associated with a much less prominent 
spiral structure compared to the thinner disk D1. These findings 
confirm that thicker disks exhibit a much weaker global dynamical 
response to accretion events compared to their thin counterparts. 


\begin{figure}[t]
\centerline{\epsfxsize=3.5in \epsffile{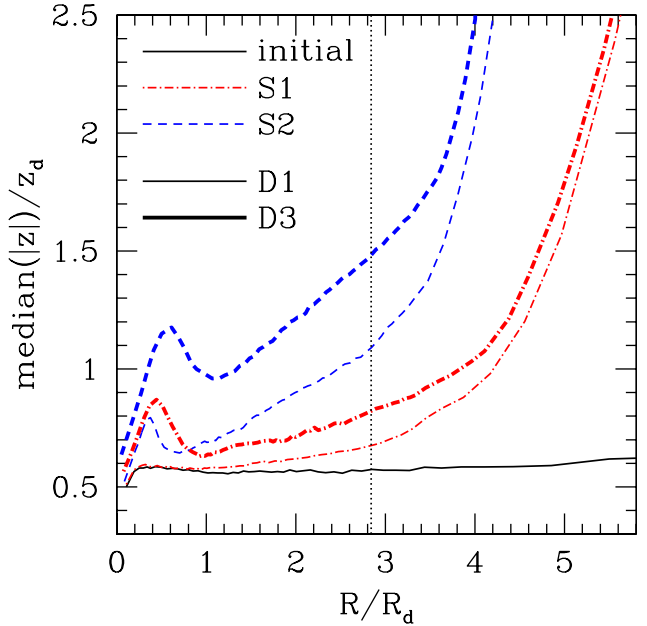}}
\caption{The effect of a bulge component on disk thickening. 
  {\it Thin} lines show thickness profiles for disk galaxy model D1 after 
  the encounters with subhalos S1 and S2. 
  {\it Thick} lines present results for the same impacts onto the bulgeless 
  disk model D3. Thicknesses and radii are normalized to the scale height, 
  $z_d$, and radial scale length, $R_d$, of the initial disk.
  {\it Dot-dashed} and {\it dashed} lines correspond 
  to satellites S1 and S2, respectively, while the {\it solid} curve shows
  results for the initial disk. A massive bulge enhances the resilience of 
  galactic disks to satellite impacts by both absorbing part 
  of the orbital energy deposited by the infalling subhalos and stabilizing 
  the disks against the development of global non-axisymmetric instabilities. 
\label{fig10}}
\end{figure}


\subsection{Presence of a Bulge Component}
\label{sub:bulge}

The results presented in Figure~\ref{fig4} demonstrate that the inner 
disk regions exhibit substantial resilience to encounters with 
subhalos. Apart from the larger binding energy of the inner 
exponential disk, the presence of a massive, central bulge with 
($M_ b \simeq 0.3 M_{\rm disk})$ in disk model D1 may be responsible 
for the robustness of the inner disk. To address the effect of 
a bulge in reducing the damage induced in galactic disks by 
infalling satellites, we repeated encounters S1 and S2 after 
adiabatically evaporating the bulge component from disk model 
D1 (see \S~\ref{sub:galaxy_models}). We refer to the 
resulting bulgeless disk galaxy model as D3. 

Figure~\ref{fig10} shows the evolution of disk thickness after 
each subhalo passage for both galaxy models D1 and D3. Not 
surprisingly, this exercise demonstrates that a massive bulge 
enhances the robustness of galactic disks to accretion events. 
Analysis of the disk velocity ellipsoids supports 
this conclusion. As expected, the relative effect of the bulge grows weaker 
as a function of radius from the center of the disk. 
The combined action of satellites S1 and S2 increases the thickness 
of model D3 at $R_\odot$ by a factor of $\sim 2.6$, compared to a 
factor of $\sim 2$ in the case of model D1. At larger radii 
($R \gtrsim 3.5 R_d$), model D3 is characterized by a more distinct 
flare compared to D1. This is interesting as the bulge is located 
in the inner regions of the disk and suggests that global instabilities 
which are curtailed by a massive bulge are efficient at driving 
significant evolution in the outer disk. Indeed, we stress that subhalo 
passages S1 and S2 excite a much stronger bar in model D3 compared to D1 
and that this bar already develops in response to the interaction with S1. 


\begin{figure*}[t]
\centerline{\epsfxsize=7.2in \epsffile{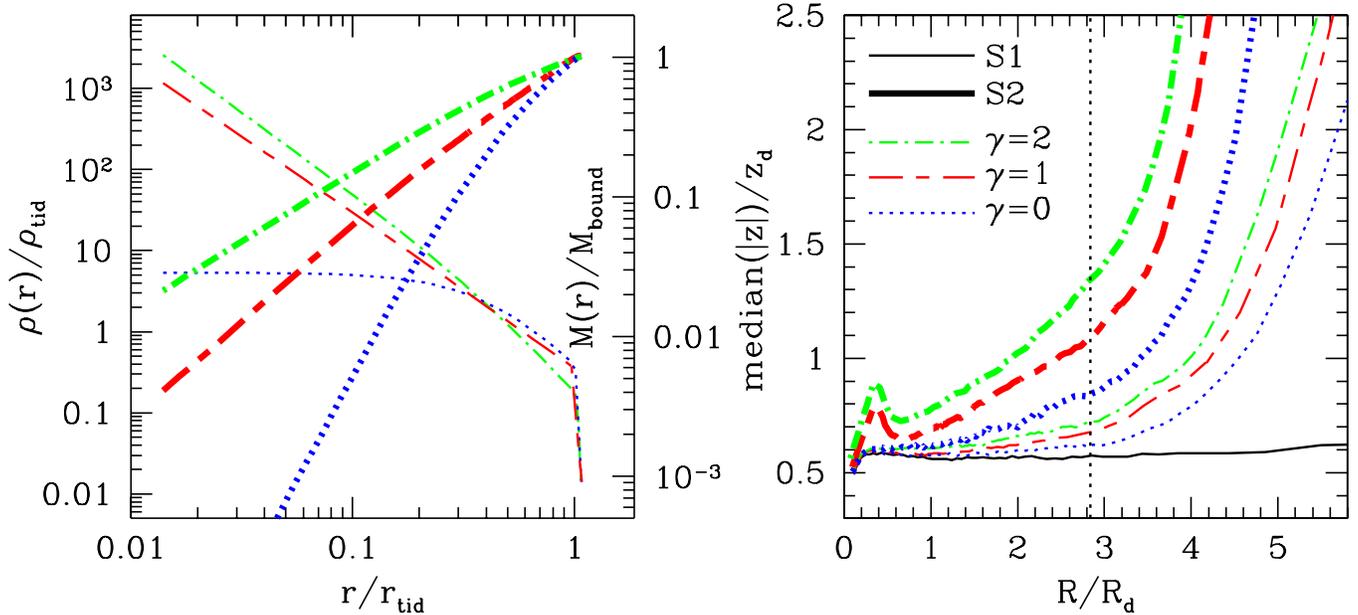}}
\caption{The effect of satellite internal density distribution on disk thickening
  {\it Left:} Spherically-averaged density, $\rho(r)$, ({\it thin lines}) and cumulative mass 
  profiles, $M(r)$, ({\it thick lines}) for infalling subhalos with different internal structure
  as a function of radius in units of the tidal radius, $r_{\rm tid}$. Densities and masses are 
  normalized to the enclosed density within the tidal radius, 
  $\rho_{\rm tid} \equiv 3 M_{\rm bound} / 4 \pi r_{\rm tid}^{3}$ and the bound mass of the system, 
  $M_{\rm bound}$ ({\it right axis}), respectively. Results are presented for 
  cosmological satellite S2. Density profiles differ only in the asymptotic
  inner slope $\gamma$ in Eq.~(\ref{general_density}). The {\it short dashed - 
  long dashed} lines show results for the standard density profile with $\gamma = 1$, while the {\it dotted} 
  and {\it dot-dashed} lines correspond to profiles with an inner power-law index equal to 
  $\gamma=0$ and $\gamma=2$, respectively. The adopted density distributions have significantly different shapes, 
  but they correspond to exactly the same bound mass.
  {\it Right:} Evolution of disk thickness induced by satellites S1 and S2 following the 
  density profiles described above. {\it Thin} and {\it thick} lines display
  results for the passages of satellites S1 and S2, respectively, and line
  types are the same as in the left panel. {\it Solid} line shows results for the initial disk model D1. 
  Thicknesses and radii are normalized to the scale height, $z_d$, and radial scale length, $R_d$, 
  of the initial disk. The vertical dotted line indicates the location of the solar radius, $R_\odot$.
  Infalling subhalos with steeper density distributions are more efficient perturbers compared 
  to their cored counterparts. 
\label{fig11}}
\end{figure*}


\subsection{Satellite Internal Density Distribution}
\label{sub:internal.density}

It is worthwhile to examine the importance of the satellite internal density 
distribution to the structural changes induced in galactic disks. 
This interest is motivated by the potential differences in subhalo structure 
in models with modified dark matter properties and density fluctuation spectra 
\citep[e.g.,][]{Hogan_Dalcanton00,Avila-Reese_etal01}, or the presence 
of baryonic components in some fraction of the accreting systems. To this end, we 
repeated the encounters between disk galaxy model D1 and satellites S1 and S2 after 
assigning different density profiles to each infalling subhalo.   

We employed density profiles that differ only in their asymptotic inner slopes, 
$\gamma$ [see Eq.~(\ref{general_density})] and required the total bound mass of each 
object to be fixed to its fiducial value. We studied two, relatively extreme options 
for the satellite density distributions with the intent that these bracket the range 
of potential subhalo structures. The first corresponds to a ``steep'' profile 
with an inner power-law index equal to $\gamma=2$, while the second follows a ``shallow'' 
profile with a constant density core, $\gamma=0$. The size of the density core was chosen 
to be approximately equal to $10\%$ of the tidal radius of each satellite, $\gtrsim 2\kpc$
(Table~\ref{table:sat_param}). The steep profile addresses the possibility of a subhalo
containing a galaxy and which may have undergone adiabatic contraction in response to 
the growth of baryons \citep{Blumenthal_etal86}, while the cored profile is designed to 
overestimate any density suppression that may result in modified dark matter models.  
The density, $\rho(r)$, and cumulative mass profiles, $M(r)$, for all initial satellite 
models are presented in the left panel of Figure~\ref{fig11}. 

The right panel of Figure~\ref{fig11} presents the evolution of disk thickness 
in these alternative models compared with the standard cases.  
Not surprisingly, the amount of disk thickening increases with the steepness of the 
inner satellite density profiles at fixed initial subhalo mass. The combined action of 
cored satellites S1 and S2 increases the disk thickness at the solar radius 
by $\sim 45\%$, compared to a factor of $\sim 2$ in the case of the standard 
satellites. The cumulative effect of the steep subhalos S1 and S2 is 
to increase the disk thickness by nearly a factor of $\sim 2.5$ at 
$R_\odot$. The relative thickness differences driven by the different 
density distributions becomes more pronounced at large radii where the disk 
self-gravity, and thus its restoring force, is weaker. It is worth noting that steep 
satellites S1 and S2 excite a stronger bar compared to their $\gamma=1$ counterparts,
whereas cored satellites S1 and S2 do not induce a bar at all on the timescales of 
these simulations. These findings indicate that satellite density structure 
as characterized by the inner slope or concentration is an important factor in 
determining the amount of damage and global dynamical evolution that subhalos 
impart upon disks. This is an important point in light of the fact that most
previous numerical investigations of satellite-disk interactions did not 
realize their satellite models with the exact density structure of cosmological 
subhalos.

The primary reason for the dependence of disk dynamical evolution on satellite 
structure is that subhalos with shallower profiles are significantly 
less tightly bound. As a consequence, they generate much smaller potential 
fluctuations when they cross the disk and correspondingly cause
less thickening and heating. In addition, cored satellites lose mass 
more efficiently {\it even} prior to the actual disk crossing. As
a result, they penetrate the disk with less bound mass, and thus 
less energy and angular momentum available to be delivered to the disk 
stars. Owing to their lower self-gravity, the outer disk regions are expected 
to be more sensitive to the amount of orbital energy deposited by the infalling 
satellites. This is also confirmed by the findings presented in the right panel 
of Figure~\ref{fig11}.


\begin{figure}[t]
\centerline{\epsfxsize=3.5in \epsffile{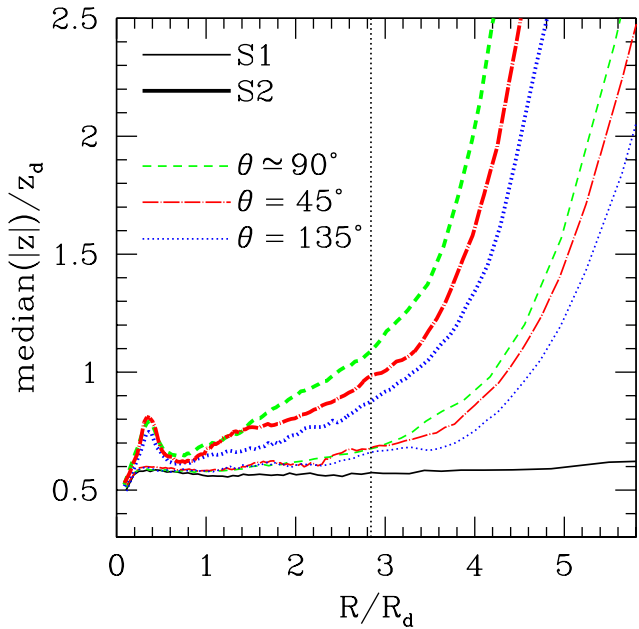}}
\caption{The effect of satellite orbital orientation on disk thickening.
  Evolution of disk thickness profiles for disk galaxy model D1 in response 
  to encounters with subhalos S1 ({\it thin lines}) and S2 ({\it thick lines}). 
  Thicknesses and radii are normalized to the scale height, $z_d$, and 
  radial scale length, $R_d$, of the initial disk. {\it Dashed} lines correspond to 
  our fiducial experiments with $\theta \simeq 90\degrees$. {\it Dot-dashed} 
  and {\it dotted} lines show results for satellites S1 and 
  S2 on a prograde ($\theta = 45\degrees$) and a retrograde orbit 
  ($\theta = 135\degrees$), respectively. {\it Solid} line shows results for the 
  initial disk model D1. The vertical dotted line indicates the location of the solar 
  radius, $R_\odot$. Prograde encounters are more efficient at thickening the disk 
  compared to their retrograde counterparts.
\label{fig12}}
\end{figure}


\subsection{Orientation of Satellite Orbits}
\label{sub:orbit.orient}

It is interesting to investigate any correlation between disk dynamical 
response and the orientation of satellite orbits. To this end, we repeated 
satellite passages S1 and S2 varying the angle $\theta$ 
between their orbital angular momentum and the initial angular momentum of the 
disk in galaxy model D1. We considered two cases with identical initial orbital 
inclinations with respect to the plane of the host disk ($i=45\degrees$), but 
exactly opposite directions with respect to its rotation. The first is a prograde 
orbit with $\theta = 45\degrees$ and the second is a retrograde orbit with 
$\theta = 135\degrees$. 

Figure~\ref{fig12} compares the results of these simulations with our fiducial 
experiments in which S1 and S2 are on nearly polar orbits with 
$\theta \simeq 90\degrees$ (Table~\ref{table:sat_param}). Encounters with subhalos 
on prograde orbits are more efficient at thickening the disk compared to their 
retrograde counterparts. This is suggestive of a resonant coupling between the 
satellite and disk angular momenta that is suppressed in retrograde encounters.
After the completion of passage S2, differences in disk thickness 
between prograde and retrograde interactions manifest at all radii. 
Prograde satellites S1 and S2 increase the disk thickness by $\sim 20\%$ more 
compared to the corresponding retrograde models. Moreover, retrograde S1 and
S2 impacts excite a weaker bar than the prograde encounters.  

Interestingly, neither prograde nor retrograde orbits cause as much thickening 
(or heating) as the nearly polar orbits of S1 and S2 in our fiducial simulations. 
However, the combined action of S1 and S2 tilts the disk by $\theta \sim 7.5 \degrees$ 
in the prograde case and $\theta \sim 9.5 \degrees$ in the retrograde case. This is to 
be compared to $\theta \lesssim 1.5\degrees$ in the fiducial case where S1 and S2 are 
on nearly polar orbits. This reinforces our interpretation in \S~\ref{sub:tilting}
of the different degrees of tilting among encounters S1-S6 as a manifestation of the 
different orbital orientations of the infalling satellites relative to the disk. 
During the encounters, part of the orbital energies and angular momenta 
of the infalling satellites are transferred into the disk. A fraction of this energy is 
converted into random vertical motions of disk stars, causing disk thickening, and 
a fraction is added to the disk coherently leading to its tilting. Because polar 
orbits do not transfer angular momentum, the fraction of the satellite orbital energy 
that thermalizes in the disk causing its thickening is larger. On the other hand,
efficient exchange of angular momentum results in a large-scale, coherent tilting 
of the disk, and thus in smaller degrees of thickening, an effect that is most 
pronounced in the retrograde orbits.  

We note that since we only consider interactions with satellites that each pass the disk 
only once, our numerical experiments cannot address the importance of satellite decay 
rate and mass loss in affecting any of the aforementioned trends. This is especially 
relevant in the case of retrograde and polar orbits which are known to suffer slower 
tidal disruption and orbital decay compared to their prograde counterparts 
\citep{Velazquez_White99}.

\section{Discussion}
\label{sec:discussion}

We have demonstrated that encounters with halo substructure in the context of 
the {\LCDM} cosmogony imprint a wealth of dynamical signatures in thin
galactic disks. These signatures include considerable thickening and heating
at all radii, surface density excesses
resembling those of observed antitruncated disks, lopsidedness at levels similar 
to those measured in observational samples of disk galaxies, and 
significant tilting. In Paper I, we also showed that the same accretion events produce 
conspicuous flares, bars, low-surface brightness ring-like features in the outskirts 
of the disk, faint filamentary structures above the disk plane, and a complex vertical 
structure that is well-described by a superposition of thin and thick disk 
components. All of these findings highlight the significant role of
encounters with CDM substructure in setting the structure of disk 
galaxies and driving galaxy evolution. 

In the present study, we have only utilized a {\it conservative} subset 
($0.2 M_{\rm disk} \lesssim M_{\rm sub} \lesssim M_{\rm disk}$, $r_{\rm peri} 
\lesssim 20\kpc$) of a typical accretion history of a Galaxy-sized host 
halo to seed our controlled satellite-disk encounter simulations, and have neglected 
interactions with extremely massive subhalos ($M_{\rm sub} \gtrsim M_{\rm
  disk}$) that could prove ruinous to thin-disk survival \citep{Purcell_etal08}. 
In this sense, our results are relevant to systems that have already 
experienced the most destructive events since $z=1$ and have re-grown 
their thin disks since. Extending our selection criteria to include 
subhalos with smaller masses ($M_{\rm sub} \lesssim 0.2 M_{\rm disk}$) 
and/or larger pericenters ($r_{\rm peri} \gtrsim 20\kpc$) will definitely 
result in many more accretion events to consider. This will perhaps change the 
detailed results for the disk dynamical response subject to a {\LCDM}
accretion history, but it will not affect any of the qualitative 
conclusions of this study. If anything, the tidal effects of halo 
substructure will be more pronounced compared to that we reported here. 

We also emphasize that our numerical experiments were not designed to elucidate 
the effect of subhalo bombardment on the structure of specific galaxies 
such as the MW or M31. Rather, the main goal of the present study was to 
investigate the most {\it generic} dynamical signatures induced in thin galactic
disks by a {\it typical} {\LCDM}-motivated satellite accretion history.
To this end, we utilized only four Galaxy-sized dark matter halos to extract
representative orbits and subhalo merger histories, and the controlled 
simulations of satellite-disk interactions were based on the accretion history 
of just one of these host halos. Nevertheless, all four of these host halos 
showed similar numbers of substantial accretion events onto their central
regions, and their accretion histories are typical of systems in this mass
range \citep[e.g.,][]{Wechsler_etal02}. In addition, the bulk of the disk
dynamical response illustrated in Figures~\ref{fig4},~\ref{fig5}, ~\ref{fig6}, 
and ~\ref{fig7} was driven by the single most massive substructure of the 
simulated accretion history ($M_{\rm sub} \sim 0.6 M_{\rm disk}$, $r_{\rm tid} 
\simeq 20 \kpc$, and $r_{\rm peri} \lesssim 10\kpc$). We identified infalling 
satellites of this kind in all four of the host halo histories studied 
(Figure~\ref{fig1}), and such accretion events should have been ubiquitous 
in the history of Galaxy-sized halos since $z \sim 1$ 
\citep[e.g.,][]{Stewart_etal08}. All of these facts suggest that our
simulation set has reasonably captured the global dynamical effects of halo 
substructure on thin galactic disks. For detailed comparisons with
specific disk galaxies it would be required to explore a range of accretion 
histories and orbital distributions of infalling objects using a larger 
sample of halos from cosmological $N$-body simulations. 

The analysis presented in Figures~\ref{fig4} and~\ref{fig5} highlight 
two distinctive signatures that infalling satellites imprint in 
the structure and kinematics of the host galactic disk.
Figure~\ref{fig4} demonstrates that the thickness of the simulated 
disk increases with galactocentric radius, while Figure~\ref{fig5} 
illustrates that the disk velocity dispersion profiles become nearly flat 
at large radii ($\gtrsim 3 R_d$). Both features appear to be a 
natural result of encounters with CDM substructure and have important
observational consequences. The absence of these signatures in a 
significant fraction of disk galaxies would be difficult to reconcile 
in the context of the present study and could be used to potentially 
falsify our proposed model.

Interestingly, disk-flaring is seen in the MW in both the 
stellar disk \citep{Lopez_etal02,Momany_etal06} and atomic hydrogen 
distribution \citep[e.g.,][]{Merrifield92,Nakanishi_Sofue03}. 
Flaring is also observed in edge-on, external galaxies, in 
their stellar light \citep[e.g.,][]{DeGrijs_Peletier97,Narayan_Jog02}
and $H_{\scriptsize I}$ gas \citep[e.g.,][]{Brinks_Burton84,0lling96,Matthews_Wood03}.
Flat vertical velocity dispersion profiles are also reported in the outskirts of 
disk galaxies. Recently, \citet{Herrmann_etal09} performed a kinematic study of planetary
nebulae in the extreme outer disks of the nearby, nearly face-on spirals M83 and
M94. In both these systems, the kinematic evidence suggests that: (1) the
stellar disks flare dramatically in their outer regions, beyond $\sim 4$ 
disk scale lengths; and (2) at these large distances, the vertical 
velocity dispersions are nearly independent of radius ($\sigma_z \sim 20\kms$) 
rather than decreasing exponentially as expected for a constant mass-to-light 
ratio, constant scale-height exponential disk. These findings are in very good 
agreement with the theoretical predictions presented in Figures~\ref{fig4} 
and~\ref{fig5}, suggesting that the flaring and higher than expected values 
of $\sigma_z$ in the disks of M83 and M94 could be due to bombardment by halo
substructure. Exploring these constraints further would require extending 
the number of kinematic surveys in the outer disks of spiral galaxies 
as well as performing an extensive series of numerical experiments to fully 
sample the statistical variation in halo accretion histories predicted by
{\LCDM}. 

In the context of distinctive signatures induced in stellar disks by halo substructure, 
another intriguing result is reported in Figure~\ref{fig6}. This figure illustrates 
that the face-on surface density distribution of the simulated disk becomes shallower 
at large projected radii ($R \gtrsim 5 R_d$) as a result of the satellite
impacts, so that there is an excess of light relative to the exponential
profile of the inner disk. Such behavior is interesting in the context 
of the so-called antitruncated disks whose surface brightness profiles display
similar excesses at large distances, beyond $4-6$ disk scale lengths
\citep[e.g.,][]{Erwin_etal05,Pohlen_Trujillo06,Pohlen_etal07,Erwin_etal08}. 
Transport of angular momentum during the subhalo bombardment
causes the migration of disk material outward in radius, leading to the excess surface 
density at large radii. Of course, satellite-disk encounters of the kind considered here 
do not constitute the only mechanism for such angular momentum transfer. 

\citet{Younger_etal07} recently investigated the origin of antitruncated disks in face-on 
spirals using hydrodynamical simulations of minor mergers of galaxies. These authors 
suggested that the antitrucation is produced by the action of two competing 
effects: (1) merger-driven gas inflows deepening the central potential of the 
primary galaxy and contracting the inner surface density profile of the disk; 
and (2) angular momentum and stellar mass transfer in the outer regions, 
causing the disk to expand. The first process acts to maintain an inner 
scale length similar to that of the initial disk of the primary, while the 
second is responsible for generating the excess surface density relative 
to the exponential profile of the inner disk. This latter process is similar to 
what occurs in our numerical experiments. 

\citet{Younger_etal07} further argued that the combination of these two mechanisms 
produces antitruncated disks only in the hydrodynamical regime.
They argued that in collisionless interactions in which the central potential is not 
as deep, the disk expands more uniformly in response to the angular momentum transfer. 
As a result, the entire surface density distribution becomes flatter and the disk 
scale length increases at all radii. Our experiments are dissipationless, but the deep 
central potential needed to maintain the inner scale length may be provided by the 
massive, central bulge. Indeed, fitting the mass surface density of the final
disk to an exponential profile yields a scale length 
only $8\%$ larger than that of the initial disk. The implication is that the 
satellite-disk interactions that should be common during the hierarchical assembly
of galaxies may have a non-negligible role in the formation of at least some of the 
antitruncated disks, even when encounters are completely dissipationless.  
Of course, the inclusion of gasdynamics, star formation, and stellar
components in the infalling systems as well as a detailed comparison of the final
disk with observed antitruncated systems, as was done in \citet{Younger_etal07},
will be important to fully explore the origin of antitruncated disks in the 
context of hierarchical CDM.

Fourier analysis of their surface density illustrates that the simulated 
stellar disks also exhibit lopsidedness spawned by the interactions with 
CDM substructure (Figure~\ref{fig7}). This has important implications as a 
significant fraction of observed spiral galaxies display lopsided asymmetries in their 
gaseous and/or stellar distributions \citep[e.g.,][]{Baldwin_etal80,Richter_Sancisi94,Rix_Zaritsky95,
Zaritsky_Rix97,Matthews_etal98,Haynes_etal98,Rudnick_Rix98,Bournaud_etal05,Angiras_etal06,
Angiras_etal07,Reichard_etal08}. For example, \citet{Richter_Sancisi94} analyzed 
the $H_{\scriptsize I}$ distributions in $\sim 1700$ disk galaxies and
concluded that the frequency of lopsided systems in their sample exceeded 
$50\%$. Lopsided stellar disks are also ubiquitous 
\citep{Rix_Zaritsky95,Zaritsky_Rix97,Rudnick_Rix98,Bournaud_etal05,Reichard_etal08}. 
Other proposed mechanisms for generating lopsidedness have included tidal interactions and mergers 
\citep{Walker_etal96,Zaritsky_Rix97,Angiras_etal06,Angiras_etal07,Mapelli_etal08}, 
asymmetric accretion of intergalactic gas into the disk
\citep{Bournaud_etal05}, offsets between the disk and the dark matter halo \citep{Levine_Sparke98}, 
dynamical instabilities/processes internal to the disk \citep{Sellwood_Merritt94,Syer_Tremaine96,
Sellwood_Valluri97,DeRijcke_Debattista04,Saha_etal07,Dury_etal08}, and halo lopsidedness
\citep{Jog97,Jog99}. It is plausible that a number of mechanisms operate in
concert to foster lopsidedness.

Further comparison of our results with observational work on stellar lopsidedness is 
worthwhile. \citet{Zaritsky_Rix97} used near-infrared photometry to study a sample 
of $60$ {\it late-type} spiral galaxies in the field. These authors defined a
quantitative measure of lopsidedness as the radially-averaged ratio of the $m=1$ 
to $m=0$ Fourier amplitudes between $1.5$ and $2.5$ disk scale lengths $R_d$,
and denoted this quantity by $\langle A_1 \rangle$. Averaging over a range of
radii reduces the effect of isolated asymmetric peaks on the results 
of the observational analysis and provides a global measure of the
lopsidedness in each galaxy. \citet{Zaritsky_Rix97} found
that $\sim 30\%$ of all disk galaxies in their sample exhibit
significant stellar lopsidedness with $\langle A_1 \rangle \geq 0.2$. 
\citet{Rudnick_Rix98} followed up on this work by using R-band photometry 
to investigate asymmetries in $54$ {\it early-type} spirals. 
These authors reported a median value of $\langle A_1 \rangle$ 
in their sample of $0.11$ and that $20\%$ of their disk galaxies 
had $\langle A_1 \rangle \geq 0.19$. The similar amplitude 
of lopsidedness in disks with very different star formation rates indicates 
that the majority of the observable asymmetries in the stellar light of galactic 
disks reflects asymmetries in the stellar mass distribution rather than asymmetric 
star formation, confirming a dynamical origin of stellar lopsidedness. 
More recently, \citet{Bournaud_etal05} analyzed a sample of $149$ galaxies
from the Ohio State University Bright Galaxy Survey observed in the
near-infrared and reported a mean $\langle A_1 \rangle$ equal to $0.11$.

We have followed these authors and computed the quantity $\langle A_1 \rangle$ 
in our simulated stellar disks. All measurements were performed after allowing 
the disks to relax from the encounter with each infalling subhalo as described 
in \S~\ref{sub:control_sims}. As shown in Figure~\ref{fig6}, the slope of
the surface density profile and hence the disk scale length does not evolve 
significantly in the relevant radial range for measuring 
$\langle A_1 \rangle$, $1.5 R_d \lesssim R \lesssim 2.5 R_d$. 
Therefore, we adopted the scale length of the initial disk 
for these calculations and computed the quantity $\langle A_1 \rangle$ 
in each case by averaging $A_1$ between $~\sim 4.2\kpc$ and $~\sim 7\kpc$.
We also confirmed that using different bin numbers in calculating $\langle A_1
\rangle$ gives similar results.

For the simulated accretion history of host halo G$_1$, we derived $\langle
A_1 \rangle$ values in the range $0.10 \lesssim \langle A_1 \rangle \lesssim
0.16$. For reference, the initial 
{\it axisymmetric} disk had $\langle A_1 \rangle \simeq 0.003$.  We stress that 
much larger values of $\langle A_1 \rangle$ are excited during the subhalo 
impacts where tidal forces are strongest, but these are shorter-lived. 
As expected, the amplitude of $\langle A_1 \rangle$ decreases steadily after 
the last satellite passage, reaching a value of $\sim 0.013$ in the 
final disk. We also found that the magnitude of the induced 
lopsided asymmetries depends sensitively on the structural properties of 
the infalling subhalos as well as those of the galactic disks. For example, 
satellites modeled with steep density profiles generated 
a significantly stronger lopsidedness compared to that of the cored counterparts. 
In addition, bulgeless disks develop more pronounced lopsided asymmetries in their
inner regions compared to disks with massive bulges. These trends can 
be readily understood as a consequence of the strength of subhalo tidal 
perturbations onto galactic disks. Extending the mass spectrum of infalling
subhalos to include the most massive systems with $M_{\rm sub} \gtrsim M_{\rm
  disk}$ \citep[e.g.,][]{Purcell_etal08} would be required to examine whether 
satellite bombardment can explain the largest lopsided asymmetries observed 
in some galaxies. Additional mechanisms for lopsidedness, as discussed above, 
may also need to be invoked in this respect. 

The above analysis demonstrates that encounters with CDM substructure can
excite lopsidedness in stellar disks at levels similar to those measured in 
observational samples of disk galaxies \citep{Rix_Zaritsky95,Zaritsky_Rix97,Rudnick_Rix98}.
It also indicates that {\LCDM}-motivated subhalo accretion histories can
maintain these lopsided asymmetries for a significant fraction of the cosmic
time ($\gtrsim 5$~Gyr). This becomes particularly relevant 
in light of the fact that several studies find no 
observed correlation between the presence of nearby companions and disk 
lopsidedness\citep[e.g.,][]{Zaritsky_Rix97,Wilcots_Prescott04,Bournaud_etal05}.

Our simulations also suggest that significant disk tilting may result in 
response to encounters with CDM subhalos (Figure~\ref{fig8}). Disk tilting 
has potentially important implications. The formation of disk 
galaxies remains poorly understood despite some recent advances 
\citep[e.g.,][]{Weil_etal98,Abadi_etal03,Sommer_Larsen_etal03,Governato_etal04,
Robertson_etal04,Governato_etal07}. The relation between the angular momenta
of galactic disks and the net angular momenta of their host halos remains
unclear. Dark matter halos typically have their net angular momenta aligned with the shortest of their 
principle axes \citep[though the degree depends on halo mass, e.g.,][]{Bailin_Steinmetz05}.  
In the Local Group, the disks of the MW and M31 seem to have angular momenta
that point along the larger-scale structure delineated by the Local Group
dwarf galaxies \citep{Majewski94,Hartwick00,Willman_etal04,Zentner_etal05b,Libeskind_etal05}.  
Several studies have attempted to test for such alignments between disk 
principle axes and large-scale structure 
\citep{Zaritsky_etal97,Yang_etal06,Azzaro_etal06,Azzaro_etal07,Bailin_etal08}, 
but little evidence for either alignment have been found in statistically-large 
samples. Our results indicate that in addition to intrinsic scatter among 
the angular momenta of halos and their large-scale environments \citep{Bailin_Steinmetz05} 
such alignment may be diluted because disks, once formed, may tilt in response to 
numerous interactions with infalling satellites. 

As far as the robustness of galactic disks to encounters with satellites is 
concerned, Figures~\ref{fig4},~\ref{fig5}, and~\ref{fig9} illustrate 
that thicker disks are relatively more resilient to subhalo bombardment 
compared to their thin counterparts (see, however, \citealt{Sellwood_etal98}).
Infalling satellites deposit energy into galactic disks via impulsively
shocking individual orbits of disk stars during their passage (``direct
heating'') as well as by exciting global collective modes in the disk. 
Collective modes include both vertical bending waves (e.g., warps)
and horizontal density waves (e.g., spiral structure and bars)
and we stress that there is coupling between planar and 
vertical modes in 3D. In this case, the energy is transferred to the 
disk causing its heating by damping of the waves via resonant coupling
\citep[e.g.,][]{Weinberg91}. Alternatively, the waves may be damped by dynamical
friction exerted by the dark matter halo, thus yielding no heating of 
the disk.

Calculations of direct heating show that during a single, random 
orientation passage of a satellite with mass $M_{\rm sat}$ moving at
relative speed $v$ and impact parameter $b$, the mean 
change in the vertical energy per unit mass of a disk star is given 
by $\Delta E_z = (h_z^2 / 3) (GM_{\rm sat} / b^3\kappa_z )^2 \beta^2 L(\beta)$ 
\citep{Spitzer58}. Here, $h_z$ is the rms thickness of the disk, $\kappa_z$ 
is the frequency of vertical oscillations, the parameter $\beta=2\kappa_z b/v$ 
is of order the characteristic passage time of the satellite divided by the
orbital period of the star, and $L(\beta)$ is a dimensionless function which is unity for
$\beta\to 0$ and exponentially small for $\beta\gg 1$. According to this
formula, direct heating should vanish as the disk becomes razor thin, both
because $h_z\rightarrow 0$ and also because $\kappa_z \rightarrow \infty$. 
This fact in conjunction with the results in Figures~\ref{fig4},~\ref{fig5},
and~\ref{fig9} suggests that the relative fragility of thinner disks 
to encounters with satellites lies in both the ability of these systems 
to support global instabilities and the effectiveness of 
collective modes to transfer energy in this case. The absence 
of a bar and the much weaker warp and spiral structure induced 
by the subhalo impacts in galaxy model D2 compared to its thinner 
counterpart D1 lends support to this conjecture. Definitive investigations 
of these issues would require a combination of targeted numerical experiments 
as well as extensions of earlier analytic work 
\citep[e.g.,][]{Weinberg98,Weinberg_Blitz06}.

A relevant issue concerns the existence of bulgeless, thin disk 
galaxies in cosmological models where accretions of massive satellites 
are as common as predicted in {\LCDM}. Figure~\ref{fig10} demonstrates 
that a bulge component reduces significantly the damage done to the disk,
so that bulgeless disk galaxies experience substantially more thickening 
by infalling satellites compared to their counterparts with bulges. 
Because most MW-sized halos are expected to have accreted numerous 
substructures that are a significant fraction of the disk mass including 
at least one system as massive as the disk since $z=1$ (Figure~\ref{fig1}; 
\citealt{Stewart_etal08}), the ubiquity of very thin, bulgeless disk 
galaxies containing dominant old stellar populations would be difficult 
to reconcile with {\LCDM}. Interestingly, using SDSS data \citet{Kautsch_etal06} 
recently compiled a uniform catalogue of $3169$ edge-on disk galaxies and found 
that $\sim 1/3$ of the galaxies in their sample were bulgeless, ``super-thin'' 
disks with extreme axial ratios. Moreover, all systems in the sample of
bulgeless, edge-on spirals of \citet{Dalcanton_Bernstein02} have pronounced 
thick disks and there are no signs of companions in the vicinity of the prototype 
thin bulgeless disk galaxy M33. Formulating a comprehensive model for the
formation and survivability of very thin, bulgeless disk galaxies in the 
context of hierarchical CDM remains challenging. 

The findings of the present study as well as those of Paper I have interesting 
implications for the formation of thick disks. Thick disks are structurally, 
chemically, and kinematically distinct from thin disks, and there is evidence
that they may assemble quite early in the history of a galaxy 
\citep{Elmegreen_Elmegreen06}. Thick-disk stars in the MW and external
galaxies are characterized by much larger scale 
heights, exhibit larger velocity dispersions and slower rotation, and are more 
metal-poor and significantly enhanced in $\alpha$-elements compared 
to thin-disk stars \citep[e.g.,][]{Reid_Majewski93,Gilmore_etal95,Wyse_Gilmore95,
Prochaska_etal00,Chiba_Beers00,Bensby_etal05,Seth_etal05,Yoachim_Dalcanton06,Allende_etal06,
Juric_etal08,Ivezic_etal08}. While our dissipationless simulations can neither
verify nor disprove any of the trends regarding metallicities, the present
study does show that encounters with infalling subhalos increase considerably
the scale heights and velocity ellipsoids of thin, galactic disks (Figures~\ref{fig4} 
and~\ref{fig5}). Furthermore, the vertical structure of the final disk is
well-described by a standard ``thin-thick'' disk decomposition (see
Figure~\ref{fig4} in Paper I) and analysis of the mean azimuthal velocity of 
disk stars at the solar radius in our simulations also reveals a vertical
gradient in rotational velocity of $\sim -20\kms \kpc^{-1}$ between $1$ and
$3$~kpc from the disk plane, which is consistent with what is inferred for the 
thick disk of the MW \citep{Allende_etal06}. These results suggest that at least 
part of a galaxy's thick-disk component may plausibly 
originate from the vertical dynamical heating of preexisting thin disks by 
CDM substructure. While this conclusion is supported by a number of observational 
studies in both the MW \citep[e.g.,][]{Robin_etal96,Chen_etal01,Bensby_etal05} 
and external galaxies \citep{Seth_etal05}, more detailed theoretical modeling
of the properties of thick disks would at least require the inclusion of gasdynamics,
star formation, and metal enrichment.

Of course, vertical dynamical heating of an existing thin disk does not constitute the 
only viable model for the origin of thick disks. Other proposed mechanisms 
include satellite accretion events that directly deposit thick-disk stars at large 
scale heights \citep[e.g.,][]{Statler88,Abadi_etal03,Yoachim_Dalcanton05,
Yoachim_Dalcanton08} as well as thick disks stars forming in situ at early
times directly from gas at large distances above the midplane 
\citep[e.g.,][]{Brook_etal04,Brook_etal05}. While there is no definitive
observational or theoretical evidence to rule out any of the 
models for the origin of thick disks conclusively, the existence of very slowly rotating or 
counter-rotating thick disks in a significant fraction of disk galaxies would 
be particularly problematic for the ``vertical heating'' mechanism 
\citep{Yoachim_Dalcanton05}. Most likely, all mechanisms do operate
simultaneously at a different degree in forming the thick disk 
components of galaxies.


\begin{figure*}[t]
\centerline{\epsfxsize=7.2in \epsffile{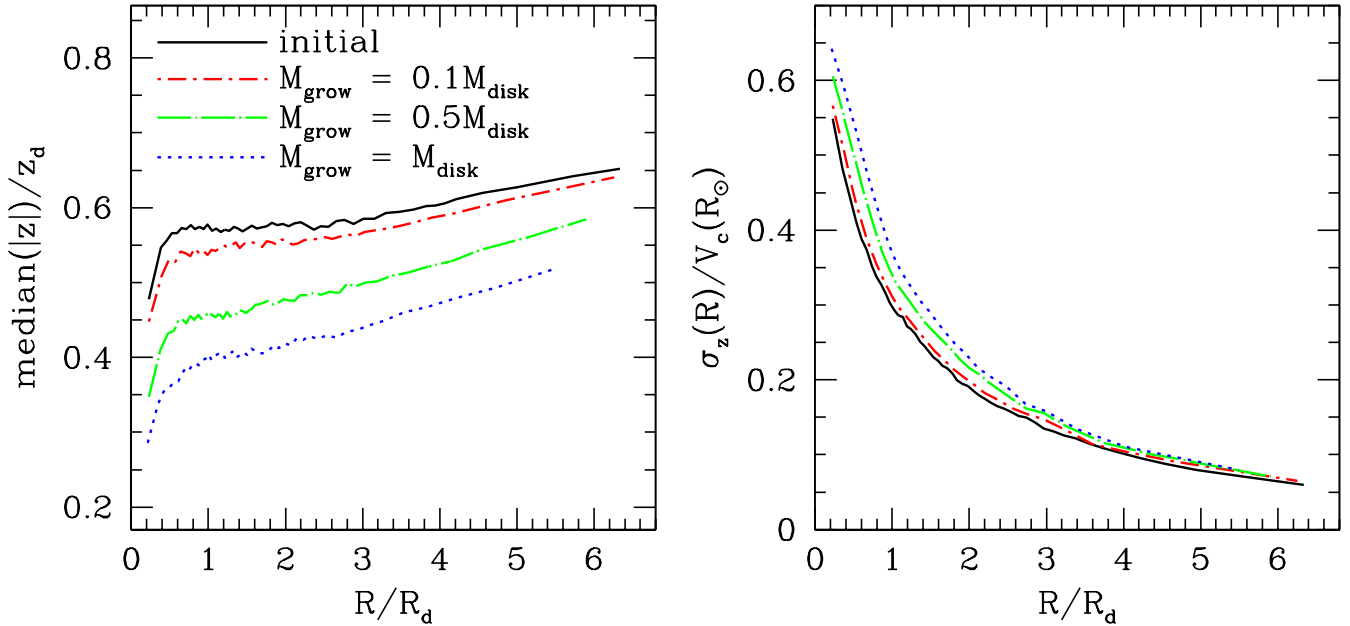}}
\caption{Effect of the adiabatic growth of a thin-disk component 
  on the vertical structure of a thick disk. For the latter, we adopt 
  disk model D2. {\it Solid lines} show results for the initial uncontracted 
  thick disk with mass $M_{\rm disk}$ while the remaining 
  lines correspond to the final state of the thick disk after the growth 
  of thin disks with various masses, $M_{\rm grow}$. All growing 
  thin disks have a sech$^2$ scale height of $z_{\rm thin} = 0.2\kpc$ 
  and the same radial scale length, $R_d$, as that of the initial thick disk. 
  {\it Left:} Disk thickness profiles. Thicknesses and radii are normalized 
  to the scale height, $z_d$, and radial scale length, $R_d$, of the 
  initial thick disk. {\it Right:} Vertical velocity dispersion profiles normalized 
  to the total circular velocity of disk model D2 at the solar radius, 
  $V_{\rm c}(R_\odot)=235.6\kms$. The slow accumulation of thin disk 
  material modifies the structure of the initial thick disk,   
  causing its vertical and radial contraction as well as an increase 
  of its vertical velocity dispersion.
\label{fig13}}
\end{figure*}


In this paper, we have focused on the gravitational interaction between galactic disks 
and CDM substructure in the collisionless regime. Given the complex interplay of effects 
relevant to the formation and evolution of galactic disks, the inclusion of gas dynamics, 
star formation, and chemical evolution would be required at the least to refine the conclusions 
presented here. Specifically, the modeling of hydrodynamics will be crucial in 
determining the extent to which the presence of gas can influence the dynamical 
response of galactic disks to satellite accretion events and affect the properties of 
the final disk. For example, both subhalos and the disk may contain gas, 
particularly at high redshift. The satellite accretion events would then trigger
bursts of star formation that may replenish a thin disk component.

Moreover, a dissipative component may alter the dynamical effects of substructure on 
stellar disks in two important ways. First, the gas itself can absorb part of the orbital 
energy deposited into the galactic disk by the infalling systems, acting as an 
energy sink. This process would lead to the heating of the gaseous component. However, the 
efficiency of this mechanism in reducing the dynamical damage done to a stellar disk 
will depend on the gas content of galactic disks at early times when most of these 
accretion events occur. Interestingly, analytical models for the evolution
of the MW disk in a cosmological context do estimate that the gaseous disk at
$z \sim 1$ should amount to $\sim 50\%$ of the mass of the stellar disk
\citep{Naab_Ostriker06}. Given that substantial gas fractions are 
expected at high redshifts, the role of gas in stabilizing the galactic 
disks against the violent gravitational encounters with satellites may 
be crucial \citep{Stewart_etal09}.

Second, owing to its dissipative nature, the gas can radiate its energy away. 
As a result, the gas that has been heated by an encounter with a subhalo can 
subsequently cool and reform a thin disk. As any gaseous component slowly
accumulates in the center of the mass distribution, it will also induce
concomitant contraction of the thickened stellar disk due to its gravity. 
Larger scale smooth gas accretion in galaxy 
formation models \citep[e.g.,][]{Murali_etal02} will also be relevant in this context. 
A full exploration of these contingencies is challenging and we defer such studies 
to future work. In what follows, we present a simple, preliminary experiment to serve 
as a crude estimate of such effects.

In particular, we investigated the response of an initial thick disk to the adiabatic 
growth of a massive, thin-disk component within it. For the former, we adopted galaxy 
model D2 with a vertical scale height of $z_d = 1\kpc$. We considered three growing, 
exponential thin disks with masses $M_{\rm grow} = [0.1, 0.5, 1] M_{\rm disk}$, where $M_{\rm disk}$ 
is the mass of the disk in model D2, and a sech$^2$ scale height of 
$z_{\rm thin} = 0.2\kpc$. The latter value is consistent with the scale heights of known, 
young, star-forming disks observed in both external galaxies \citep[e.g.,][]{Wainscoat_etal89,
Matthews00} and in the MW \citep[e.g.,][]{Bahcall_Soneira80,Reid_Majewski93}. 
All growing disks were treated as rigid potentials and had 
the same radial scale lengths as disk D2. Furthermore, the scale length of each thin disk 
was kept constant while its mass was slowly increased from zero to its final value linearly 
over a timescale of $1$~Gyr. Such timescales are in general accordance with disk formation 
models \citep[e.g.,][]{Fall_Efstathiou80} and ensure that the process of disk growth is 
approximately adiabatic.

Figure~\ref{fig13} presents the results of these experiments. In all cases, the initial 
thick disk contracts vertically as well as radially in response to the growth of the 
thin-disk component. Moreover, its vertical velocity dispersion increases, reflecting 
the deepening of the potential well due to the slow accumulation of thin-disk material. 
As expected, these changes in the structure of the thick disk depend 
sensitively on the total mass of the growing disk. In the most dramatic case with 
$M_{\rm grow} = M_{\rm disk}$, the decrease in the thickness of disk model D2 is 
$\sim 30\%$ in the solar neighborhood. Furthermore, these changes do not occur
uniformly as a function of radius. The disk potentials are centrally concentrated, 
so the evolution of the inner disk is much more pronounced compared to that of the 
outer disk. Overall, the results presented in Figure~\ref{fig13} suggest that
the mass of the growing disk would need to be many times that of the thickened 
stellar disk to reduce its thickness appreciably, and even then, there would
be stability issues.

It is important to emphasize that the aforementioned experiments were not designed to 
elucidate the importance of adiabatic thick-disk contraction in the specific 
case of the MW whose midplane density ratio of thick-to-thin disk is only 
$12\%$ \citep[e.g.,][]{Juric_etal08}. In a study of thick disks in the Hubble Space 
Telescope Ultra Deep Field (UDF), \citet{Elmegreen_Elmegreen06} did examine whether 
such a process could have determined the present-day scale height of the thick 
disk of the MW. Their calculations showed that if the present thick-disk component 
of the MW began as an equilibrium pure-thick disk at a young age, and if subsequent 
accretion of the entire thin disk was adiabatic, then the initial thick-disk scale 
height had to be $\sim 3\kpc$. This is considerably larger than that observed for 
young thick disks in the UDF, where the average scale height is $1.0 \pm 0.4\kpc$.

\section{Comparison to Previous Work}
\label{sec:comparison}

The response of disks to encounter with infalling satellites 
has received a great deal of attention owing to the numerous 
implications that it entails for theories of galaxy formation 
and evolution. Here we discuss the main differences between our 
results and those reported in a subset of previous studies. 

\citet{Font_etal01} and \citet{Gauthier_etal06} carried out numerical
simulations of the gravitational interaction between galactic disks 
and a large ensemble of dark matter subhalos. Both of these investigations 
considered the $z=0$ satellite populations present in a MW-sized 
CDM halo, and both studies reached the conclusion that halo substructure 
has an insignificant effect on the global structure of a galactic disk.
In particular, \citet{Font_etal01} showed similar tidal heating rates 
in their $N$-body stellar disk with and without substructure, while 
\citet{Gauthier_etal06} reported appreciable heating in the inner disk 
regions due to the formation of a bar and only mild heating of the disk 
at intermediate and large radii. These results are explained by the fact 
that the strategies of \citet{Font_etal01} and \citet{Gauthier_etal06} 
excepted those systems that are most capable of strongly perturbing the disk. 
Massive subhalos on highly eccentric orbits at early epochs
suffer substantial mass loss or become preferentially disrupted 
during their orbital evolution in the host potential
prior to $z=0$ \citep{Zentner_Bullock03,Gao_etal04,Zentner_etal05a,Benson05}, 
and so they are more likely to be absent from the present-day 
subhalo populations. Indeed, in the \citet{Font_etal01} experiments 
only subhalos with masses below $10^{9}\Mo$ had pericenters within 
the solar radius, $R_\odot$. A primary improvement in our modeling 
is that we have followed the formation history of the host halo 
since $z \sim 1$, and consequently we have accounted for a larger 
number of important satellite-disk interactions than that based
on the $z=0$ substructure. As a result, we report significantly 
more damage to the structure of the galactic disk than that 
demonstrated by either \citet{Font_etal01} or \citet{Gauthier_etal06}. 

Past numerical investigations into the resilience of galactic disks 
to infalling satellites have also suffered from numerical limitations 
and/or from assumptions which curtail their ability to accurately capture 
the degree of global dynamical evolution that accreting subhalos 
can induce in cold, stellar disks in a cosmological context. 
For example, the modeling of various components in the primary disk galaxy 
and/or the satellites as rigid potentials \citep[e.g.,][]{Quinn_Goodman86,Quinn_etal93,
Sellwood_etal98,Ardi_etal03,Hayashi_Chiba06}, the initialization of a
disk much thicker compared to typical thin disks such as the old, thin
stellar disk of the MW \citep[e.g.,][]{Quinn_Goodman86,Quinn_etal93,Walker_etal96,
Huang_Carlberg97,Velazquez_White99,Font_etal01,Villalobos_Helmi08}, 
and the infall of satellites with orbital parameters and/or structural
properties inconsistent with {\LCDM} expectations \citep[e.g.,][]
{Quinn_etal93,Walker_etal96,Huang_Carlberg97,Velazquez_White99}.
For example, \citet{Quinn_Goodman86}, \citet{Quinn_etal93}, and \citet{Walker_etal96} 
performed the first numerical explorations of the interaction of single 
satellites with larger disk galaxies. These studies unanimously found disks to
be quite fragile, reporting that an encounter with a satellite of 
only $10\%$ of the disk mass could increase the disk thickness by a factor of $\sim
2$ at the solar radius \citep{Quinn_Goodman86,Quinn_etal93}. In contrast,
we find galactic disks to be generally more robust to accretion events than
these earlier investigations have indicated. 

These differences may be due to a variety of factors. When a satellite is 
modeled as a distribution of interacting particles as opposed to a rigid
potential, the efficiency with which it can heat a galactic disk is suppressed 
for two reasons. First, a self-gravitating 
satellite suffers mass loss due to tidal stripping and shocking. Second, a 
responsive satellite can absorb part of its orbital energy, decreasing the 
amount of energy deposited into random motions of disk stars. Moreover, 
live halos are needed to treat the effect of an accreting satellite on disk 
structure properly. Representing DM halos as rigid potentials leads to 
an over-estimate of disk thickening by a factor of $\sim 1.5-2$, whereas 
a self-gravitating halo will respond to both the disk and satellite and 
aid in stabilizing the disk \citep{Nelson_Tremaine95,Velazquez_White99}.  
Moreover, the focus on prograde circular or nearly circular orbits in some of 
the aforementioned studies is also likely to have overestimated the typical 
amount of disk damage. The orbit with the most damaging effect for a thin disk
is a coplanar, prograde circular orbit, since this causes the satellite
force to be in near resonance with the disk stars. More eccentric
orbits are likely to cause less damage to a disk. This is immediately apparent
from the fact that, in the impulsive limit, the energy transfer scales as 
$\propto v^{-2}$. To a first order, the high-speed pericentric passages of 
CDM subhalos on highly eccentric orbits are thus expected to cause less 
thickening than circular ones. However, it is unclear to what extent highly
eccentric orbits can excite global modes in the disk. This requires
a detailed series of numerical experiments to evaluate. 

\citet{Velazquez_White99} conducted a large number of $N$-body experiments 
in which they investigated the dependence of disk heating on, among other variables, 
the inclination and eccentricity of the initial satellite orbit. 
These authors showed that prograde encounters are more efficient at thickening
the disk than retrograde ones, arguing for a resonant coupling between the
satellite and disk angular momenta that is suppressed in retrograde
interactions. \citet{Velazquez_White99} found that a satellite with an initial 
mass of $M_{\rm sub}=0.2 M_{\rm disk}$ on a prograde orbit causes an
additional increase in the scale height of the initial disk at the solar
radius of $\sim 35\%$ compared to its retrograde counterpart. 

While our simulations indicate qualitatively similar differences between 
prograde and retrograde interactions (Figure~\ref{fig12}), they show a less 
significant differential effect compared to \citet{Velazquez_White99}. This 
discrepancy has two plausible sources. First, we study
cosmologically-motivated, comparably high-speed passages of substructures.  
In such rapid (``impulsive'') interactions, the resonant coupling between 
the satellite and disk angular momenta is expected to be weaker, explaining 
why the relative effect of prograde versus retrograde interactions is smaller 
in our experiments. Second, we study single subhalo passages, whereas 
\citet{Velazquez_White99} followed the infalling satellites over many orbits 
until they merge with the galactic disk or become disrupted in the process.
This is relevant because \citet{Velazquez_White99} found that during 
the early stages of the interaction when the satellite 
orbit is still eccentric the difference between the amount of disk thickening 
caused by prograde and retrograde encounters was fairly small. Satellites on
prograde orbits constituted much more efficient perturbers compared to their 
retrograde counterparts only during the late stages of the interaction when, 
owing to dynamical friction, the satellite speed has been reduced, and its
orbit has both circularized and become coplanar with the 
disk (H. Velazquez 2008, private communication).

\citet{Toth_Ostriker92}, \citet{Benson_etal04}, and \citet{Hopkins_etal08}
quantified the fragility of galactic disks to infalling satellites by 
using semi-analytic models of different degrees of sophistication. 
Analytical approaches have the advantage of not being limited by numerical 
resolution, allowing the calculation of a statistically-large number of 
model realizations, but also have the drawback that they cannot account 
fully for the non-linear interaction between satellites and disks.
Yet, as both our results in Figure~\ref{fig11} suggest and other more targeted 
studies have indicated \citep[e.g.,][]{Weinberg91,Sellwood_etal98}, 
the excitation of global collective modes in the disk is an essential 
mechanism for energy deposition by an accreting satellite.

\citet{Toth_Ostriker92} concluded that sinking satellites with only 
a few percent of the disk mass could lead to substantial disk 
thickening. Specific comparisons with the work 
of \citet{Toth_Ostriker92} is very difficult. In a sense, 
the fact that their disk suffers substantial damage strengthens 
our findings. However, their results are averaged all possible 
orbital inclinations making specific model comparisons cumbersome.
Another difficulty lies in their assumption that the orbital 
energy of the satellite is deposited locally at the point 
of impact. Our results do indicate that the energy imparted by 
typical cosmological substructures will be deposited globally 
across the entire disk. Lastly, \citet{Toth_Ostriker92} assumed 
that the total thickening and heating scale with satellite mass 
and that they are the same regardless of the initial thickness 
and velocity ellipsoid. Figures~\ref{fig4} and~\ref{fig9} show 
that this conjecture was also incorrect.
The implications of most of these assumptions are unclear.

In contrast, \citet{Benson_etal04} suggested that the observed 
thickness of stellar disks is entirely compatible with the 
abundance of substructure in CDM halos. In addition to not 
accounting for global collective modes, \citet{Benson_etal04} 
also adopted satellite orbits that spanned only a limited range of 
orbital energies and angular momenta in spherical halo potentials 
rather than the richer variety of impacts experienced over the 
course of halo formation \citep[e.g.,][]{Zentner_etal05a,
Zentner_etal05b,Benson05}. Consequently, this study
suffers from a flaw similar to that of \citet{Font_etal01} and 
\citet{Gauthier_etal06}, explaining possibly why we report 
more significant damage to the disk structure by subhalo 
bombardment.  

More direct comparison can be performed with the recent work of
\citet{Hopkins_etal08}. These authors have argued that deposition
of orbital energy into the disk in the context of realistically 
radial subhalo orbits and {\LCDM}-motivated accretion histories 
would yield a much less dramatic impact on the disk structure than 
previously thought, with or without the presence of gas. 
In particular, \citet{Hopkins_etal08} derived that 
the disk heating efficiency is nonlinear in mass ratio, 
$\propto (M_{\rm sub} /M_{\rm disk})^2$, instead of the linear scaling 
of \citet{Toth_Ostriker92}, implying that the fractional change in disk 
scale height should be very small even for the very massive subhalos 
we considered in the present study 
($0.2 M_{\rm disk} \lesssim M_{\rm sub} \lesssim M_{\rm disk}$). 
\citet{Hopkins_etal08} defined a disk thickening parameter
$\Delta H/R$, where $H$ is the median disk scale height 
and $R$ is the radius where the scale height is measured, typically 
within a factor of two of the disk half-mass radius, $R_h$. 

For the simulated accretion history of host halo G$_1$, the 
\citet{Hopkins_etal08} formula predicts $\Delta H/R \simeq 0.01$, 
if we simply add the masses of all cosmological subhalos S1-S6.
Our numerical experiments indicate significantly more thickening. 
Indeed, at $R=R_h=1.7 Rd$, we measure $\Delta H/R \simeq 0.02$, 
and because of the pronounced flaring in response to the accretion events, 
we report even larger $\Delta H/R \simeq 0.04$, at $R = 2R_h$. 
A variety of reasons may be responsible for these discrepancies. 
As before, global collective modes which may 
dominate the disk response to accretion events were not included in these
simple analytic scalings. Furthermore, \citet{Hopkins_etal08} calibrated 
their results to numerical simulations \citep{Velazquez_White99,Villalobos_Helmi08}
with initial disks that were significantly thicker compared to the thin, 
galactic disk we employed here, and were therefore intrinsically
more robust to encounters with satellites (Figure~\ref{fig11}).

Lastly, special emphasis should be placed on the recent study by 
\citet{Purcell_etal08}. These authors performed collisionless $N$-body
simulations to study the response of an initially-thin ($z_d \simeq 400$~pc), 
MW type disk galaxy to $\sim 1:10$ satellite impacts. Such accretion events 
represent the primary concern for disk survival in the {\LCDM} cosmological 
model and should have been commonplace in the history of Galaxy-sized halos
\citep{Stewart_etal08}. \citet{Purcell_etal08} quantitatively demonstrated 
for the first time the destruction of a thin, stellar disk by these 
cosmologically motivated common events.  They find that regardless of orbital 
configuration, the impacts transform the disks into structures that are
roughly three times as thick and more than twice as kinematically 
hot as the observed dominant old thin disk component of the MW. 
On the other hand, our work shows that a thin disk component may survive, 
even strongly perturbed, the encounters with halo substructure (Paper I).
This is because we have focused on infalling systems with masses in the range 
$0.2 M_{\rm disk} \lesssim M_{\rm sub} \lesssim M_{\rm disk}$, ignoring
the most massive accretion events that could prove ruinous 
to thin-disk survival. Our findings are thus relevant to systems that have already 
experienced the most destructive events simulated by \citet{Purcell_etal08}
since $z=1$ and have re-grown their thin disks since. In this sense, our
simulation set and results are complementary to those of
\citet{Purcell_etal08}.

\vspace{2cm}
\section{Summary and Future Directions}
\label{sec:summary}

Using a suite of high-resolution, fully self-consistent dissipationless
$N$-body simulations we have examined the dynamical effects of halo
substructure on thin galactic disks in the context of the {\LCDM} paradigm 
of structure formation. Our simulation campaign utilizes cosmological 
simulations of the formation of Galaxy-sized CDM halos to derive the
properties of infalling subhalo populations and controlled numerical 
experiments of consecutive satellite encounters with an initially-thin,
fully-formed disk galaxy. As a corollary, we have quantified the importance 
of various physical effects that could influence the response of a galactic disk 
to substructure accretion events. The properties we address are the 
initial disk thickness, the presence of a bulge component in the
primary galaxy, the internal density distribution of the infalling systems, 
and the relative orientation of disk and satellite angular momenta.

Our work expands upon past numerical investigations into the 
dynamical response of galactic disks to merging satellites in 
several ways. One improvement concerns the more realistic treatment 
of the infalling subhalo populations. Previous studies of the interaction 
between disks and ensembles of subhalos considered only the $z=0$ surviving 
substructure present in a CDM halo \citep{Font_etal01,Gauthier_etal06}. 
This leads to estimates of the damage done to the disk that are biased low, 
because massive subhalos with small orbital pericenters that are most capable 
of strongly perturbing the disk are preferential removed from the satellite 
populations over time. We have accounted for satellite-disk interactions with merging
subhalos that typically do not survive to the present day but nevertheless 
cause significant damage to the disk. This is the major conceptual advantage 
of our work and this difference drives our initial disks to be more
dramatically affected by halo substructure than those in earlier studies.  
Second, the primary disk galaxy models we utilize are fully self-consistent
particle realizations derived from explicit DFs and are designed 
to satisfy a broad range of observational constraints available for 
actual disk galaxies. Finally, we represent satellite systems by equilibrium 
numerical realizations, whose properties (mass functions, orbital 
parameters, internal structures, and accretion times) are extracted 
directly from cosmological simulations of the formation of Galaxy-sized 
CDM halos.

Our main conclusions can be summarized as follows.

\begin{itemize}

\item[1.] Close encounters between massive satellites and galactic 
  disks since $z=1$ are common occurrences in the {\LCDM} cosmological model. 
  Statistics of four Galaxy-sized CDM halos indicate that, 
  on average, $\sim 5$ subhalos more massive than $0.2M_{\rm disk}$, 
  where $M_{\rm disk}$ is the present-day mass of the stellar disk 
  of the MW, pass within $\sim 20\kpc$ from the centers of their 
  host halos in the past $\sim 8$~Gyr. In contrast, very few 
  satellites in present-day substructure populations are 
  likely to have a significant impact on the structure of a galactic disk. 
  This is because massive subhalos on potentially damaging orbits 
  suffer severe mass loss or become tidally disrupted prior to 
  $z=0$ as a result of penetrating deeply into the central regions of their 
  hosts (\S~\ref{sec:sat_disk_encounters}).

\item[2.] A conservative subset of one host halo accretion history was used 
  to seed controlled subhalo-disk encounter simulations. The specific merger 
  history involved the accretion of six satellites with masses, orbital pericenters, 
  and tidal radii of $7.4 \times 10^{9} \lesssim M_{\rm sub}/M_{\odot} \lesssim 2 
  \times 10^{10}$, $r_{\rm peri} \lesssim 20\kpc$, and $r_{\rm tid} \gtrsim 20
  \kpc$, respectively, since $z=1$. These events severely perturb an initially-thin 
  ($z_d = 0.4\kpc$), MW type disk galaxy ($M_{\rm disk} \approx 3.5 \times 10^{10} \Mo$) 
  and imprint the following distinctive dynamical signatures on its structure 
  and kinematics:

  \begin{itemize}

      \item[$\bullet$] The development of non-axisymmetric structures
        including a warp, a moderately strong bar, and extended ring-like 
        features in the outskirts of the disk (\S~\ref{sub:morphology}).

      \item[$\bullet$] Considerable thickening and heating at all radii, 
        with a factor of $\sim 2$ increase in disk thickness and velocity 
        ellipsoid at the solar radius (\S~\ref{sub:thickening} 
        and \S~\ref{sub:heating}).

      \item[$\bullet$] Prominent flaring associated with an increase of
        disk thickness greater than a factor of $4$ in the disk outskirts
        (\S~\ref{sub:thickening}).
           
      \item[$\bullet$] Surface density excesses at large radii, beyond
        $\sim 5$ disk scale lengths, as observed in antitruncated disks 
	(\S~\ref{sub:surface.density}).

      \item[$\bullet$] Long-lived, lopsidedness at levels similar 
        to those measured in observational samples of disk galaxies
        (\S~\ref{sub:lopsidednes}).

      \item[$\bullet$] Substantial tilting of the disk from its initial
        orientation in the host halo, resulting in a misalignment between halo 
        and disk principal axes and angular momenta (\S~\ref{sub:tilting}).

   \end{itemize}

\item[3.] Detailed predictions for the dynamical response of galactic disks
         to subhalo bombardment are subject to a variety of 
         assumptions including the initial structures of the disk and
         infalling systems, the prominence of the bulge component in the 
         primary disk, and the relative orientation of disk and satellite 
	angular momenta (\S~\ref{sec:effects}).

\end{itemize}

We close with a few words of caution and a discussion of fruitful directions 
for future work that may lead to more conclusive statements about the 
detailed structure of disk galaxies. We reiterate that we have only addressed the 
gravitational interaction of galactic disks and halo substructure in the 
collisionless regime. A full consideration to the rich structure of perturbed 
galactic disks is challenging and would require detailed knowledge of galaxy 
star formation histories, gas cooling and feedback, among other things. 
However, such studies that treat both the gaseous components of the disk and 
subhalos, and accreted stars will be fundamental in refining our understanding 
of disk galaxy evolution and we plan to extend the present investigation 
in this direction. 

Specifically, spiral galaxies contain atomic and molecular 
gas in their disks which can absorb and subsequently radiate away
some of the orbital energy deposited by the sinking satellites.  
Thus, including a dissipative component in the primary galaxy will
allow for an estimate of the extent to which the presence of gas
can reduce the damage done to disks and stabilize them against
the violent gravitational encounters with satellites. Adding star 
formation as a further ingredient will provide the opportunity to 
analyze any reformed thin disk and establish the degree to which contraction
subsequent to gas cooling can decrease the thickness of a heated 
disk (Figure~\ref{fig13}). The effects of gas dynamics
and star formation will offer the possibility to determine the magnitude of 
starbursts induced in the disk as a result of subhalo bombardment, while
gaining a deeper understanding of the build-up of the inner stellar halos 
of galaxies \citep{Bullock_Johnston05}. Lastly, the inclusion of baryonic
components in the satellites will enable studies whereby, original disk stars, 
newly formed stars, and accreted stars can be located and studied in their final 
configurations. 

Such predictions will be vital as instruments and surveys like SDSS III, 
RAVE, GAIA, SIM, Pan-STARRS, LSST, and TMT are poised to provide spatial 
and kinematic maps of the MW and other local volume galaxies to unprecedented 
detail and depth. Our simulations suggest that these experiments should
uncover detailed disk structure that is substantially perturbed via 
interactions with infalling satellites. Our ability to interpret these data 
sets will rely on a comprehensive set of theoretical predictions for how 
galactic disks respond to accretion events and how this process is convolved 
with direct stellar and gaseous accretions (either via the same encounters or 
other delivery process). In this sense, detailed probes of the local volume 
offer a valuable and unique avenue for constraining the process of disk 
galaxy formation and galaxy formation in general.

\acknowledgments

The authors are grateful to Andrew Benson, Simone Callegari, Jack Clompus, Annette Ferguson, 
Andreea Font, Kimberly Herrmann, Kathryn Johnston, Mario Juri{\'c}, Lucio Mayer, Ben Moore,
Heather Morrison, Panos Patsis, Chris Purcell, Tom Quinn, Joachim Stadel, 
Octavio Valenzuela, Hector Velazquez, and David Weinberg for many stimulating 
discussions. SK would like to thank Frank van den Bosch for communicating unpublished 
results, and John Dubinski and Larry Widrow for kindly making available the software 
used to set up the primary galaxy models. SK, JSB, and AVK acknowledge the Aspen Center 
for Physics for hosting the summer workshop ``Deconstructing the Local Group -- 
Dissecting Galaxy Formation in our Own Background'' during the initial stages 
of this work. SK is also grateful to the Kavli Institute for Theoretical Physics 
for organizing the stimulating workshop ``Building the Milky Way'' and for its 
hospitality while this work was in progress. SK is supported by the Center for Cosmology 
and Astro-Particle Physics (CCAPP) at The Ohio State University. 
ARZ is funded by the University of Pittsburgh and by the National Science Foundation (NSF) 
through grant AST 0806367. JSB is supported by NSF grants AST 05-07916 
and AST 06-07377. AVK is supported by the NSF grants AST-0239759 and AST-0507596, and by KICP.
The numerical simulations were performed on the zBox supercomputer at the
University of Z\"urich and on the Cosmos cluster at the Jet Propulsion
Laboratory (JPL). Cosmos was provided by funding from the JPL Office of the 
Chief Information Officer. This research made use of the NASA Astrophysics 
Data System.

\bibliography{disk2}

\begin{thebibliography}{187}
\expandafter\ifx\csname natexlab\endcsname\relax\def\natexlab#1{#1}\fi

\bibitem[{{Abadi} {et~al.}(2003){Abadi}, {Navarro}, {Steinmetz}, \&
  {Eke}}]{Abadi_etal03}
{Abadi}, M.~G., {Navarro}, J.~F., {Steinmetz}, M., \& {Eke}, V.~R. 2003, \apj,
  591, 499

\bibitem[{{Allende Prieto} {et~al.}(2006){Allende Prieto}, {Beers}, {Wilhelm},
  {Newberg}, {Rockosi}, {Yanny}, \& {Lee}}]{Allende_etal06}
{Allende Prieto}, C., {Beers}, T.~C., {Wilhelm}, R., {Newberg}, H.~J.,
  {Rockosi}, C.~M., {Yanny}, B., \& {Lee}, Y.~S. 2006, \apj, 636, 804

\bibitem[{{Angiras} {et~al.}(2007){Angiras}, {Jog}, {Dwarakanath}, \&
  {Verheijen}}]{Angiras_etal07}
{Angiras}, R.~A., {Jog}, C.~J., {Dwarakanath}, K.~S., \& {Verheijen}, M.~A.~W.
  2007, \mnras, 378, 276

\bibitem[{{Angiras} {et~al.}(2006){Angiras}, {Jog}, {Omar}, \&
  {Dwarakanath}}]{Angiras_etal06}
{Angiras}, R.~A., {Jog}, C.~J., {Omar}, A., \& {Dwarakanath}, K.~S. 2006,
  \mnras, 369, 1849

\bibitem[{{Ardi} {et~al.}(2003){Ardi}, {Tsuchiya}, \& {Burkert}}]{Ardi_etal03}
{Ardi}, E., {Tsuchiya}, T., \& {Burkert}, A. 2003, \apj, 596, 204

\bibitem[{{Avila-Reese} {et~al.}(2001){Avila-Reese}, {Col{\'{\i}}n},
  {Valenzuela}, {D'Onghia}, \& {Firmani}}]{Avila-Reese_etal01}
{Avila-Reese}, V., {Col{\'{\i}}n}, P., {Valenzuela}, O., {D'Onghia}, E., \&
  {Firmani}, C. 2001, \apj, 559, 516

\bibitem[{{Azzaro} {et~al.}(2007){Azzaro}, {Patiri}, {Prada}, \&
  {Zentner}}]{Azzaro_etal07}
{Azzaro}, M., {Patiri}, S.~G., {Prada}, F., \& {Zentner}, A.~R. 2007, \mnras,
  376, L43

\bibitem[{{Azzaro} {et~al.}(2006){Azzaro}, {Zentner}, {Prada}, \&
  {Klypin}}]{Azzaro_etal06}
{Azzaro}, M., {Zentner}, A.~R., {Prada}, F., \& {Klypin}, A.~A. 2006, \apj,
  645, 228

\bibitem[{{Bahcall} \& {Soneira}(1980)}]{Bahcall_Soneira80}
{Bahcall}, J.~N. \& {Soneira}, R.~M. 1980, \apjs, 44, 73

\bibitem[{{Bailin} {et~al.}(2008){Bailin}, {Power}, {Norberg}, {Zaritsky}, \&
  {Gibson}}]{Bailin_etal08}
{Bailin}, J., {Power}, C., {Norberg}, P., {Zaritsky}, D., \& {Gibson}, B.~K.
  2008, \mnras, 390, 1133

\bibitem[{{Bailin} \& {Steinmetz}(2005)}]{Bailin_Steinmetz05}
{Bailin}, J. \& {Steinmetz}, M. 2005, \apj, 627, 647

\bibitem[{{Baldwin} {et~al.}(1980){Baldwin}, {Lynden-Bell}, \&
  {Sancisi}}]{Baldwin_etal80}
{Baldwin}, J.~E., {Lynden-Bell}, D., \& {Sancisi}, R. 1980, \mnras, 193, 313

\bibitem[{{Belokurov} {et~al.}(2006)}]{Belokurov_etal06}
{Belokurov}, V. {et~al.} 2006, \apjl, 642, L137

\bibitem[{{Bensby} {et~al.}(2005){Bensby}, {Feltzing}, {Lundstr{\"o}m}, \&
  {Ilyin}}]{Bensby_etal05}
{Bensby}, T., {Feltzing}, S., {Lundstr{\"o}m}, I., \& {Ilyin}, I. 2005, \aap,
  433, 185

\bibitem[{{Benson}(2005)}]{Benson05}
{Benson}, A.~J. 2005, \mnras, 358, 551

\bibitem[{{Benson} {et~al.}(2004){Benson}, {Lacey}, {Frenk}, {Baugh}, \&
  {Cole}}]{Benson_etal04}
{Benson}, A.~J., {Lacey}, C.~G., {Frenk}, C.~S., {Baugh}, C.~M., \& {Cole}, S.
  2004, \mnras, 351, 1215

\bibitem[{{Bissantz} \& {Gerhard}(2002)}]{Bissantz_Gerhard02}
{Bissantz}, N. \& {Gerhard}, O. 2002, \mnras, 330, 591

\bibitem[{{Bizyaev} \& {Mitronova}(2002)}]{Bizyaev_Mitrova02}
{Bizyaev}, D. \& {Mitronova}, S. 2002, \aap, 389, 795

\bibitem[{{Blumenthal} {et~al.}(1986){Blumenthal}, {Faber}, {Flores}, \&
  {Primack}}]{Blumenthal_etal86}
{Blumenthal}, G.~R., {Faber}, S.~M., {Flores}, R., \& {Primack}, J.~R. 1986,
  \apj, 301, 27

\bibitem[{{Blumenthal} {et~al.}(1984){Blumenthal}, {Faber}, {Primack}, \&
  {Rees}}]{Blumenthal_etal84}
{Blumenthal}, G.~R., {Faber}, S.~M., {Primack}, J.~R., \& {Rees}, M.~J. 1984,
  \nat, 311, 517

\bibitem[{{Bournaud} {et~al.}(2005){Bournaud}, {Combes}, {Jog}, \&
  {Puerari}}]{Bournaud_etal05}
{Bournaud}, F., {Combes}, F., {Jog}, C.~J., \& {Puerari}, I. 2005, \aap, 438,
  507

\bibitem[{{Brinks} \& {Burton}(1984)}]{Brinks_Burton84}
{Brinks}, E. \& {Burton}, W.~B. 1984, \aap, 141, 195

\bibitem[{{Brook} {et~al.}(2005){Brook}, {Gibson}, {Martel}, \&
  {Kawata}}]{Brook_etal05}
{Brook}, C.~B., {Gibson}, B.~K., {Martel}, H., \& {Kawata}, D. 2005, \apj, 630,
  298

\bibitem[{{Brook} {et~al.}(2004){Brook}, {Kawata}, {Gibson}, \&
  {Freeman}}]{Brook_etal04}
{Brook}, C.~B., {Kawata}, D., {Gibson}, B.~K., \& {Freeman}, K.~C. 2004, \apj,
  612, 894

\bibitem[{{Bullock} \& {Johnston}(2005)}]{Bullock_Johnston05}
{Bullock}, J.~S. \& {Johnston}, K.~V. 2005, \apj, 635, 931

\bibitem[{{Chen} {et~al.}(2001)}]{Chen_etal01}
{Chen}, B. {et~al.} 2001, \apj, 553, 184

\bibitem[{{Chiba} \& {Beers}(2000)}]{Chiba_Beers00}
{Chiba}, M. \& {Beers}, T.~C. 2000, \aj, 119, 2843

\bibitem[{{Choi} {et~al.}(2007){Choi}, {Park}, \& {Vogeley}}]{Choi_etal07}
{Choi}, Y.-Y., {Park}, C., \& {Vogeley}, M.~S. 2007, \apj, 658, 884

\bibitem[{{Dalcanton} \& {Bernstein}(2002)}]{Dalcanton_Bernstein02}
{Dalcanton}, J.~J. \& {Bernstein}, R.~A. 2002, \aj, 124, 1328

\bibitem[{{de Grijs}(1998)}]{de_Grijs98}
{de Grijs}, R. 1998, \mnras, 299, 595

\bibitem[{{de Grijs} \& {Peletier}(1997)}]{DeGrijs_Peletier97}
{de Grijs}, R. \& {Peletier}, R.~F. 1997, \aap, 320, L21

\bibitem[{{De Rijcke} \& {Debattista}(2004)}]{DeRijcke_Debattista04}
{De Rijcke}, S. \& {Debattista}, V.~P. 2004, \apjl, 603, L25

\bibitem[{{Debattista} {et~al.}(2006){Debattista}, {Mayer}, {Carollo}, {Moore},
  {Wadsley}, \& {Quinn}}]{Debattista_etal06}
{Debattista}, V.~P., {Mayer}, L., {Carollo}, C.~M., {Moore}, B., {Wadsley}, J.,
  \& {Quinn}, T. 2006, \apj, 645, 209

\bibitem[{{Debattista} \& {Sellwood}(2000)}]{Debattista_Sellwood00}
{Debattista}, V.~P. \& {Sellwood}, J.~A. 2000, \apj, 543, 704

\bibitem[{{Dehnen} \& {Binney}(1998)}]{Dehnen_Binney98}
{Dehnen}, W. \& {Binney}, J. 1998, \mnras, 294, 429

\bibitem[{{Dury} {et~al.}(2008){Dury}, {de Rijcke}, {Debattista}, \&
  {Dejonghe}}]{Dury_etal08}
{Dury}, V., {de Rijcke}, S., {Debattista}, V.~P., \& {Dejonghe}, H. 2008,
  \mnras, 611

\bibitem[{{Elmegreen} \& {Elmegreen}(2006)}]{Elmegreen_Elmegreen06}
{Elmegreen}, B.~G. \& {Elmegreen}, D.~M. 2006, \apj, 650, 644

\bibitem[{{Erwin} {et~al.}(2005){Erwin}, {Beckman}, \& {Pohlen}}]{Erwin_etal05}
{Erwin}, P., {Beckman}, J.~E., \& {Pohlen}, M. 2005, \apjl, 626, L81

\bibitem[{{Erwin} {et~al.}(2008){Erwin}, {Pohlen}, \& {Beckman}}]{Erwin_etal08}
{Erwin}, P., {Pohlen}, M., \& {Beckman}, J.~E. 2008, \aj, 135, 20

\bibitem[{{Fall} \& {Efstathiou}(1980)}]{Fall_Efstathiou80}
{Fall}, S.~M. \& {Efstathiou}, G. 1980, \mnras, 193, 189

\bibitem[{{Ferguson} {et~al.}(2002){Ferguson}, {Irwin}, {Ibata}, {Lewis}, \&
  {Tanvir}}]{Ferguson_etal02}
{Ferguson}, A.~M.~N., {Irwin}, M.~J., {Ibata}, R.~A., {Lewis}, G.~F., \&
  {Tanvir}, N.~R. 2002, \aj, 124, 1452

\bibitem[{{Ferguson} {et~al.}(2005)}]{Ferguson_etal05}
{Ferguson}, A.~M.~N. {et~al.} 2005, \apjl, 622, L109

\bibitem[{{Font} {et~al.}(2001){Font}, {Navarro}, {Stadel}, \&
  {Quinn}}]{Font_etal01}
{Font}, A.~S., {Navarro}, J.~F., {Stadel}, J., \& {Quinn}, T. 2001, \apjl, 563,
  L1

\bibitem[{{Forbes} {et~al.}(2003){Forbes}, {Beasley}, {Bekki}, {Brodie}, \&
  {Strader}}]{Forbes_etal03}
{Forbes}, D.~A., {Beasley}, M.~A., {Bekki}, K., {Brodie}, J.~P., \& {Strader},
  J. 2003, Science, 301, 1217

\bibitem[{{Gao} {et~al.}(2004){Gao}, {White}, {Jenkins}, {Stoehr}, \&
  {Springel}}]{Gao_etal04}
{Gao}, L., {White}, S.~D.~M., {Jenkins}, A., {Stoehr}, F., \& {Springel}, V.
  2004, \mnras, 355, 819

\bibitem[{{Gauthier} {et~al.}(2006){Gauthier}, {Dubinski}, \&
  {Widrow}}]{Gauthier_etal06}
{Gauthier}, J.-R., {Dubinski}, J., \& {Widrow}, L.~M. 2006, \apj, 653, 1180

\bibitem[{{Ghigna} {et~al.}(1998){Ghigna}, {Moore}, {Governato}, {Lake},
  {Quinn}, \& {Stadel}}]{Ghigna_etal98}
{Ghigna}, S., {Moore}, B., {Governato}, F., {Lake}, G., {Quinn}, T., \&
  {Stadel}, J. 1998, \mnras, 300, 146

\bibitem[{{Ghigna} {et~al.}(2000){Ghigna}, {Moore}, {Governato}, {Lake},
  {Quinn}, \& {Stadel}}]{Ghigna_etal00}
---. 2000, \apj, 544, 616

\bibitem[{{Gilmore} {et~al.}(1995){Gilmore}, {Wyse}, \&
  {Jones}}]{Gilmore_etal95}
{Gilmore}, G., {Wyse}, R.~F.~G., \& {Jones}, J.~B. 1995, \aj, 109, 1095

\bibitem[{{Gnedin} \& {Kravtsov}(2006)}]{Gnedin_Kravtsov06}
{Gnedin}, N.~Y. \& {Kravtsov}, A.~V. 2006, \apj, 645, 1054

\bibitem[{{Governato} {et~al.}(2004)}]{Governato_etal04}
{Governato}, F. {et~al.} 2004, \apj, 607, 688

\bibitem[{{Governato} {et~al.}(2007)}]{Governato_etal07}
---. 2007, \mnras, 374, 1479

\bibitem[{{Grillmair} \& {Dionatos}(2006)}]{Grillmair_Dionatos06}
{Grillmair}, C.~J. \& {Dionatos}, O. 2006, \apjl, 643, L17

\bibitem[{{Grosbol} \& {Patsis}(1998)}]{Grosbol_Patsis98}
{Grosbol}, P.~J. \& {Patsis}, P.~A. 1998, \aap, 336, 840

\bibitem[{{Hartwick}(2000)}]{Hartwick00}
{Hartwick}, F.~D.~A. 2000, \aj, 119, 2248

\bibitem[{{Hayashi} {et~al.}(2003){Hayashi}, {Navarro}, {Taylor}, {Stadel}, \&
  {Quinn}}]{Hayashi_etal03}
{Hayashi}, E., {Navarro}, J.~F., {Taylor}, J.~E., {Stadel}, J., \& {Quinn}, T.
  2003, \apj, 584, 541

\bibitem[{{Hayashi} \& {Chiba}(2006)}]{Hayashi_Chiba06}
{Hayashi}, H. \& {Chiba}, M. 2006, \pasj, 58, 835

\bibitem[{{Haynes} {et~al.}(1998){Haynes}, {van Zee}, {Hogg}, {Roberts}, \&
  {Maddalena}}]{Haynes_etal98}
{Haynes}, M.~P., {van Zee}, L., {Hogg}, D.~E., {Roberts}, M.~S., \&
  {Maddalena}, R.~J. 1998, \aj, 115, 62

\bibitem[{{Hernquist}(1990)}]{Hernquist90}
{Hernquist}, L. 1990, \apj, 356, 359

\bibitem[{{Herrmann} {et~al.}(2009){Herrmann}, {Ciardullo}, \&
  {Sigurdsson}}]{Herrmann_etal09}
{Herrmann}, K.~A., {Ciardullo}, R., \& {Sigurdsson}, S. 2009, ApJL accepted
  (astro-ph/0901.3798)

\bibitem[{{Hogan} \& {Dalcanton}(2000)}]{Hogan_Dalcanton00}
{Hogan}, C.~J. \& {Dalcanton}, J.~J. 2000, PRD, 62, 063511

\bibitem[{{Hopkins} {et~al.}(2008){Hopkins}, {Hernquist}, {Cox}, {Younger}, \&
  {Besla}}]{Hopkins_etal08}
{Hopkins}, P.~F., {Hernquist}, L., {Cox}, T.~J., {Younger}, J.~D., \& {Besla},
  G. 2008, \apj, 688, 757

\bibitem[{{Huang} \& {Carlberg}(1997)}]{Huang_Carlberg97}
{Huang}, S. \& {Carlberg}, R.~G. 1997, \apj, 480, 503

\bibitem[{{Ibata} {et~al.}(2001{\natexlab{a}}){Ibata}, {Irwin}, {Lewis},
  {Ferguson}, \& {Tanvir}}]{Ibata_etal01b}
{Ibata}, R., {Irwin}, M., {Lewis}, G., {Ferguson}, A.~M.~N., \& {Tanvir}, N.
  2001{\natexlab{a}}, \nat, 412, 49

\bibitem[{{Ibata} {et~al.}(2001{\natexlab{b}}){Ibata}, {Lewis}, {Irwin},
  {Totten}, \& {Quinn}}]{Ibata_etal01a}
{Ibata}, R., {Lewis}, G.~F., {Irwin}, M., {Totten}, E., \& {Quinn}, T.
  2001{\natexlab{b}}, \apj, 551, 294

\bibitem[{{Ibata} {et~al.}(2007){Ibata}, {Martin}, {Irwin}, {Chapman},
  {Ferguson}, {Lewis}, \& {McConnachie}}]{Ibata_etal07}
{Ibata}, R., {Martin}, N.~F., {Irwin}, M., {Chapman}, S., {Ferguson}, A.~M.~N.,
  {Lewis}, G.~F., \& {McConnachie}, A.~W. 2007, \apj, 671, 1591

\bibitem[{{Ibata} {et~al.}(1994){Ibata}, {Gilmore}, \& {Irwin}}]{Ibata_etal94}
{Ibata}, R.~A., {Gilmore}, G., \& {Irwin}, M.~J. 1994, \nat, 370, 194

\bibitem[{{Ivezi{\'c}} {et~al.}(2008)}]{Ivezic_etal08}
{Ivezi{\'c}}, {\v Z}. {et~al.} 2008, \apj, 684, 287

\bibitem[{{Jog}(1997)}]{Jog97}
{Jog}, C.~J. 1997, \apj, 488, 642

\bibitem[{{Jog}(1999)}]{Jog99}
---. 1999, \apj, 522, 661

\bibitem[{{Juri{\'c}} {et~al.}(2008)}]{Juric_etal08}
{Juri{\'c}}, M. {et~al.} 2008, \apj, 673, 864

\bibitem[{{Kalirai} {et~al.}(2006){Kalirai}, {Guhathakurta}, {Gilbert},
  {Reitzel}, {Majewski}, {Rich}, \& {Cooper}}]{Kalirai_etal06}
{Kalirai}, J.~S., {Guhathakurta}, P., {Gilbert}, K.~M., {Reitzel}, D.~B.,
  {Majewski}, S.~R., {Rich}, R.~M., \& {Cooper}, M.~C. 2006, \apj, 641, 268

\bibitem[{{Kautsch} {et~al.}(2006){Kautsch}, {Grebel}, {Barazza}, \&
  {Gallagher}}]{Kautsch_etal06}
{Kautsch}, S.~J., {Grebel}, E.~K., {Barazza}, F.~D., \& {Gallagher}, III, J.~S.
  2006, \aap, 445, 765

\bibitem[{{Kazantzidis} {et~al.}(2008){Kazantzidis}, {Bullock}, {Zentner},
  {Kravtsov}, \& {Moustakas}}]{Kazantzidis_etal08}
{Kazantzidis}, S., {Bullock}, J.~S., {Zentner}, A.~R., {Kravtsov}, A.~V., \&
  {Moustakas}, L.~A. 2008, \apj, 688, 254

\bibitem[{{Kazantzidis} {et~al.}(2004{\natexlab{a}}){Kazantzidis}, {Magorrian},
  \& {Moore}}]{Kazantzidis_etal04a}
{Kazantzidis}, S., {Magorrian}, J., \& {Moore}, B. 2004{\natexlab{a}}, \apj,
  601, 37

\bibitem[{{Kazantzidis} {et~al.}(2004{\natexlab{b}}){Kazantzidis}, {Mayer},
  {Mastropietro}, {Diemand}, {Stadel}, \& {Moore}}]{Kazantzidis_etal04b}
{Kazantzidis}, S., {Mayer}, L., {Mastropietro}, C., {Diemand}, J., {Stadel},
  J., \& {Moore}, B. 2004{\natexlab{b}}, \apj, 608, 663

\bibitem[{{Kazantzidis} {et~al.}(2006){Kazantzidis}, {Zentner}, \&
  {Nagai}}]{Kazantzidis_etal06}
{Kazantzidis}, S., {Zentner}, A.~R., \& {Nagai}, D. 2006, in EAS Publications
  Series, ed. G.~A. {Mamon}, F.~{Combes}, C.~{Deffayet}, \& B.~{Fort}, 65

\bibitem[{{Kent} {et~al.}(1991){Kent}, {Dame}, \& {Fazio}}]{Kent_etal91}
{Kent}, S.~M., {Dame}, T.~M., \& {Fazio}, G. 1991, \apj, 378, 131

\bibitem[{{Klypin} {et~al.}(1999{\natexlab{a}}){Klypin}, {Gottl{\"o}ber},
  {Kravtsov}, \& {Khokhlov}}]{Klypin_etal99a}
{Klypin}, A., {Gottl{\"o}ber}, S., {Kravtsov}, A.~V., \& {Khokhlov}, A.~M.
  1999{\natexlab{a}}, \apj, 516, 530

\bibitem[{{Klypin} {et~al.}(1999{\natexlab{b}}){Klypin}, {Kravtsov},
  {Valenzuela}, \& {Prada}}]{Klypin_etal99b}
{Klypin}, A., {Kravtsov}, A.~V., {Valenzuela}, O., \& {Prada}, F.
  1999{\natexlab{b}}, \apj, 522, 82

\bibitem[{{Klypin} {et~al.}(2001){Klypin}, {Kravtsov}, {Bullock}, \&
  {Primack}}]{Klypin_etal01}
{Klypin}, A.~A., {Kravtsov}, A.~V., {Bullock}, J.~S., \& {Primack}, J.~R. 2001,
  \apj, 554, 903

\bibitem[{{Kravtsov}(1999)}]{Kravtsov99}
{Kravtsov}, A.~V. 1999, PhD thesis, New Mexico State University

\bibitem[{{Kravtsov} {et~al.}(2004){Kravtsov}, {Gnedin}, \&
  {Klypin}}]{Kravtsov_etal04}
{Kravtsov}, A.~V., {Gnedin}, O.~Y., \& {Klypin}, A.~A. 2004, \apj, 609, 482

\bibitem[{{Kravtsov} {et~al.}(1998){Kravtsov}, {Klypin}, {Bullock}, \&
  {Primack}}]{Kravtsov_etal98}
{Kravtsov}, A.~V., {Klypin}, A.~A., {Bullock}, J.~S., \& {Primack}, J.~R. 1998,
  \apj, 502, 48

\bibitem[{{Kravtsov} {et~al.}(1997){Kravtsov}, {Klypin}, \&
  {Khokhlov}}]{Kravtsov_etal97}
{Kravtsov}, A.~V., {Klypin}, A.~A., \& {Khokhlov}, A.~M. 1997, \apjs, 111, 73

\bibitem[{{Kregel} {et~al.}(2002){Kregel}, {van der Kruit}, \& {de
  Grijs}}]{Kregel_etal02}
{Kregel}, M., {van der Kruit}, P.~C., \& {de Grijs}, R. 2002, \mnras, 334, 646

\bibitem[{{Kuhlen} {et~al.}(2007){Kuhlen}, {Diemand}, \&
  {Madau}}]{Kuhlen_etal07}
{Kuhlen}, M., {Diemand}, J., \& {Madau}, P. 2007, \apj, 671, 1135

\bibitem[{{Lacey} \& {Cole}(1993)}]{Lacey_Cole93}
{Lacey}, C. \& {Cole}, S. 1993, \mnras, 262, 627

\bibitem[{{Larsen} \& {Humphreys}(2003)}]{Larsen_Humphreys03}
{Larsen}, J.~A. \& {Humphreys}, R.~M. 2003, \aj, 125, 1958

\bibitem[{{Levine} \& {Sparke}(1998)}]{Levine_Sparke98}
{Levine}, S.~E. \& {Sparke}, L.~S. 1998, \apjl, 496, L13

\bibitem[{{Libeskind} {et~al.}(2007){Libeskind}, {Cole}, {Frenk}, {Okamoto}, \&
  {Jenkins}}]{Libeskind_etal07}
{Libeskind}, N.~I., {Cole}, S., {Frenk}, C.~S., {Okamoto}, T., \& {Jenkins}, A.
  2007, \mnras, 374, 16

\bibitem[{{Libeskind} {et~al.}(2005){Libeskind}, {Frenk}, {Cole}, {Helly},
  {Jenkins}, {Navarro}, \& {Power}}]{Libeskind_etal05}
{Libeskind}, N.~I., {Frenk}, C.~S., {Cole}, S., {Helly}, J.~C., {Jenkins}, A.,
  {Navarro}, J.~F., \& {Power}, C. 2005, \mnras, 363, 146

\bibitem[{{L{\'o}pez-Corredoira} {et~al.}(2002){L{\'o}pez-Corredoira},
  {Cabrera-Lavers}, {Garz{\'o}n}, \& {Hammersley}}]{Lopez_etal02}
{L{\'o}pez-Corredoira}, M., {Cabrera-Lavers}, A., {Garz{\'o}n}, F., \&
  {Hammersley}, P.~L. 2002, \aap, 394, 883

\bibitem[{{Majewski}(1994)}]{Majewski94}
{Majewski}, S.~R. 1994, ApJL, 431, L17

\bibitem[{{Majewski} {et~al.}(2003){Majewski}, {Skrutskie}, {Weinberg}, \&
  {Ostheimer}}]{Majewski_etal03}
{Majewski}, S.~R., {Skrutskie}, M.~F., {Weinberg}, M.~D., \& {Ostheimer}, J.~C.
  2003, \apj, 599, 1082

\bibitem[{{Malin} \& {Hadley}(1997)}]{Malin_Hadley97}
{Malin}, D. \& {Hadley}, B. 1997, Publications of the Astronomical Society of
  Australia, 14, 52

\bibitem[{{Mapelli} {et~al.}(2008){Mapelli}, {Moore}, \&
  {Bland-Hawthorn}}]{Mapelli_etal08}
{Mapelli}, M., {Moore}, B., \& {Bland-Hawthorn}, J. 2008, \mnras, 388, 697

\bibitem[{{Martin} {et~al.}(2004){Martin}, {Ibata}, {Bellazzini}, {Irwin},
  {Lewis}, \& {Dehnen}}]{Martin_etal04}
{Martin}, N.~F., {Ibata}, R.~A., {Bellazzini}, M., {Irwin}, M.~J., {Lewis},
  G.~F., \& {Dehnen}, W. 2004, \mnras, 348, 12

\bibitem[{{Mart{\'{\i}}nez-Delgado} {et~al.}(2005){Mart{\'{\i}}nez-Delgado},
  {Butler}, {Rix}, {Franco}, {Pe{\~n}arrubia}, {Alfaro}, \&
  {Dinescu}}]{Martinez-Delgado_etal05}
{Mart{\'{\i}}nez-Delgado}, D., {Butler}, D.~J., {Rix}, H.-W., {Franco}, V.~I.,
  {Pe{\~n}arrubia}, J., {Alfaro}, E.~J., \& {Dinescu}, D.~I. 2005, \apj, 633,
  205

\bibitem[{{Matthews}(2000)}]{Matthews00}
{Matthews}, L.~D. 2000, \aj, 120, 1764

\bibitem[{{Matthews} {et~al.}(1998){Matthews}, {van Driel}, \&
  {Gallagher}}]{Matthews_etal98}
{Matthews}, L.~D., {van Driel}, W., \& {Gallagher}, III, J.~S. 1998, \aj, 116,
  1169

\bibitem[{{Matthews} \& {Wood}(2003)}]{Matthews_Wood03}
{Matthews}, L.~D. \& {Wood}, K. 2003, \apj, 593, 721

\bibitem[{{Mayer} {et~al.}(2008){Mayer}, {Governato}, \&
  {Kaufmann}}]{Mayer_etal08}
{Mayer}, L., {Governato}, F., \& {Kaufmann}, T. 2008, Invited Review in
  "Advanced Science Letters" (astro-ph/0801.3845)

\bibitem[{{Mayer} {et~al.}(2007){Mayer}, {Kazantzidis}, {Mastropietro}, \&
  {Wadsley}}]{Mayer_etal07}
{Mayer}, L., {Kazantzidis}, S., {Mastropietro}, C., \& {Wadsley}, J. 2007,
  \nat, 445, 738

\bibitem[{{Mendez} \& {Guzman}(1998)}]{Mendez_Guzman98}
{Mendez}, R.~A. \& {Guzman}, R. 1998, \aap, 333, 106

\bibitem[{{Merrifield}(1992)}]{Merrifield92}
{Merrifield}, M.~R. 1992, \aj, 103, 1552

\bibitem[{{Meza} {et~al.}(2005){Meza}, {Navarro}, {Abadi}, \&
  {Steinmetz}}]{Meza_etal05}
{Meza}, A., {Navarro}, J.~F., {Abadi}, M.~G., \& {Steinmetz}, M. 2005, \mnras,
  359, 93

\bibitem[{{Momany} {et~al.}(2006){Momany}, {Zaggia}, {Gilmore}, {Piotto},
  {Carraro}, {Bedin}, \& {de Angeli}}]{Momany_etal06}
{Momany}, Y., {Zaggia}, S., {Gilmore}, G., {Piotto}, G., {Carraro}, G.,
  {Bedin}, L.~R., \& {de Angeli}, F. 2006, \aap, 451, 515

\bibitem[{{Moore} {et~al.}(1999){Moore}, {Ghigna}, {Governato}, {Lake},
  {Quinn}, {Stadel}, \& {Tozzi}}]{Moore_etal99}
{Moore}, B., {Ghigna}, S., {Governato}, F., {Lake}, G., {Quinn}, T., {Stadel},
  J., \& {Tozzi}, P. 1999, \apjl, 524, L19

\bibitem[{{Moore} {et~al.}(1996){Moore}, {Katz}, \& {Lake}}]{Moore_etal96}
{Moore}, B., {Katz}, N., \& {Lake}, G. 1996, \apj, 457, 455

\bibitem[{{Moore} {et~al.}(2004){Moore}, {Kazantzidis}, {Diemand}, \&
  {Stadel}}]{Moore_etal04}
{Moore}, B., {Kazantzidis}, S., {Diemand}, J., \& {Stadel}, J. 2004, \mnras,
  354, 522

\bibitem[{{Murali} {et~al.}(2002){Murali}, {Katz}, {Hernquist}, {Weinberg}, \&
  {Dav{\'e}}}]{Murali_etal02}
{Murali}, C., {Katz}, N., {Hernquist}, L., {Weinberg}, D.~H., \& {Dav{\'e}}, R.
  2002, \apj, 571, 1

\bibitem[{{Naab} \& {Ostriker}(2006)}]{Naab_Ostriker06}
{Naab}, T. \& {Ostriker}, J.~P. 2006, \mnras, 366, 899

\bibitem[{{Nakanishi} \& {Sofue}(2003)}]{Nakanishi_Sofue03}
{Nakanishi}, H. \& {Sofue}, Y. 2003, \pasj, 55, 191

\bibitem[{{Narayan} \& {Jog}(2002)}]{Narayan_Jog02}
{Narayan}, C.~A. \& {Jog}, C.~J. 2002, \aap, 390, L35

\bibitem[{{Navarro} {et~al.}(1996){Navarro}, {Frenk}, \&
  {White}}]{Navarro_etal96}
{Navarro}, J.~F., {Frenk}, C.~S., \& {White}, S.~D.~M. 1996, \apj, 462, 563

\bibitem[{{Nelson} \& {Tremaine}(1995)}]{Nelson_Tremaine95}
{Nelson}, R.~W. \& {Tremaine}, S. 1995, \mnras, 275, 897

\bibitem[{{Newberg} {et~al.}(2002)}]{Newberg_etal02}
{Newberg}, H.~J. {et~al.} 2002, \apj, 569, 245

\bibitem[{{Nordstr{\"o}m} {et~al.}(2004)}]{Nordstrom_etal04}
{Nordstr{\"o}m}, B. {et~al.} 2004, \aap, 418, 989

\bibitem[{{Olling}(1996)}]{0lling96}
{Olling}, R.~P. 1996, \aj, 112, 457

\bibitem[{{Pe{\~n}arrubia} {et~al.}(2002){Pe{\~n}arrubia}, {Kroupa}, \&
  {Boily}}]{Penarrubia_etal02}
{Pe{\~n}arrubia}, J., {Kroupa}, P., \& {Boily}, C.~M. 2002, \mnras, 333, 779

\bibitem[{{Peng} {et~al.}(2002){Peng}, {Ford}, {Freeman}, \&
  {White}}]{Peng_etal02}
{Peng}, E.~W., {Ford}, H.~C., {Freeman}, K.~C., \& {White}, R.~L. 2002, \aj,
  124, 3144

\bibitem[{{Pohlen} {et~al.}(2004){Pohlen}, {Mart{\'{\i}}nez-Delgado},
  {Majewski}, {Palma}, {Prada}, \& {Balcells}}]{Pohlen_etal04}
{Pohlen}, M., {Mart{\'{\i}}nez-Delgado}, D., {Majewski}, S., {Palma}, C.,
  {Prada}, F., \& {Balcells}, M. 2004, in Astronomical Society of the Pacific
  Conference Series, Vol. 327, Satellites and Tidal Streams, ed. F.~{Prada},
  D.~{Martinez Delgado}, \& T.~J. {Mahoney}, 288

\bibitem[{{Pohlen} \& {Trujillo}(2006)}]{Pohlen_Trujillo06}
{Pohlen}, M. \& {Trujillo}, I. 2006, \aap, 454, 759

\bibitem[{{Pohlen} {et~al.}(2007){Pohlen}, {Zaroubi}, {Peletier}, \&
  {Dettmar}}]{Pohlen_etal07}
{Pohlen}, M., {Zaroubi}, S., {Peletier}, R.~F., \& {Dettmar}, R.-J. 2007,
  \mnras, 378, 594

\bibitem[{{Prada} {et~al.}(2006){Prada}, {Klypin}, {Simonneau},
  {Betancort-Rijo}, {Patiri}, {Gottl{\"o}ber}, \&
  {Sanchez-Conde}}]{Prada_etal06}
{Prada}, F., {Klypin}, A.~A., {Simonneau}, E., {Betancort-Rijo}, J., {Patiri},
  S., {Gottl{\"o}ber}, S., \& {Sanchez-Conde}, M.~A. 2006, \apj, 645, 1001

\bibitem[{{Prochaska} {et~al.}(2000){Prochaska}, {Naumov}, {Carney},
  {McWilliam}, \& {Wolfe}}]{Prochaska_etal00}
{Prochaska}, J.~X., {Naumov}, S.~O., {Carney}, B.~W., {McWilliam}, A., \&
  {Wolfe}, A.~M. 2000, \aj, 120, 2513

\bibitem[{{Purcell} {et~al.}(2007){Purcell}, {Bullock}, \&
  {Zentner}}]{Purcell_etal07}
{Purcell}, C.~W., {Bullock}, J.~S., \& {Zentner}, A.~R. 2007, \apj, 666, 20

\bibitem[{{Purcell} {et~al.}(2008){Purcell}, {Kazantzidis}, \&
  {Bullock}}]{Purcell_etal08}
{Purcell}, C.~W., {Kazantzidis}, S., \& {Bullock}, J.~S. 2008, ApJL submitted
  (astro-ph/0810.2785)

\bibitem[{{Quillen} \& {Garnett}(2000)}]{Quillen_Garnet01}
{Quillen}, A.~C. \& {Garnett}, D.~R. 2000, ArXiv Astrophysics e-prints

\bibitem[{{Quinn} \& {Goodman}(1986)}]{Quinn_Goodman86}
{Quinn}, P.~J. \& {Goodman}, J. 1986, \apj, 309, 472

\bibitem[{{Quinn} {et~al.}(1993){Quinn}, {Hernquist}, \&
  {Fullagar}}]{Quinn_etal93}
{Quinn}, P.~J., {Hernquist}, L., \& {Fullagar}, D.~P. 1993, \apj, 403, 74

\bibitem[{{Read} {et~al.}(2008){Read}, {Lake}, {Agertz}, \&
  {Debattista}}]{Read_etal08}
{Read}, J.~I., {Lake}, G., {Agertz}, O., \& {Debattista}, V.~P. 2008, \mnras,
  389, 1041

\bibitem[{{Reichard} {et~al.}(2008){Reichard}, {Heckman}, {Rudnick},
  {Brinchmann}, \& {Kauffmann}}]{Reichard_etal08}
{Reichard}, T.~A., {Heckman}, T.~M., {Rudnick}, G., {Brinchmann}, J., \&
  {Kauffmann}, G. 2008, \apj, 677, 186

\bibitem[{{Reid} \& {Majewski}(1993)}]{Reid_Majewski93}
{Reid}, N. \& {Majewski}, S.~R. 1993, \apj, 409, 635

\bibitem[{{Richter} \& {Sancisi}(1994)}]{Richter_Sancisi94}
{Richter}, O.-G. \& {Sancisi}, R. 1994, \aap, 290, L9

\bibitem[{{Rix} \& {Zaritsky}(1995)}]{Rix_Zaritsky95}
{Rix}, H.-W. \& {Zaritsky}, D. 1995, \apj, 447, 82

\bibitem[{{Robertson} {et~al.}(2004){Robertson}, {Yoshida}, {Springel}, \&
  {Hernquist}}]{Robertson_etal04}
{Robertson}, B., {Yoshida}, N., {Springel}, V., \& {Hernquist}, L. 2004, \apj,
  606, 32

\bibitem[{{Robin} {et~al.}(1996){Robin}, {Haywood}, {Creze}, {Ojha}, \&
  {Bienayme}}]{Robin_etal96}
{Robin}, A.~C., {Haywood}, M., {Creze}, M., {Ojha}, D.~K., \& {Bienayme}, O.
  1996, \aap, 305, 125

\bibitem[{{Rudnick} \& {Rix}(1998)}]{Rudnick_Rix98}
{Rudnick}, G. \& {Rix}, H.-W. 1998, \aj, 116, 1163

\bibitem[{{Saha} {et~al.}(2007){Saha}, {Combes}, \& {Jog}}]{Saha_etal07}
{Saha}, K., {Combes}, F., \& {Jog}, C.~J. 2007, \mnras, 382, 419

\bibitem[{{Schommer} {et~al.}(1992){Schommer}, {Suntzeff}, {Olszewski}, \&
  {Harris}}]{Schommer_etal92}
{Schommer}, R.~A., {Suntzeff}, N.~B., {Olszewski}, E.~W., \& {Harris}, H.~C.
  1992, \aj, 103, 447

\bibitem[{{Seabroke} \& {Gilmore}(2007)}]{Seabroke_Gilmore07}
{Seabroke}, G.~M. \& {Gilmore}, G. 2007, \mnras, 380, 1348

\bibitem[{{Sellwood} \& {Merritt}(1994)}]{Sellwood_Merritt94}
{Sellwood}, J.~A. \& {Merritt}, D. 1994, \apj, 425, 530

\bibitem[{{Sellwood} {et~al.}(1998){Sellwood}, {Nelson}, \&
  {Tremaine}}]{Sellwood_etal98}
{Sellwood}, J.~A., {Nelson}, R.~W., \& {Tremaine}, S. 1998, \apj, 506, 590

\bibitem[{{Sellwood} \& {Valluri}(1997)}]{Sellwood_Valluri97}
{Sellwood}, J.~A. \& {Valluri}, M. 1997, \mnras, 287, 124

\bibitem[{{Seth} {et~al.}(2005){Seth}, {Dalcanton}, \& {de Jong}}]{Seth_etal05}
{Seth}, A.~C., {Dalcanton}, J.~J., \& {de Jong}, R.~S. 2005, \aj, 130, 1574

\bibitem[{{Shang} {et~al.}(1998)}]{Shang_etal98}
{Shang}, E. {et~al.} 1998, \apjl, 504, L23

\bibitem[{{Shen} \& {Sellwood}(2006)}]{Shen_Sellwood06}
{Shen}, J. \& {Sellwood}, J.~A. 2006, \mnras, 370, 2

\bibitem[{{Sommer-Larsen} {et~al.}(2003){Sommer-Larsen}, {G{\"o}tz}, \&
  {Portinari}}]{Sommer_Larsen_etal03}
{Sommer-Larsen}, J., {G{\"o}tz}, M., \& {Portinari}, L. 2003, \apj, 596, 47

\bibitem[{{Soubiran} {et~al.}(2003){Soubiran}, {Bienaym{\'e}}, \&
  {Siebert}}]{Soubiran_etal03}
{Soubiran}, C., {Bienaym{\'e}}, O., \& {Siebert}, A. 2003, \aap, 398, 141

\bibitem[{{Spitzer}(1958)}]{Spitzer58}
{Spitzer}, L.~J. 1958, \apj, 127, 17

\bibitem[{{Stadel}(2001)}]{Stadel01}
{Stadel}, J.~G. 2001, Ph.D.~Thesis, Univ. of Washington

\bibitem[{{Statler}(1988)}]{Statler88}
{Statler}, T.~S. 1988, \apj, 331, 71

\bibitem[{{Stewart} {et~al.}(2009){Stewart}, {Bullock}, {Wechsler}, \&
  {Maller}}]{Stewart_etal09}
{Stewart}, K.~R., {Bullock}, J.~S., {Wechsler}, R.~H., \& {Maller}, A.~H. 2009,
  ApJ submitted (astro-ph/0901.4336)

\bibitem[{{Stewart} {et~al.}(2008){Stewart}, {Bullock}, {Wechsler}, {Maller},
  \& {Zentner}}]{Stewart_etal08}
{Stewart}, K.~R., {Bullock}, J.~S., {Wechsler}, R.~H., {Maller}, A.~H., \&
  {Zentner}, A.~R. 2008, \apj, 683, 597

\bibitem[{{Syer} \& {Tremaine}(1996)}]{Syer_Tremaine96}
{Syer}, D. \& {Tremaine}, S. 1996, \mnras, 281, 925

\bibitem[{{Tormen} {et~al.}(1998){Tormen}, {Diaferio}, \&
  {Syer}}]{Tormen_etal98}
{Tormen}, G., {Diaferio}, A., \& {Syer}, D. 1998, \mnras, 299, 728

\bibitem[{{Toth} \& {Ostriker}(1992)}]{Toth_Ostriker92}
{Toth}, G. \& {Ostriker}, J.~P. 1992, \apj, 389, 5

\bibitem[{{Velazquez} \& {White}(1999)}]{Velazquez_White99}
{Velazquez}, H. \& {White}, S.~D.~M. 1999, \mnras, 304, 254

\bibitem[{{Villalobos} \& {Helmi}(2008)}]{Villalobos_Helmi08}
{Villalobos}, {\'A}. \& {Helmi}, A. 2008, \mnras, 391, 1806

\bibitem[{{Wainscoat} {et~al.}(1989){Wainscoat}, {Freeman}, \&
  {Hyland}}]{Wainscoat_etal89}
{Wainscoat}, R.~J., {Freeman}, K.~C., \& {Hyland}, A.~R. 1989, \apj, 337, 163

\bibitem[{{Walker} {et~al.}(1996){Walker}, {Mihos}, \&
  {Hernquist}}]{Walker_etal96}
{Walker}, I.~R., {Mihos}, J.~C., \& {Hernquist}, L. 1996, \apj, 460, 121

\bibitem[{{Wechsler} {et~al.}(2002){Wechsler}, {Bullock}, {Primack},
  {Kravtsov}, \& {Dekel}}]{Wechsler_etal02}
{Wechsler}, R.~H., {Bullock}, J.~S., {Primack}, J.~R., {Kravtsov}, A.~V., \&
  {Dekel}, A. 2002, \apj, 568, 52

\bibitem[{{Weil} {et~al.}(1998){Weil}, {Eke}, \& {Efstathiou}}]{Weil_etal98}
{Weil}, M.~L., {Eke}, V.~R., \& {Efstathiou}, G. 1998, \mnras, 300, 773

\bibitem[{{Weinberg}(1989)}]{Weinberg98}
{Weinberg}, M.~D. 1989, \mnras, 239, 549

\bibitem[{{Weinberg}(1991)}]{Weinberg91}
---. 1991, \apj, 373, 391

\bibitem[{{Weinberg} \& {Blitz}(2006)}]{Weinberg_Blitz06}
{Weinberg}, M.~D. \& {Blitz}, L. 2006, \apjl, 641, L33

\bibitem[{{Weinmann} {et~al.}(2006){Weinmann}, {van den Bosch}, {Yang}, \&
  {Mo}}]{Weinmann_etal06}
{Weinmann}, S.~M., {van den Bosch}, F.~C., {Yang}, X., \& {Mo}, H.~J. 2006,
  \mnras, 366, 2

\bibitem[{{White} \& {Rees}(1978)}]{White_Rees78}
{White}, S.~D.~M. \& {Rees}, M.~J. 1978, \mnras, 183, 341

\bibitem[{{Widrow} \& {Dubinski}(2005)}]{Widrow_Dubinski05}
{Widrow}, L.~M. \& {Dubinski}, J. 2005, \apj, 631, 838

\bibitem[{{Wilcots} \& {Prescott}(2004)}]{Wilcots_Prescott04}
{Wilcots}, E.~M. \& {Prescott}, M.~K.~M. 2004, \aj, 127, 1900

\bibitem[{{Willman} {et~al.}(2004){Willman}, {Governato}, {Dalcanton}, {Reed},
  \& {Quinn}}]{Willman_etal04}
{Willman}, B., {Governato}, F., {Dalcanton}, J.~J., {Reed}, D., \& {Quinn}, T.
  2004, \mnras, 353, 639

\bibitem[{{Wyse}(2001)}]{Wyse01}
{Wyse}, R.~F.~G. 2001, in ASP Conf. Ser. 230: Galaxy Disks and Disk Galaxies,
  ed. J.~G. {Funes} \& E.~M. {Corsini}, 71--80

\bibitem[{{Wyse} \& {Gilmore}(1995)}]{Wyse_Gilmore95}
{Wyse}, R.~F.~G. \& {Gilmore}, G. 1995, \aj, 110, 2771

\bibitem[{{Yang} {et~al.}(2006){Yang}, {van den Bosch}, {Mo}, {Mao}, {Kang},
  {Weinmann}, {Guo}, \& {Jing}}]{Yang_etal06}
{Yang}, X., {van den Bosch}, F.~C., {Mo}, H.~J., {Mao}, S., {Kang}, X.,
  {Weinmann}, S.~M., {Guo}, Y., \& {Jing}, Y.~P. 2006, \mnras, 369, 1293

\bibitem[{{Yanny} {et~al.}(2000)}]{Yanny_etal00}
{Yanny}, B. {et~al.} 2000, \apj, 540, 825

\bibitem[{{Yoachim} \& {Dalcanton}(2005)}]{Yoachim_Dalcanton05}
{Yoachim}, P. \& {Dalcanton}, J.~J. 2005, \apj, 624, 701

\bibitem[{{Yoachim} \& {Dalcanton}(2006)}]{Yoachim_Dalcanton06}
---. 2006, \aj, 131, 226

\bibitem[{{Yoachim} \& {Dalcanton}(2008)}]{Yoachim_Dalcanton08}
---. 2008, \apj, 682, 1004

\bibitem[{{Younger} {et~al.}(2007){Younger}, {Cox}, {Seth}, \&
  {Hernquist}}]{Younger_etal07}
{Younger}, J.~D., {Cox}, T.~J., {Seth}, A.~C., \& {Hernquist}, L. 2007, \apj,
  670, 269

\bibitem[{{Zaritsky} \& {Rix}(1997)}]{Zaritsky_Rix97}
{Zaritsky}, D. \& {Rix}, H.-W. 1997, \apj, 477, 118

\bibitem[{{Zaritsky} {et~al.}(1997){Zaritsky}, {Smith}, {Frenk}, \&
  {White}}]{Zaritsky_etal97}
{Zaritsky}, D., {Smith}, R., {Frenk}, C.~S., \& {White}, S.~D.~M. 1997, \apjl,
  478, L53

\bibitem[{{Zentner} {et~al.}(2005{\natexlab{a}}){Zentner}, {Berlind},
  {Bullock}, {Kravtsov}, \& {Wechsler}}]{Zentner_etal05a}
{Zentner}, A.~R., {Berlind}, A.~A., {Bullock}, J.~S., {Kravtsov}, A.~V., \&
  {Wechsler}, R.~H. 2005{\natexlab{a}}, \apj, 624, 505

\bibitem[{{Zentner} \& {Bullock}(2003)}]{Zentner_Bullock03}
{Zentner}, A.~R. \& {Bullock}, J.~S. 2003, \apj, 598, 49

\bibitem[{{Zentner} {et~al.}(2005{\natexlab{b}}){Zentner}, {Kravtsov},
  {Gnedin}, \& {Klypin}}]{Zentner_etal05b}
{Zentner}, A.~R., {Kravtsov}, A.~V., {Gnedin}, O.~Y., \& {Klypin}, A.~A.
  2005{\natexlab{b}}, \apj, 629, 219

\bibitem[{{Zhao}(1996)}]{Zhao96}
{Zhao}, H. 1996, \mnras, 278, 488

\end{thebibliography}

\end{document}